\documentclass[11pt,a4paper]{article}
\pdfoutput=1

\usepackage[english]{babel} 
\usepackage[left=0.8in,right=0.8in,top=1.0in,bottom=1.2in]{geometry}

\usepackage[labelfont=bf]{caption}

\usepackage{dsfont}
\usepackage{amsfonts}
\usepackage{amsmath}
\usepackage{amssymb}
\usepackage{amsthm}
\usepackage{bbold}
\usepackage{calc}
\usepackage{cancel}
\usepackage{float}
\usepackage{slashed}
\usepackage{mathrsfs} 
\usepackage{pdfpages}
\usepackage{hyperref}
\usepackage{multirow}
\usepackage{mathtools}
\usepackage{braket}
\usepackage{cite}
\usepackage{tikz}
\usepackage{tikzsymbols}
\usepackage{array}
\usepackage{colortbl} 	
\definecolor{Gray}{gray}{0.95}
\definecolor{RGray}{gray}{0.90}
\definecolor{CGray}{gray}{0.92}
\usepackage{arydshln}
\usepackage{soul}
\usepackage{booktabs}

\numberwithin{equation}{section}
\numberwithin{figure}{section}
\numberwithin{table}{section}

\allowdisplaybreaks

\newcommand{\abs}[1]{\left| #1 \right|}

\newcommand{\be}{\begin{equation}}
\newcommand{\ee}{\end{equation}}
\newcommand{\bea}{\begin{eqnarray}}
\newcommand{\eea}{\end{eqnarray}}  
\newcommand{\no}{\nonumber}
\newcommand{\gsim}{\lower.7ex\hbox{$\;\stackrel{\textstyle>}{\sim}\;$}}
\newcommand{\lsim}{\lower.7ex\hbox{$\;\stackrel{\textstyle<}{\sim}\;$}}
\newcommand{\cO}{{\mathcal O}}
\newcommand{\cC}{{\mathcal C}}
\newcommand{\cL}{{\mathcal L}}

\newcommand{\cB}{{\mathcal B}}
\newcommand{\bsll}{b\to s\hspace{0.3mm}\ell^+\ell^-}
\newcommand{\bctnu}{b\to c \tau \bar \nu}
\newcommand{\bculnu}{b\to c(u) \tau \bar \nu}
\newcommand{\cLL}{{\mathcal C}_{LL}}
\newcommand{\cLR}{{\mathcal C}_{LR}}
\newcommand{\cRR}{{\mathcal C}_{RR}}
\newcommand{\betaR}{\beta_{R}^{b \tau}}

\makeatletter
\g@addto@macro\bfseries{\boldmath}
\makeatother

\begin{document}

\begin{flushright}
MITP-21-016\\
ZU-TH-13/21\\
\end{flushright}

\begin{center}
\vspace{0.7cm}
{\Large\bf Reading the footprints of the $B$-meson flavor anomalies}\\[1.0cm] 
{Claudia Cornella$^a$, Darius A.~Faroughy$^a$, Javier Fuentes-Mart\'{\i}n$^b$,\\
Gino Isidori$^a$ and Matthias Neubert$^{a,b,c}$}\\[0.2cm]
{\em ${}^a$Physik-Institut, Universit\"at Z\"urich, CH-8057 Z\"urich, Switzerland\\
${}^b$PRISMA$^+$\! Cluster of Excellence {\em \&} MITP, 
Johannes Gutenberg University, 55099 Mainz, Germany\\
${}^c$Department of Physics {\em \&} LEPP, 
Cornell University, Ithaca, NY 14853, U.S.A.}
\end{center}
\vspace{0.5 cm}

\centerline{\large\bf Abstract}
\begin{quote}
Motivated by the recent LHCb announcement of a $3.1\sigma$ violation of lepton-flavor universality in the ratio $R_K=\Gamma(B\to K\mu^+\mu^-)/\Gamma(B\to K e^+ e^-)$, we present an updated, comprehensive analysis of the flavor anomalies seen in both neutral-current ($\bsll$) and charged-current ($\bctnu$) decays of $B$ mesons. Our study starts from a model-independent effective field-theory approach and then considers both a simplified model and a UV-complete extension of the Standard Model featuring a vector leptoquark $U_1$ as the main mediator of the anomalies. We show that the new LHCb data corroborate the emerging pattern of a new, predominantly left-handed, semileptonic current-current interaction with a flavor structure respecting a (minimally) broken $U(2)^5$ flavor symmetry. New aspects of our analysis include a combined analysis of the semileptonic operators involving tau leptons, including in particular the important constraint from $B_s$--$\bar B_s$ mixing, a systematic study of the effects of right-handed leptoquark couplings and of deviations from minimal  flavor-symmetry breaking, a
detailed analysis of various rare $B$-decay modes which would provide smoking-gun signatures of this non-standard framework (LFV decays, di-tau modes, and 
$B\to K^{(*)}\nu\bar\nu$), and finally an updated analysis of collider bounds on the leptoquark mass and couplings.
\end{quote}
\thispagestyle{empty}

\newpage
\thispagestyle{empty}
\setcounter{page}{0}

\tableofcontents

\newpage

\section{Introduction}
 
Since 2012, several deviations from the predictions of the Standard Model (SM) have been observed in a series of semileptonic decays of $B$ mesons, providing strong evidence 
of a new short-distance interaction violating Lepton Flavor Universality (LFU).
The recent result by the LHCb collaboration on the ratio $R_K=\cB(B^+\to K^+\mu^+\mu^-)/\cB(B^+\to K^+e^+e^-)$~\cite{Aaij:2021vac} marks an important milestone 
in the study of these phenomena, which in the literature are often referred to as the ``$B$ anomalies''.
The evidence collected so far can naturally be grouped into two categories,
according to the underlying partonic process: $i$)~deviations from $\mu/e$ universality 
in $b\to s \ell^+ \ell^-$ neutral-current  transitions~\cite{Aaij:2014ora,Aaij:2017vbb,Aaij:2019wad} together with deviations from the SM predictions 
in observables involving $\mu^+\mu^-$ pairs only~\cite{Aaij:2013qta,Aaij:2015oid,CMS:2014xfa}, and $ii$)~deviations from 
$\tau/\mu$ (and $\tau/e$) universality in  $b\to c \ell \bar\nu$
charged-current transitions~\cite{Lees:2012xj,Lees:2013uzd,Huschle:2015rga,Aaij:2015yra,Hirose:2016wfn,Aaij:2017deq,Aaij:2017uff}.
With the recent result reported in~\cite{Aaij:2021vac}, for the first time a single observable affected by negligible theoretical uncertainties  
exhibits a deviation from the SM exceeding the $3\sigma$ level.
Equally striking is the overall coherence of the picture that emerges, especially in 
$\bsll$ transitions. As we shall show in this paper, combining all the 
$\bsll$ observables in a very conservative way, the significance of the 
New Physics (NP) hypothesis formulated in 2014--2015 of a purely left-handed 
LFU-violating contact interaction has now reached a significance  of~4.6$\sigma$.

Attempts to explain one or both sets of anomalies have stimulated an intense theoretical activity, 
which ranges from pure Effective Field Theory (EFT) approaches to the formulation of motivated ultraviolet (UV) completions 
of the SM. In the latter category, models containing a TeV-scale vector leptoquark,  
$U_1\sim(\mathbf{3},\mathbf{1},2/3)$, as the main mediator are particularly appealing. Besides addressing both sets of anomalies, such models can connect them to an underlying theory of flavor. 
The purpose of this paper is to reanalyze this class of models, taking into account the new data and, importantly, 
distinguishing the robust predictions of the models from those sensitive to the details of the UV completion, which are still largely unknown.
The goal is to identify a set of observables which could allow us to validate or disprove these models in the future, 
possibly answering the questions that are still open.

Before introducing the class of models we are interested in, it is worth recalling the key steps that led 
to their formulation. 
The first important observation was the identification of a purely left-handed contact interaction involving 
muons as the most natural candidate 
to explain the $\bsll$ data available in 2014~\cite{Hiller:2014yaa}.
The link between $R_K$, originally proposed in~\cite{Bobeth:2007dw} as a clean NP probe, 
and the various anomalies in $b\to s \mu^+ \mu^-$ observables, and in particular the non-standard behavior of the 
angular distribution in $B \to K^* \mu^+\mu^-$~\cite{Descotes-Genon:2013wba,Altmannshofer:2013foa,Hurth:2013ssa}, was then established more firmly by a series of global analyses 
(see e.g.~\cite{Hurth:2014vma,Altmannshofer:2014rta,Descotes-Genon:2015uva,Ciuchini:2017mik,Alguero:2019ptt,Ciuchini:2020gvn}).
A second important step was the hypothesis that such a weak contact interaction could be the result 
of a stronger interaction, which violates lepton flavor and involves mainly the third-generation fermions~\cite{Glashow:2014iga}.  
Soon after, it was realized that the two sets of anomalies (in $b\to s$ and $b\to c$ transitions) can be linked in an EFT approach involving mainly left-handed fields~\cite{Bhattacharya:2014wla,Alonso:2015sja,Greljo:2015mma,Calibbi:2015kma}.
Meanwhile, models involving leptoquarks started to emerge as the most promising candidates for a UV completion of these EFTs, 
both in the case of  $\bsll$ transitions~\cite{Hiller:2014yaa,Gripaios:2014tna} 
-- for which the possibility of a leptoquark-induced effect in motivated NP models was proposed even prior to the existence of the anomalies~\cite{Gripaios:2009dq} -- and especially in the case of combined 
 explanations~\cite{Alonso:2015sja,Fajfer:2015ycq,Calibbi:2015kma,Bauer:2015knc,Barbieri:2015yvd,Becirevic:2016yqi,Becirevic:2018afm}, which necessarily require a lower NP scale.  
 The reason is simple: leptoquarks
 can contribute at the tree-level to the semileptonic transitions exhibiting anomalies, while they contribute 
 only at the loop level to four-quark or four-lepton contact interactions, which so far do not exhibit significant deviations from the SM. 
It also emerged more clearly that the flavor structure of the new interaction is quite constrained in the case of combined 
explanations, and naturally follows the hypothesis of a minimally broken $U(2)^5$ flavor symmetry~\cite{Greljo:2015mma, Barbieri:2015yvd} -- a hypothesis formulated well before the anomalies appeared~\cite{Barbieri:2011ci}, which links them 
to the origin of the hierarchies observed in the SM Yukawa couplings.  The simplified model proposed in~\cite{Barbieri:2015yvd} at the end of 2015 already contained the two key  
ingredients of the UV models we are interested in, since it was
based on a TeV-scale $U_1$ mediator and exhibited a $U(2)^5$ flavor symmetry. At that time other options, such as colorless mediators, were still open, 
but it was soon realized that electroweak precision observables and LFU tests in $\tau$ decays~\cite{Feruglio:2016gvd,Feruglio:2017rjo}, 
 collider bounds~\cite{Faroughy:2016osc,Greljo:2017vvb}, and other flavor observables~\cite{Calibbi:2015kma,Buttazzo:2017ixm} imply very strong constraints.
The detailed EFT analysis presented in~\cite{Buttazzo:2017ixm} has shown that the $U_1$ case is the simplest option 
at the level of simplified models.

A massive vector field necessarily requires a UV completion and, as pointed out in~\cite{Barbieri:2015yvd},  
the $U_1$ field naturally points toward the $SU(4)$ group unifying quarks and leptons, which was proposed by Pati and Salam in 1974~\cite{Pati:1974yy}.
The $SU(4)$ group could either be realized as a global symmetry of some new strongly-interacting sector~\cite{Barbieri:2016las,Barbieri:2017tuq} or be part of a local symmetry broken above the electroweak scale to the SM gauge group.  The original Pati-Salam model offers a very elegant
possibility, but it does not match the flavor structure required to explain the anomalies, because the massive $U_1$ (and associated $Z^\prime$) arising from the spontaneous symmetry breaking 
${\rm PS} \to {\rm SM}$ is flavor blind. An interesting proposal to overcome this problem has been put forward 
in \cite{DiLuzio:2017vat,Diaz:2017lit}, following the idea laid out in  \cite{Georgi:2016xhm} that color could appear as a diagonal 
subgroup of a larger $SU(3 + N)\times SU(3)$ local symmetry valid at high energies. In the model of \cite{DiLuzio:2017vat}, 
a suitable mixing between the SM fermions and new heavy fermions allows one to adjust the effective $U_1$ couplings so as to obtain the desired flavor structure.
Alternative UV models based on variations of  the original Pati-Salam model have been proposed in~\cite{Blanke:2018sro,Fornal:2018dqn}.
A more structural  way of addressing the flavor structure of the model is the idea of implementing Pati-Salam unification 
in a flavor non-universal manner, originally proposed in~\cite{Bordone:2017bld} and further developed 
in~\cite{Greljo:2018tuh,Fuentes-Martin:2020bnh,Fuentes-Martin:2020pww}.
This proposal is quite appealing, since third-family quark-lepton unification close to the electroweak scale is phenomenologically allowed,
and the setup naturally accommodates an accidental $U(2)^5$ global flavor symmetry at the TeV scale~\cite{Bordone:2017bld}.
Remarkably, this setup can also provide a coherent description of neutrino masses~\cite{Fuentes-Martin:2020pww}.
Moreover, its rich field content and the three-scale structure behind the origin of the flavor hierarchies finds a simple interpretation 
in terms of a compact extra dimension of spacetime, with the flavor index of the four-dimensional fields being in one-to-one correspondence with the location of four-dimensional branes along the fifth dimension. 
 
There are different ways of implementing Pati-Salam unification in a flavor non-universal manner. However, the low-energy dynamics of this class of models is characterized by a flavor non-universal $SU(4) \times SU(3) \times SU(2)\times U(1)$ gauge symmetry (the ``4321 group''), 
where third-generation fermions are charged under $SU(4)$, the light SM fermions are charged under $SU(3)$, 
and the $U_1$ field acquires mass  from the symmetry breaking $SU(4) \times SU(3)  \times SU(2)\times U(1) \to {\rm SM}$ occurring near the TeV scale.
This is the class of UV completions we are most interested in.  However, as we shall discuss, many of the phenomenological 
conclusions we can derive from the present data are valid for the wider class of models where the  $U_1$ is the main mediator of the anomalies,
including in particular the well motivated case of models of compositeness.  The gauge-theory extension offers the advantage of allowing for a systematic estimate of quantum corrections~\cite{Fuentes-Martin:2019ign,Fuentes-Martin:2020luw,Fuentes-Martin:2020hvc}. We will thus refer specifically to this option when discussing UV-sensitive observables.

Following the bottom-up approach that led to the formulation of these models, the analysis presented in this paper is structured as follows. 
In Section~\ref{sect:EFT} we present a model-independent EFT analysis of the $B$ anomalies. The $U_1$ hypothesis enters only indirectly, via the selection of the relevant set of semileptonic operators at the TeV scale. In Section~\ref{sect:U1} we 
first present an analysis of the effective 
 $U_1$ couplings to SM fermions, taking into account all  relevant low-energy constraints (including also loop-induced  observables). We then confront the results of the fit to low-energy data with the collider constraints on the $U_1$ leptoquark.
 In Section~\ref{sect:UV} we present predictions for a series of low-energy observables, which in the future could allow us to 
 resolve those model-building aspects that are currently still open.
 To do so, we go beyond the simplified model, analyzing the effects of vector-like fermions. We further analyze high--$p_T$ constraints 
 on the heavy TeV-scale color-octet boson $G^\prime$, which necessarily accompanies the $U_1$ leptoquark in 4321 models. 
The results presented in this work represent a substantial step forward in the study of a combined solution 
of the $B$-meson flavor anomalies on all the three fronts: the model-independent EFT analysis, the analysis within a simplified model for a $U_1$ leptoquark, and the  phenomenology of the 4321 framework. In addition to important updates of the experimental inputs on the low-energy side,
 the main innovative points 
of the present analysis with respect to previous studies can be summarized as follows:
\begin{itemize}
\item{} In Section~\ref{sect:EFT}, we present for the first time a combined analysis of the semileptonic 
operators involving tau leptons, at the pure EFT level, taking into account both sets of flavor anomalies, 
collider constraints and $\Delta F=2$ bounds.
\item{} In Section~\ref{sect:U1}, we analyze and compare two benchmark scenarios for the right-handed couplings of the $U_1$ leptoquark to the third-generation fermions. We also relax and validate the hypothesis of minimal breaking of the $U(2)^5$ flavor symmetry for the subleading mixing terms involving first-generation quarks. Most importantly, we update the analysis of the high-energy constraints on the $U_1$ leptoquark by taking into account ATLAS and CMS results with full Run-II statistics. 
\item{} The analysis in Section~\ref{sect:UV} takes into account, for the first time, complete NLO corrections in the leptoquark coupling $\alpha_4$, which were computed in \cite{Fuentes-Martin:2019ign,Fuentes-Martin:2020luw,Fuentes-Martin:2020hvc} within the non-universal 4321 model. The latter play a key role in the predictions we obtain for $B$-meson mixing and the rare decay $B\to K\nu\bar\nu$. The $\cO(\alpha_4)$ corrections are also implemented for the first time in the analysis of the $G^\prime$ constraints from $pp\to$~dijet and $pp\to t\bar t$, which turn out to be the most relevant constraints on the overall mass scale of the model, once recent ATLAS and CMS results are taken into account. 
\end{itemize}

\section{EFT analysis of the \texorpdfstring{$B$}{B} anomalies}\label{sect:EFT}

\subsection{Operator basis and general flavor structure}
\label{sect:OPbasis}

The goal of this section is to provide a general analysis of the flavor anomalies in terms of semileptonic four-fermion operators.
We start by analyzing the two sets of anomalies separately and then discuss the consequences of a combined analysis within the EFT. 

Rather than considering all possible dimension-6 operators that can describe  a single set of measurements, 
we focus on the operators which have been identified in previous studies as the relevant set necessary 
for a combined explanation of both anomalies, once all constraints (including high-$p_T$ data, electroweak precision tests and other flavor observables) are taken into account~\cite{Buttazzo:2017ixm}. In practice, this set coincides with the operators generated at the tree-level by the exchange of a spin-1 $SU(2)_L$-singlet leptoquark $U_1$, i.e.\ (a contraction of color and $SU(2)_L$ indices between fermions inside parenthesis  is implied)
\begin{align}
\begin{aligned}
 \cO^{ij\alpha\beta}_{LL}  &= (\bar q_{L}^{\,i} \gamma_{\mu}  \ell_{L}^{\alpha}) (\bar \ell_{L}^{\beta} \gamma^{\mu}  q_{L}^{\,j} ) \!\!\!
	&=~& \frac{1}{2}  \left[ Q_{lq}^{(1)} + Q_{lq}^{(3)} \right]^{\beta\alpha ij}  ,  \\
 \cO^{ij\alpha\beta}_{LR} &=   (\bar q_{L}^{\,i} \gamma_{\mu}  \ell_{L}^{\alpha}) (\bar e_{R}^{\beta} \gamma^{\mu} d_{R}^{\,j} ) \!\!\!
	&=~& -  2\,\big[ Q^\dagger_{ledq}\big]^{\beta\alpha ij}   \,,  \\[1mm]
	 \cO^{ij\alpha\beta}_{RR} &= (\bar d_{R}^{\,i} \gamma_{\mu}  e_{R}^{\alpha}) (\bar e_{R}^{\beta} \gamma^{\mu}  d_{R}^{\,j}) \!\!\!
	&=~&   \big[ Q_{ed}\big]^{\beta\alpha ij}  \,,
\label{eq:opbasis}
\end{aligned}
\end{align}
where $Q_{lq}^{(1,3)}$, $Q_{ledq}$ and $Q_{ed}$ are defined as in the so-called Warsaw basis \cite{Grzadkowski:2010es}
of dimension-6 SMEFT operators built out of SM fields.
We normalize the effective Lagrangian describing the NP contributions as
\be
\label{eq:SMEFTLag}
\cL^{\rm NP}_{\rm EFT} = -\frac{2}{v^2}  \left[\cLL^{ij\alpha\beta}\, \cO^{ij\alpha\beta}_{LL}  +
\cRR^{ij\alpha\beta}\, \cO^{ij\alpha\beta}_{RR}  +
 \left(  \cLR^{ij\alpha\beta}\,
\cO^{ij\alpha\beta}_{LR}+{\rm h.c.} \right) \right] ,
\ee
where $v=(\sqrt{2}\,G_F)^{-1/2} \approx 246$~GeV, and the Wilson coefficients are inversely proportional to the square of the NP scale $\Lambda$. 
The coefficients of the hermitian operators $\cO_{LL}$ and $\cO_{RR}$ satisfy the relations
\begin{align}
\cLL^{ji\beta\alpha} &= \big( \cLL^{ij\alpha\beta} \big)^* \,,&
\cRR^{ji\beta\alpha} &= \big( \cRR^{ij\alpha\beta} \big)^* \,.    
\end{align}

We assume that these coefficients respect an approximate $U(2)^5$ flavor symmetry, 
with non-negligible breaking terms 
only in the left-handed 
quark and lepton sectors (see Appendix~\ref{app:U2}).
This assumption implies that the leading couplings in $\cL^{\rm NP}_{\rm EFT}$ are those with third-generation 
indices, while all other couplings 
are suppressed. More specifically, we make the following two assumptions:
\begin{itemize}
\item 
Wilson coefficients of operators containing first- or second-generation right-handed fields are negligibly small, i.e.\ $\cRR^{ij \alpha\beta} \approx 0$ unless $i=j=3$ and $\alpha=\beta=\tau$, and 
 $\cLR^{ij \alpha\beta} \approx 0$ unless $j=3$ and $\beta=\tau$.
\item 
Wilson coefficients associated with second-generation left-handed particles are suppressed (relative to those for third-generation particles) by factors of  $\epsilon_q,\epsilon_\ell\sim10^{-1}$ for each second-generation quark or lepton, e.g.~$\cLL^{23 \tau\tau} \sim \epsilon_q\,\cLL^{33 \tau\tau}$, 
 $\cLL^{23 \mu\mu} \sim \epsilon_q\, \epsilon_\ell^2\,\cLL^{33 \tau\tau}$ etc., and a further suppression arises in the case of operators involving first-generation fields.
\end{itemize}
As explicitly indicated by the labels, the flavor basis of the lepton fields is taken to be the charged-lepton mass basis.
The flavor basis of the quark fields is identified with the mass basis of the down-type quarks. However, we use numerical indices to stress that this choice is a 
model-dependent assumption (the implications of a possible small misalignment are briefly discussed in Section~\ref{sect:UV}). Note that a change of basis from 
down-type to up-type quarks would not invalidate the scaling discussed above, and would suggest that --at least in the quark sector-- first-generation indices bring an additional $\epsilon_q$ suppression compared to second-generation indices.
The flavor structure specified by the two assumptions stated above is the rationale behind the combined explanation of the two sets of anomalies and their possible connection to the 
dynamics underlying the structure of the SM Yukawa matrices. As we shall show, these scaling rules are clearly supported by the present data. 

As pointed out in~\cite{Fuentes-Martin:2019mun}, one can  obtain the same set of relevant operators and flavor structure by 
starting from the full set of SMEFT operators and imposing the assumption of a minimally-broken $U(2)^5$ flavor symmetry, without any hypothesis about the mediator. 
The only relevant difference under this more general hypothesis is that the operators $Q_{lq}^{(1)}$ and $Q_{lq}^{(3)}$  
can appear in a different linear combination than in (\ref{eq:opbasis}). An EFT analysis leaving their coefficients as free
parameters has been performed in~\cite{Buttazzo:2017ixm}, where it was  shown that 
the combination orthogonal to $\cO_{LL}$ 
is tightly constrained by data on $b\to s \bar \nu \nu$ transitions and electroweak precision tests (at least for the leading flavor structures).
Since this combination is not generated by $U_1$ tree-level exchange, and ignoring it does not lead to a qualitative change 
in the description of the two sets of anomalies, we shall not consider it further in this section.
We will, however, come back to this term in Section~\ref{sect:UV}, when discussing the effects generated by the exchange of the $U_1$ leptoquark beyond tree level.

\subsection{The \texorpdfstring{$\bsll$}{b->sll} anomalies}
\label{sect:bsll}

In $\bsll$ transitions ($\ell =e,\mu$), the NP effects induced by $\cL^{\rm NP}_{\rm EFT}$ in (\ref{eq:SMEFTLag}) amount to a modification 
of the Wilson coefficients already present in the SM below the electroweak scale. The latter are usually normalized as~\cite{Buchalla:1995vs}
\be
 \cL_{\bsll}= \frac{4 G_F}{\sqrt{2}}\, V_{ts}^* V_{tb}\,\sum_i\, \cC_i^\ell\,\cO_i^\ell  \,, 
\ee 
where $V_{ij}$ denote the elements of the Cabibbo-Kobayashi-Maskawa (CKM) matrix
and the relevant semileptonic operators are defined as
\be
\cO^\ell_{9}=  \frac{\alpha}{4\pi}\, (\bar{s}_L\gamma_\mu b_L)(\bar\ell \gamma^\mu\ell)  \,,  \qquad 
\cO^\ell_{10}=  \frac{\alpha}{4\pi}\, (\bar{s}_L\gamma_\mu b_L)(\bar\ell\gamma^\mu\gamma_5\ell) \, .
\ee
The Wilson coefficients of operators involving
a right-handed quark current are, by assumption, negligibly small in our approach. 
A key prediction of the SM is that the Wilson coefficients of these operators are lepton-flavor universal.
In order to analyze in general terms NP effects that violate this prediction but preserve the SM operator basis, 
it is convenient to distinguish LFU-breaking contributions from universal NP corrections. We choose to define the latter using the 
Wilson coefficients of the electron modes as reference, i.e.\
\be
\Delta \cC_{i}^U \equiv \cC^e_i -\cC^{\rm SM}_i \,,
\ee
such that the LFU-breaking terms can be defined as
\be
\Delta \cC_i^{\mu} \equiv \cC_i^\mu - \cC_i^e  =
 \cC^\mu_i - ( \cC^{\rm SM}_i +\Delta \cC_{i}^U ) \,.
\ee
In general terms, this amounts to introducing four
complex NP parameters for $i=9,10$. 

From a tree-level matching with $\cL^{\rm NP}_{\rm EFT}$ in (\ref{eq:SMEFTLag}), it is straightforward to derive the expressions for the LFU-breaking contributions in our setup. We find
\be
\Delta \cC_9^\mu = - \Delta \cC_{10}^\mu 
\equiv \Delta \cC_L^\mu \,, 
\label{eq:C9C10CV}
\ee
where
\be
 \Delta \cC_L^\mu =  - \frac{2 \pi }{\alpha V_{ts}^* V_{tb}}\, 
 \big[ \cLL^{23 \mu\mu} 
 -  \cLL^{23 ee} \big] \approx -\frac{2 \pi}{\alpha V_{ts}^* V_{tb}} \,\cLL^{23 \mu\mu} \,.
 \label{eq:CLmu}
\ee
The last relation follows from the assumption that $|\cLL^{23 ee} | \ll |\cLL^{23 \mu\mu} |$, which is a key hypothesis of our framework.
The multiplicative correction of $\cC_{LL}^{23 \mu\mu}$ due to renormalization-group (RG) evolution from the NP scale $\Lambda$ 
to the electroweak scale is at the percent level and can be safely neglected.  As a result, the LFU-violating corrections $\Delta \cC_{9,10}^\mu$ are scale independent to good accuracy. On the other hand, since $|\cC_{LL}^{23 \tau\tau} | \gg |\cC_{LL}^{23 \mu\mu} |$, the mixing of $\cO_{LL}^{23 \tau\tau}$ into operators containing light leptons is an important RG effect. These loop-induced contributions are responsible for the flavor-universal corrections $\Delta\cC_i^U$, which are sizable in the case of  $\cC_9$ only. One finds~\cite{Crivellin:2018yvo}
 \be
 \begin{aligned}
\Delta \cC_{9}^U (m_b) &= \frac{1}{V_{ts}^* V_{tb}}\,\frac{2}{3} \sum_{\ell=e,\mu,\tau}  \cC_{LL}^{23 \ell\ell}(\Lambda) \, 
 \ln\frac{\Lambda^2}{m_b^2} \,  \approx \frac{1}{V_{ts}^* V_{tb}}\,  \frac{2}{3} \,\cC_{LL}^{23 \tau\tau}(\Lambda) \,
 \ln\frac{\Lambda^2}{m_b^2} \,,  \\
\Delta \cC_{10}^U (m_b) &\approx 0 \,.
\label{eq:C9U} 
\end{aligned}
\ee
The smallness of $\Delta \cC_{10}^U$ is a dynamical feature of the setup we are considering. Since flavor-violating 
effects at the high scale are encoded in semileptonic operators only, the loop-induced 
flavor-violating couplings of the $Z$ boson are suppressed by the square of the tau-lepton Yukawa coupling. 
The result for $\Delta \cC_9^U$ reported above is obtained in the leading logarithmic approximation. The resummed RG contribution,
which is included in our numerical analysis, leads to a relative decrease of the effect by about 10\% for $\Lambda = 2\,\mathrm{TeV}$.

Within our framework, the final expressions of $\cC_{9,10}^{\ell}$ including the effects described by $\cL^{\rm NP}_{\rm EFT}$ 
can thus be written in terms of two independent parameters ($\Delta \cC^\mu_L$ and $\Delta \cC^U_9$) rather than four, which we further assume to be real
after factoring out the CKM matrix elements, as in the SM case:\footnote{A justification of this assumption is given in Section~\ref{sect:UVsensitive} in terms of the
UV completion of the model.} 
\begin{align}
\begin{aligned}
& \cC_9^e= \cC_{9, \mathrm{SM}} + \Delta \cC_9^U\,, &  & \cC^e_{10} = \cC_{10, \mathrm{SM}}  \,,  \\ 
& \cC^\mu_9 = \cC_{9, \mathrm{SM}}  +\Delta \cC_9^U + \Delta \cC_L^\mu  \,, & \quad  & \cC^\mu_{10} = \cC_{10, \mathrm{SM}} - \Delta \cC_L^\mu   \,. 
\label{eq:C9C10mue} 
\end{aligned} 
\end{align} 
It is worth mentioning that the relation (\ref{eq:C9C10CV}) among the  LFU-breaking terms as well as the absence of 
operators with right-handed quark currents
in the effective Lagrangian (\ref{eq:SMEFTLag}) 
are direct consequences of the assumption of a minimally-broken $U(2)^5$ flavor symmetry~\cite{Barbieri:2011ci,Fuentes-Martin:2019mun}. On the other hand, as mentioned earlier, 
the relations (\ref{eq:C9C10mue}) and, in particular, the absence of a universal correction to $\cC_{10}$, are consequences of the dynamical assumptions 
we are making.
An additional specific feature 
of the dynamical model we are considering  is the smallness of the 
coefficient of the scalar operator  $(\bar s_L  b_R)\, (\bar \mu_R\mu_L)$; the minimally-broken $U(2)^5$ flavor symmetry
alone, while implying a strong  suppression for this operator, would not forbid a relevant 
contribution to the helicity-suppressed  $B_s\to \mu^+\mu^-$ rate~\cite{Bordone:2018nbg,Fuentes-Martin:2019mun}.

\begin{table}[t]
\centering
\renewcommand{\arraystretch}{1.2} 
\begin{tabular}{ccc}
\toprule
{\bf Observable} & {\bf Experiment}   & {\bf SM}    \\
\specialrule{0.75pt}{1pt}{5pt}
$R_{K^*}^{[0.045,1.1]}$ &     $0.66^{+0.11}_{-0.07}\pm0.03$~\cite{Aaij:2017vbb}   &   $0.906\pm0.028$~\cite{Bordone:2016gaq} \\[5pt]  
$R_{K^*}^{[1.1,6.0]} $   &     $0.69^{+0.11}_{-0.07}\pm0.05$~\cite{Aaij:2017vbb}   &   $1.00\pm0.01$~\cite{Bordone:2016gaq}  \\[5pt] 
$R_{K^{\phantom{*}}}^{[1.1,6.0]}$  &    $0.846^{+0.042+0.013}_{-0.039-0.012}$~\cite{Aaij:2021vac}    &   $1.00\pm0.01$~\cite{Bordone:2016gaq} \\[2pt] 
$\mathcal{B}(B_s \to \mu^+ \mu^-)$  &  $(2.85^{+0.32}_{-0.31}) \times 10^{-9}$~\cite{Aaboud:2018mst,Sirunyan:2019xdu,LHCbBsmumu}    & 
$(3.66\pm0.14)\times 10^{-9}$~\cite{Beneke:2019slt} \\
\bottomrule
\end{tabular}
\caption{\label{tab:bslldata}
Experimental results and SM predictions for the clean observables in $\bsll$ decays.}
\end{table}

\paragraph{Data analysis.}
In order to extract the values of the NP Wilson coefficients from data, we distinguish two sets of observables:
\begin{itemize}
\item {\em Clean observables}. This set contains the observables with high sensitivity to short-distance dynamics and small, controllable theoretical uncertainties. Among the quantities measured so far, we include in this category only the LFU-testing ratios measured by LHCb,
$R_K$~\cite{Aaij:2021vac} and $R_{K^*}$~\cite{Aaij:2017vbb} (defined in \eqref{eq:RKRKs_def}) in all $q^2$ bins, and the branching ratio $\cB(B_s \to \mu^+\mu^-)$, where we perform our own combination of ATLAS~\cite{Aaboud:2018mst}, CMS~\cite{Sirunyan:2019xdu} and the recent LHCb~\cite{LHCbBsmumu} measurements.\footnote{More precisely, each of the two-dimensional profile likelihoods provided by the experimental collaborations is fitted to a two-dimensional variable-width Gaussian~\cite{Barlow:2004wg}, which are then summed and fitted to a new variable-width Gaussian. We note, however, that CMS and ATLAS measurements do not include the recent LHCb update of the ratio of $B_s^0$ and $B^0$ fragmentation fractions $f_s/f_d$~\cite{Aaij:2021nyr}, which is expected to 
modify the reported measurements slightly.} Within the SM, 
the theory uncertainty on $R_K$ and $R_{K^*}$ is only due to QED effects and does not exceed $1\%$ in $R_K$ and in the 
high-$q^2$ bin of $R_{K^*}$~\cite{Bordone:2016gaq,Isidori:2020acz}. 
The theory uncertainty on $\cB(B_s \to \mu^+\mu^-)$, which is due to a combination of parametric uncertainties 
(the leading contribution) and QCD corrections that are difficult to estimate, amounts to about $4\%$~\cite{Beneke:2019slt}.
In both cases, these uncertainties are well below the current experimental precision, as shown in Table~1.
 \item {\em Other $b \to s  \mu^+ \mu^-$ observables}.  This set includes all the other relevant observables, namely the differential branching ratios and angular distributions of the  semileptonic decays $B\to K^{(*)}\mu^+\mu^-$~\cite{CDF:2012qwd,Aaij:2014pli,Aaij:2016flj,Khachatryan:2015isa,CMS:2017ivg,Aaboud:2018krd,Aaij:2020nrf,Aaij:2020ruw}, $B_s\to\phi\mu^+\mu^-$~\cite{CDF:2012qwd,Aaij:2015esa,Aaij:2021nyr} and $\Lambda_b\to\Lambda\mu^+\mu^-$~\cite{Aaij:2015xza,Aaij:2018gwm}, including the recent LHCb experimental updates. These observables retain some sensitivity to short-distance dynamics, but are afflicted by sizable theoretical uncertainties, in particular due to the possibility of $c\bar c$ re-scattering effects in the final state, which are difficult to quantify reliably. 
\end{itemize}

\begin{figure}[t]
\centering
\includegraphics[width=0.45\textwidth]{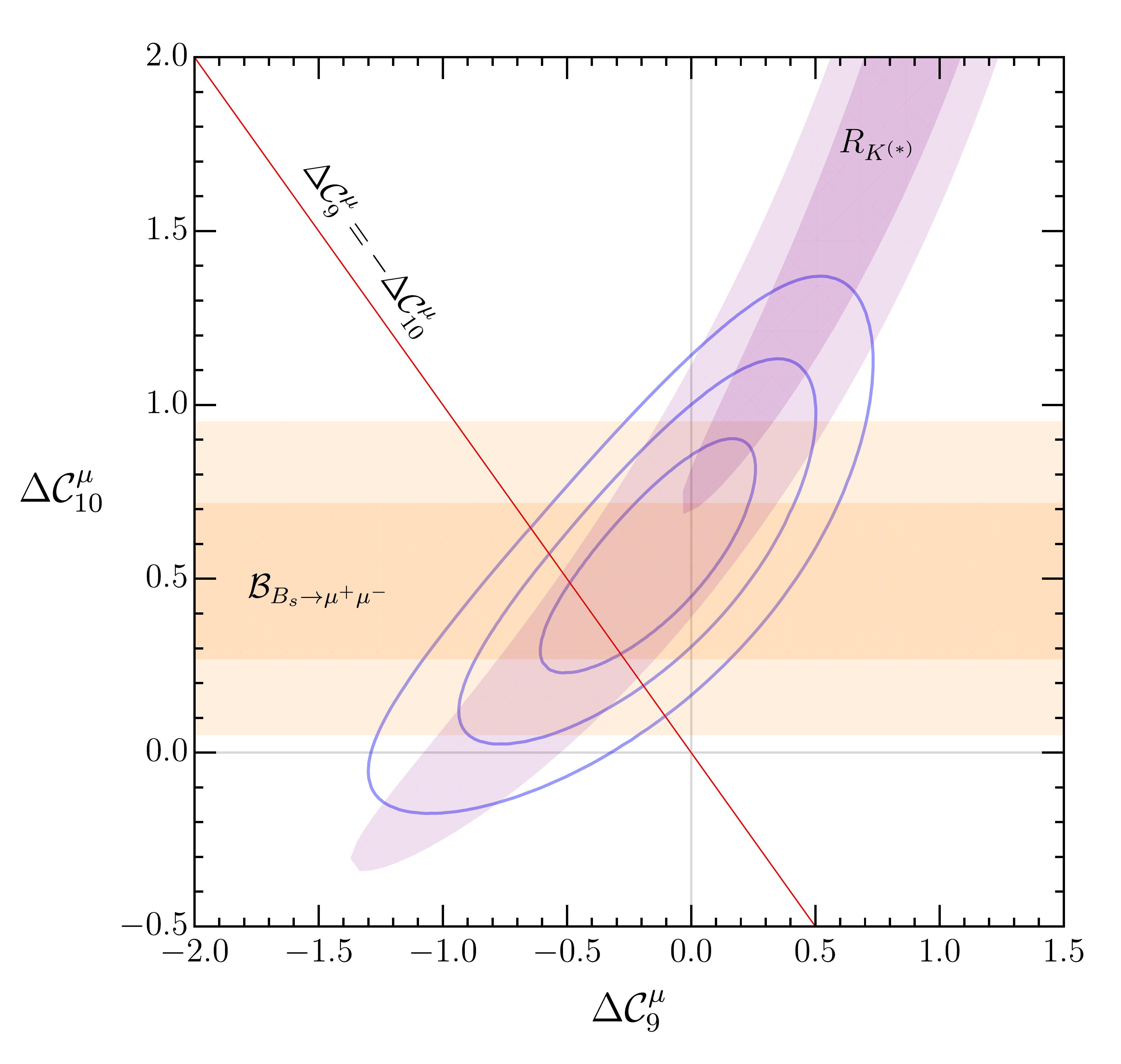}  \quad
\includegraphics[width=0.51\textwidth]{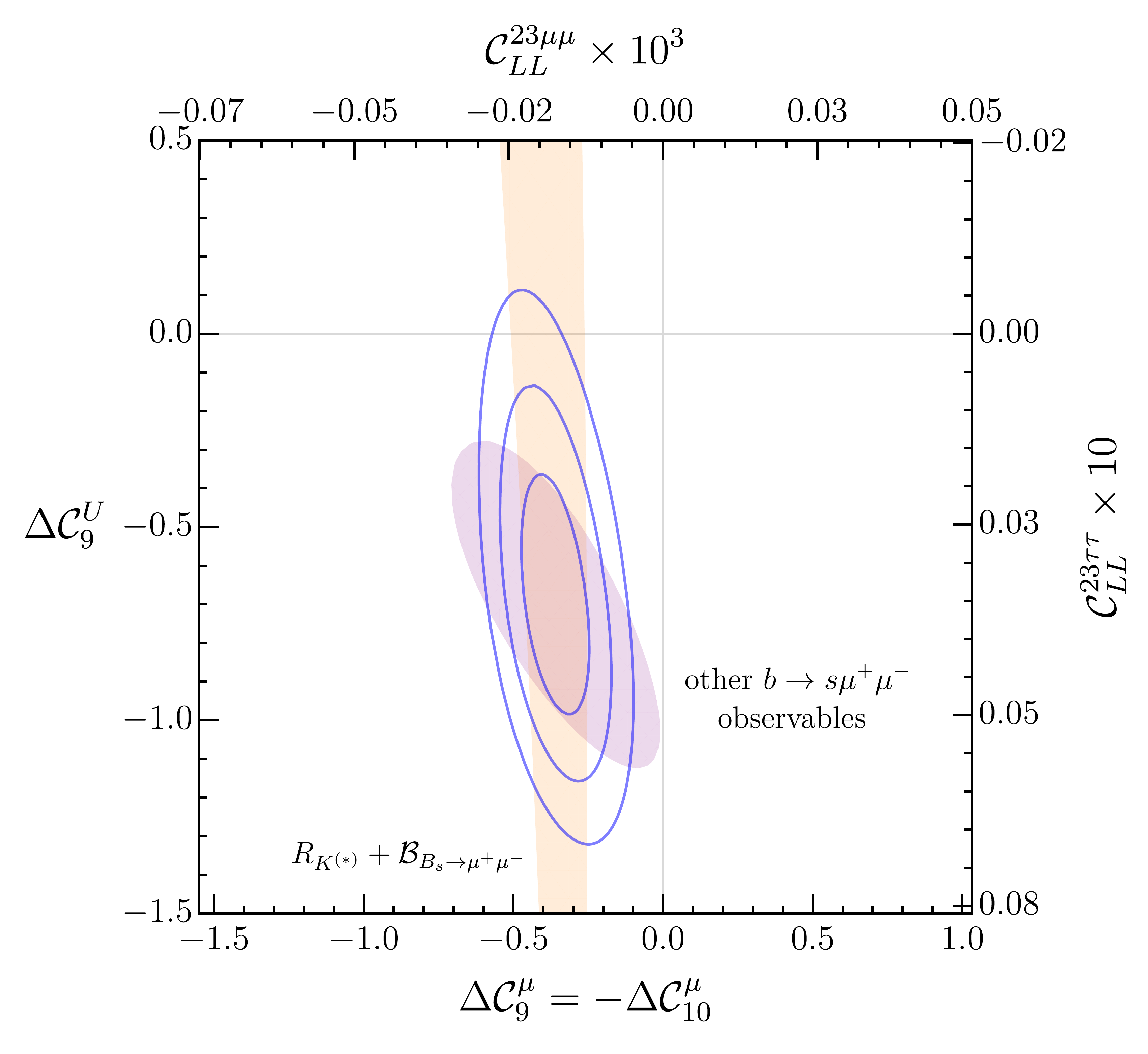} 
\caption{  \label{fig:bsll}
EFT constraints from the $\bsll$ anomalies. {\em Left:} Results of the two-dimensional fit 
$\Delta \cC^\mu_9$ vs.~$\Delta \cC^\mu_{10}$ using clean observables only ($1\sigma$, $2\sigma$ and $3 \sigma$ intervals). Also shown are the 
$1\sigma$ and $2 \sigma$ intervals from $R_{K^{(*)}}$ and $\cB(B_s\to \mu^+\mu^-)$, the latter under the hypothesis  $\Delta \cC_{10}^{U}=0$.
{\em Right:} Results of the two-dimensional fit $\Delta \cC^\mu_9 =-\Delta \cC^\mu_{10}$ vs.~$\Delta \cC_9^U$
using all $\bsll$ observables. The vertical band shows the result using clean observables only ($1\sigma$ interval), while
the ellipse denote the contribution of all the other observables, estimated  
using Flavio ($1\sigma$ interval).
The upper and right axes show the corresponding constraint on the high-scale EFT coefficients
(see main text).
}
\end{figure}

The clean observables are sensitive to $\Delta \cC_{9}^\mu$ and $\Delta \cC_{10}^\mu$ via $R_K$ and $R_{K^*}$ (in the limit where we neglect small threshold effects in the low-$q^2$ bin of $R_{K^*}$), and to $\Delta \cC_{10}^U$ via the $B_s \to \mu^+\mu^-$ branching ratio. Explicit phenomenological expressions of the observables in terms of the Wilson coefficients are collected in Appendix~\ref{app:Obs}.
According to (\ref{eq:C9C10mue}), in our setup these observables are described by the single NP parameter $\Delta C_L^\mu$.
Importantly, however, the data allow us to check the consistency of this hypothesis. In the left panel of Figure~\ref{fig:bsll}, we show the result of a fit to the clean observables (blue contour lines) performed under the assumption that $\Delta \cC_{10}^U=0$. 
We also show separately the parameter regions preferred by $R_K$ and $R_{K^*}$ (purple regions) and $\cB(B_s \to \mu^+\mu^-)$ (orange bands), as well as the line where $\Delta\cC_9^\mu=-\Delta\cC_{10}^\mu$. The inclusion of a non-vanishing $\Delta \cC_{10}^U$ parameter would amount to a translation of the $\cB(B_s \to \mu^+\mu^-)$ bands in the vertical direction. The plot shows that the two fit regions overlap in a region compatible with $\Delta \cC^\mu_{10}=-\Delta \cC^\mu_{9}$, without the need for an extra shift due to a non-zero $\Delta \cC_{10}^U$
(or a non-vanishing Wilson coefficient for the scalar operator).
 This observation provides a strong consistency check that all NP effects in the clean observables can be described by a single parameter. More quantitatively,
the $p$-value of the fit to the clean observables only, assuming $\Delta \cC^\mu_{10} = - \Delta \cC^\mu_{9}$ 
(single-parameter fit) is 12\%.  The significance of the NP hypothesis we are considering with respect to the SM 
(again based on a single-parameter fit) is 4.6$\sigma$.
It should be stressed that this estimate of the significance is a {\em conservative} one, because it does not 
include the contributions of the other $\bsll$ observables. Including them would further reinforce the NP hypothesis, but at the expense of introducing larger hadronic uncertainties.

 Besides the clean observables,  
 whose contribution is considered separately in the fit described above, 
 the effect of the other $\bsll$ observables is taken into account using the public code Flavio~\cite{Straub:2018kue}. For these observables, we follow the same prescription as in~\cite{Aebischer:2018iyb},  including only $q^2$-bins below the $J/\psi$ resonance (extending up to $6~\mathrm{GeV}^2$) and bins above the $\psi(2S)$ resonance that are at least $4~\mathrm{GeV}^2$ wide. In the right panel of Figure~\ref{fig:bsll}, we show the result of a combined fit of all observables in terms of 
$\Delta \cC_9^\mu$ (which equals $-\Delta\cC_{10}^\mu$ in our framework) and $\Delta \cC_9^U$.
As can be seen from this figure, once the strong constraint on $\Delta \cC_9^\mu$
arising from the clean observables
is implemented (orange band), the other $\bsll$ observables (purple region) can be used to constrain $\cC_9^U$, which is found to differ from zero by more than $2\sigma$ (see also~\cite{Alguero:2018nvb}). In the same plot, we also show the results interpreted as constraints on the high-scale 
Wilson coefficients $\cC_{LL}^{23 \mu\mu}$ and $\cC_{LL}^{23 \tau\tau}$,which follow from (\ref{eq:CLmu}) and (\ref{eq:C9U}). In the latter case, the contribution from RG evolution is estimated setting $\Lambda = 2$~TeV. As can be seen, the hierarchy of these Wilson coefficients is perfectly compatible with the scaling rule $\cC_{LL}^{23 \mu\mu}\sim\epsilon_\ell^2\,\cC_{LL}^{23 \tau\tau}$ discussed in Section~\ref{sect:OPbasis}.

\begin{table}[t]
\centering
\renewcommand{\arraystretch}{1.2} 
\begin{tabular}{ccc}
\toprule
{\bf Observable} &  {\bf Experiment} & {\bf SM}     \\
\specialrule{0.75pt}{1pt}{1pt}
\{$R_D$, $R_{D^*}$\}  &  $\begin{matrix}\{0.337(30),0.298(14)\}\\[-2pt]\rho=-0.42\\\end{matrix}$~\cite{Bernlochner:2021vlv}  & $\{0.299(3),0.258(5)\}$\,\cite{Amhis:2019ckw}       \\[10pt]  
$\cB(B^- \to \tau \bar\nu)$  & $1.09(24)\times10^{-4}$~\cite{Zyla:2020zbs}   & $0.812(54)\times10^{-4}$~\cite{Alpigiani:2017lpj}  \\
\bottomrule
\end{tabular}
\caption{\label{tab:bclnudata}
Experimental results and SM predictions for $\bculnu$ decays.
In the first entry, we provide the present combined experimental average, with $\rho$ denoting the correlation among the two observables.}
\end{table}

\subsection{The \texorpdfstring{$\bctnu$}{b->c tau nu} anomalies}

In $b \to c\tau \bar\nu$ and $b \to u\tau \bar\nu$ charged-current transitions, the NP effects induced by the effective Lagrangian $\cL^{\rm NP}_{\rm EFT}$ in (\ref{eq:SMEFTLag}) not only amount to a simple 
rescaling of the SM contribution but also introduce new (scalar-current) operators not present in the SM. More precisely, we find that in our approach the low-energy effective Lagrangian for these transitions takes the form
\be
\begin{aligned}
   \cL_{b\to u_i\tau\bar\nu}
   = - \frac{4 G_F}{\sqrt{2}}\,\sum_{i=1,2}\,\bigg[
   & \Big( V_{i b} + \sum_{k=1}^3\,V_{i k}\,\cC_{LL}^{k3\tau\tau} \Big)
    (\bar u^i_L \gamma_\mu b_L) (\bar\tau_L \gamma^\mu \nu_L) \\
   &- 2\,\sum_{k=1}^3\,V_{i k}\,\cC_{LR}^{k3\tau\tau}\,
    (\bar u^i_L b_R) (\bar\tau_R\,\nu_L) \bigg] \,.
\end{aligned}
\ee 

We recall that the flavor basis for the NP operators is the down-quark and charged-lepton mass basis, i.e.\
\begin{align}\label{eq:DownBasis}
q_L^i=
\begin{pmatrix}
V_{ji}^*\,u^j_L\\
d_L^i
\end{pmatrix}
,\qquad 
\ell_L^i=
\begin{pmatrix}
\nu^i_L\\
e_L^i
\end{pmatrix}
.
\end{align}
This implies that different Wilson coefficients contribute to a given $b\to u^i$ transition. This is conveniently taken into account
defining new coefficients (for $i=1,2$)
\be
   \cC_{LL}^{u_i} \equiv \cC^{33\tau\tau}_{LL} 
    \left[ 1 + \frac{\cC^{23\tau\tau}_{LL}}{\cC^{33\tau\tau}_{LL}}\,
    \frac{V_{i s}}{V_{i b}} \left( 1 
    + \frac{\cC^{13\tau\tau}_{LL}}{\cC^{23\tau\tau}_{LL}}\,
    \frac{V_{u d}}{V_{i s}} \right) \right] , 
\label{eq:CVSc}
\ee
and similarly for $\cC_{LR}^{u_i}$ with $LL\to LR$ everywhere, such that $\cC_{LL}^{u_i}$ and $\cC_{LR}^{u_i}$ are the only two effective combinations appearing in $b\to u_i$ transitions. With these definitions, the low-energy effective Lagrangian reads
\be
   \cL_{b\to u_i\tau\nu}
   = - \frac{4 G_F}{\sqrt{2}}\,\sum_{i=1,2}\,V_{i b}\,\bigg[
    \Big( 1 + \cC_{LL}^{u_i} \Big)
    (\bar u^i_L \gamma_\mu b_L) (\bar\tau_L \gamma^\mu \nu_L) 
   - 2\,\cC_{LR}^{u_i} \,
    (\bar u^i_L b_R) (\bar\tau_R\,\nu_L) \bigg] \,.
\ee 

Under the generic assumption $|\cC^{13\tau\tau}_{LL,LR}| \ll | \cC^{23\tau\tau}_{LL,LR}|$ stated in Section~\ref{sect:OPbasis}, 
the hierarchy of the relevant CKM matrix elements implies that the coefficients $\cC^{13\tau\tau}_{LL,LR}$ play a negligible role in $b\to c\tau\bar\nu$ transitions,  
while they might be relevant in $b\to u\tau\bar\nu$ decays. Hence, in this generic case the NP effects in the two processes are not strictly correlated. Under the stronger assumption of a minimal breaking of the $U(2)^5$ flavor symmetry
see Appendix~\ref{app:U2}) we however expect that
\be
   \frac{\cC^{13\tau\tau}_{LL}}{\cC^{23\tau\tau}_{LL}}
   = \frac{\cC^{13\tau\tau}_{LR}}{\cC^{23\tau\tau}_{LR}}
   = \frac{V_{td}^*}{V^*_{ts}} \,. 
\label{eq:U2min}
\ee
This implies a complete correlation of the NP effects in $b\to c\tau\bar\nu$ and $b\to u\tau\bar\nu$ transitions, i.e.\
\be
   \frac{\cC^{23\tau\tau}_{LL}}{\cC^{33\tau\tau}_{LL}}\,
    \frac{V_{is}}{V_{i b}} \left( 1
    + \frac{\cC^{13\tau\tau}_{LL}}{\cC^{23\tau\tau}_{LL}}\,
     \frac{V_{id}}{V_{i s}} \right) 
   = - \frac{\cC^{23\tau\tau}_{LL}}{\cC^{33\tau\tau}_{LL}}\,
    \frac{V_{tb}^*}{V_{ts}^*} \,,
\ee 
and similarly for the LR coefficients. This in turn implies that
\begin{align}\label{eq:CcCu}
   \cC_{LL}^u & = \cC_{LL}^c \,, &
   \cC_{LR}^u & = \cC_{LR}^c \,.
\end{align}

\begin{figure}[t]
\centering
\includegraphics[width=0.5\textwidth]{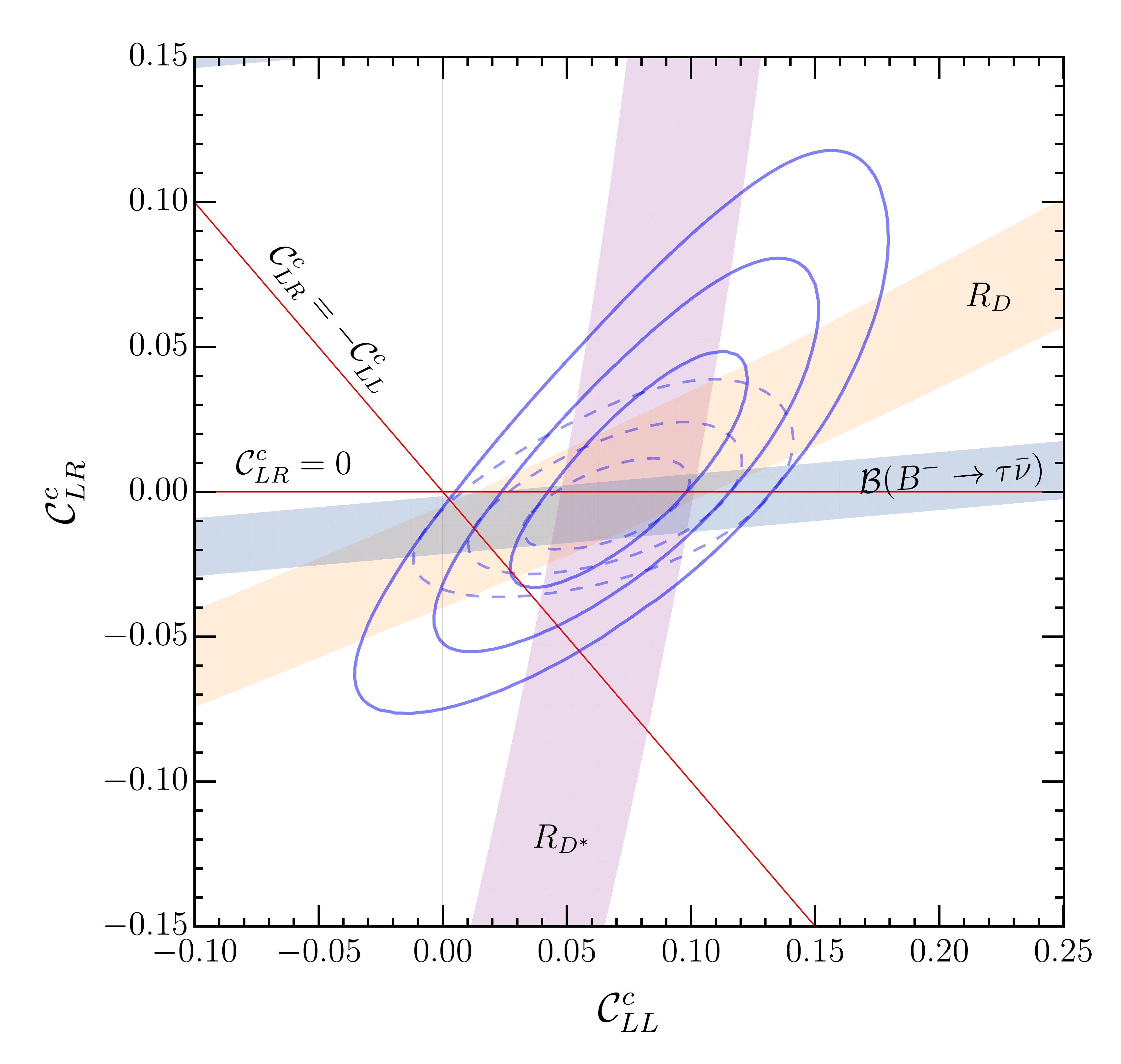} 
\caption{\label{fig:bctn}
EFT constraints from the $\bctnu$ anomalies. The solid blue ellipses denote the $1\sigma$, $2\sigma$ and $3\sigma$ intervals of the two-dimensional fit to $R_D$ and $R_{D^*}$ in the $\cC_{LL}^{c}$--\,$\cC_{LR}^{c}$ plane (coefficients evaluated at $\Lambda=2~\mathrm{TeV}$). The dashed contours denote 
the fit results taking also the constraint from $\cB(B^-\to \tau\bar\nu)$ into account, under the hypothesis of minimal $U(2)^5$ breaking
(i.e.\ for $\cC_{LL}^{u}=\cC_{LL}^{c}$, $\cC_{LR}^{u}=\cC_{LR}^{c}$). The bands correspond to  $1\sigma$ regions. The red lines show the benchmark scenarios we consider in our analysis.}
\end{figure}

\paragraph{Data analysis.}
At present, the observables constraining NP effects in  $b \to c \tau \bar\nu$ transitions are the LFU ratios $R_D$ and $R_{D^*}$ (defined in \eqref{eq:RDRDs_def}), which can be used to probe the NP coefficients $\cC_{LL}^c$ and $\cC_{LR}^c$.
In the situation where we correlate $b\to c\tau\bar\nu$ and $b\to u\tau\bar\nu$ transitions via the flavor-symmetry assumption in   (\ref{eq:CcCu}), also the branching ratio for the decay $B^-\to\tau\bar\nu$ plays an important role.
The data used for the fit are reported in Table~\ref{tab:bclnudata}, while the explicit expressions of the various observables in terms of the Wilson coefficients are collected in Appendix~\ref{app:Obs}. 

In Figure~\ref{fig:bctn}, we show the allowed regions for the coefficients $\cC_{LL}^c$ and $\cC_{LR}^c$ obtained from the measurements of $R_D$, $R_{D^*}$ and $\cB(B^-\to\tau\bar\nu)$ (colored bands), where in the latter case we assume the validity of the relations (\ref{eq:CcCu}). The solid contour lines show the result of a fit to $R_D$ and $R_{D^*}$ only, while the dashed lines refer to a fit including also $\cB(B^-\to\tau\bar\nu)$ under the hypothesis of minimal $U(2)^5$ breaking. In the first case, which is more conservative,
the significance of the NP hypothesis compared to the SM case (two-parameter fit) is 3.2$\sigma$. 
In the figure, we report the results in terms of the effective coefficients $\cC_{LL,LR}^c$ evaluated at the high scale $\Lambda=2$~TeV. While RG evolution effects are negligible for $\cC_{LL}^c$, the mixed-chirality coefficients $\cC_{LR}^{c}$ exhibits a sizable scale variation due to QCD corrections. 

In Figure~\ref{fig:bctn}, we also show as red lines the relations $\cC_{LR}^{c}=0$ and $\cC_{LR}^{c}=-\cC_{LL}^{c}$. The latter is consistent with the expectation $|\cC^{i3\tau\tau}_{LR}|=|\cC^{i3\tau\tau}_{LL}|$ that, in turn, is a natural benchmark for models in which these two coefficients are generated by the tree-level exchange of a Pati-Salam-like massive leptoquark (see Section~\ref{sect:U1} for a detailed discussion). From the figure we draw the following conclusions: 
\begin{itemize}
\item 
Without the inclusion of $\cB(B^-\to\tau\bar\nu)$, the present data are compatible with both $\cC_{LR}^c=0$ and $\cC_{LR}^{c}=-\cC_{LL}^{c}$, and likewise with any intermediate case. 
\item 
The inclusion of $\cB(B^-\to\tau\bar\nu)$ under the hypothesis of minimal $U(2)^5$ breaking is perfectly consistent with the other constraints. However, it slightly disfavors (by less than $2\sigma$) a scenario where $|\cC_{LR}^{c}|\approx|\cC_{LL}^{c}|$.
\end{itemize}
In our analysis below we study the cases $\cC^{i3\tau\tau}_{LR}=0$ and $\cC^{i3\tau\tau}_{LR}=-\cC^{i3\tau\tau}_{LL}$ as two  representative benchmark scenarios. To keep the discussion general, we allow in both cases for non-minimal, subleading $U(2)^5$-breaking terms, which modify the relation (\ref{eq:U2min}), and provide an {\em a posteriori} validation of it.

\subsection{Combined analysis of the semileptonic couplings involving \texorpdfstring{$\tau$}{tau} leptons}
\label{sect:DF2}
    
\begin{figure}[t]
\centering
\includegraphics[width=0.49\textwidth]{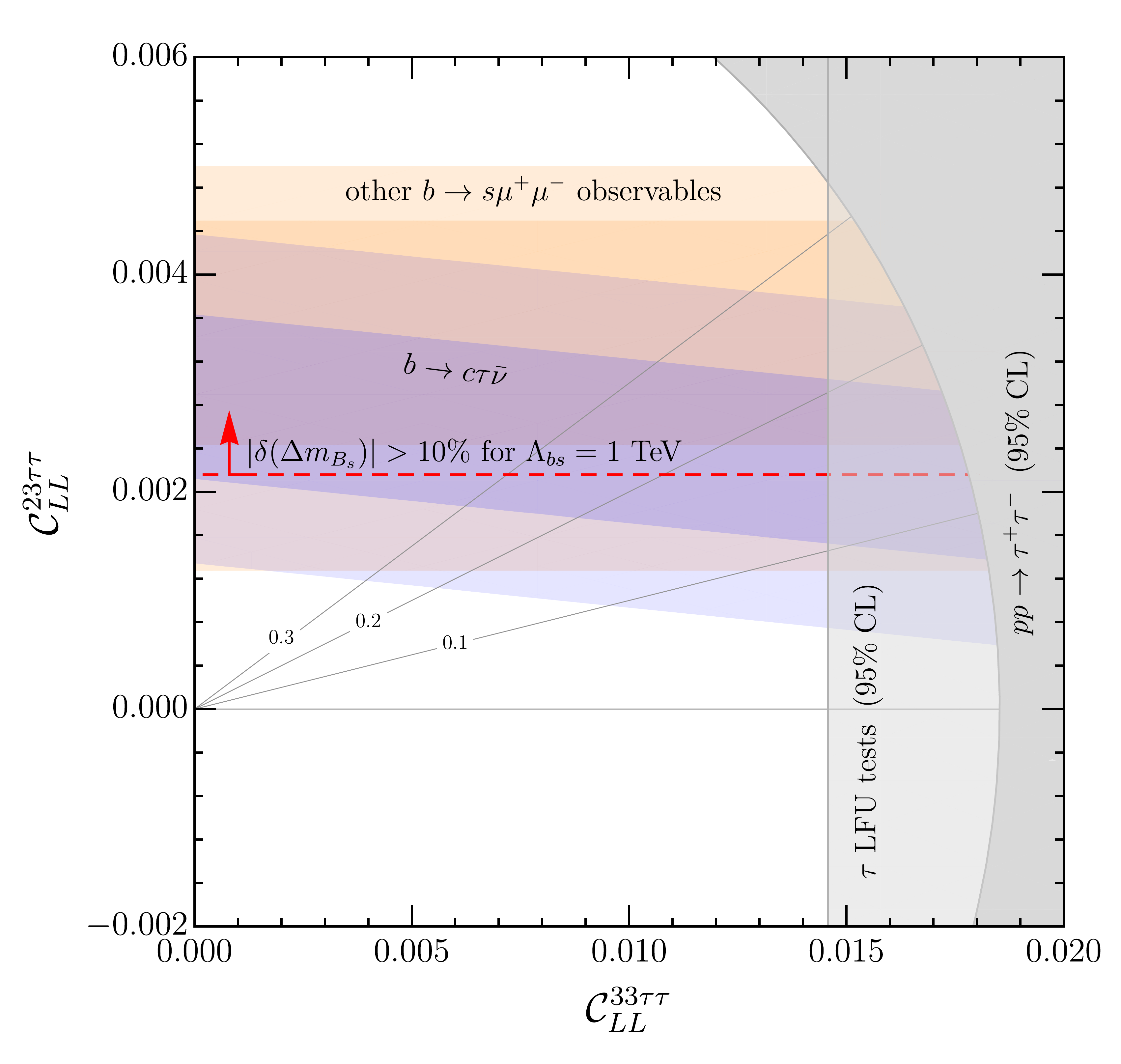} 
\includegraphics[width=0.49\textwidth]{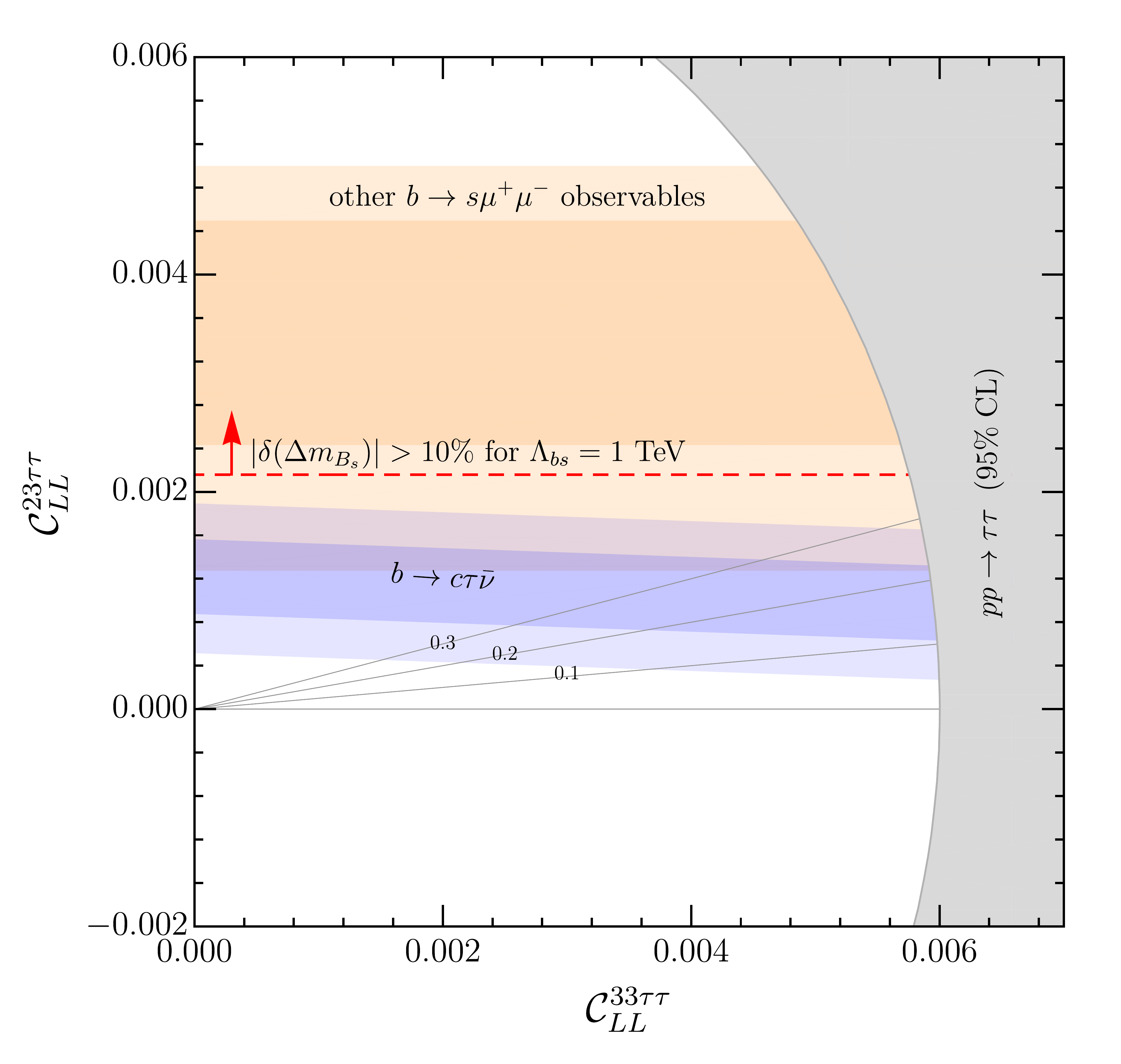} 
\caption{\label{fig:EFTcomb} Combined analysis of the coefficients $\cC_{LL}^{33\tau\tau}$ and $\cC_{LL}^{23\tau\tau}$ in the two benchmark scenarios defined by $\cC_{LR}^{i3\tau\tau}=0$ (left panel) and $\cC_{LR}^{i3\tau\tau}=-\cC_{LL}^{i3\tau\tau}$ (right panel). 
The blue bands denote the $1\sigma$ and $2 \sigma$ regions preferred by $\bctnu$ data, while the gray bands show the exclusion regions derived from $\sigma(pp\to\tau^+\tau^-+X)$. The preferred values of $\cC^{23\tau\tau}_{LL}$ derived from $b\to s\mu^+\mu^-$ data (at $1\sigma$ and $2 \sigma$) are indicated by the horizontal orange bands. The dashed red lines provide a qualitative indication of the bound from $B_s$--$\bar B_s$ mixing (see text for more details). The gray lines indicate reference values of the ratio $\cC^{23\tau\tau}_{LL}/\cC^{33\tau\tau}_{LL}\sim\epsilon_q$.}
\end{figure}

We will now study the overall consistency of the EFT description of the two sets of anomalies by focusing on the couplings $\cC^{33\tau\tau}_{LL}$ and $\cC^{23\tau\tau}_{LL}$ involving tau leptons. In the approximation where the very small contribution to $\cC_{LL}^c$ proportional to $\cC^{13\tau\tau}_{LL}$ in (\ref{eq:CVSc}) is neglected, the observables $R_D$ and $R_{D^*}$ are sensitive to NP effects described by both $\cC^{33\tau\tau}_{LL}$ and $\cC^{23\tau\tau}_{LL}$. 
The blue band in Figure~\ref{fig:EFTcomb} shows the allowed 
$1\sigma$ and $2 \sigma$ regions in the $\cC^{33\tau\tau}_{LL}$--$\,\cC^{23\tau\tau}_{LL}$ plane in the two benchmark scenarios defined above, in which the corresponding mixed-chirality coefficients $\cC^{33\tau\tau}_{LR}$ and $\cC^{23\tau\tau}_{LR}$ are fixed. The $b\to s\mu^+ \mu^- $ observables, on the other hand, are sensitive to NP effects parameterized by $\cC^{23\tau\tau}_{LL}$ alone, after we marginalize over $\cC^{33\mu\mu}_{LL}$ (see the right panel of Figure~\ref{fig:bsll}). The corresponding allowed region 
(at $1\sigma$ and $2 \sigma$) is shown by the horizontal orange bands. 

The EFT approach considered so far does not allow us to take into account in a precise way all constraints on the couplings $\cC^{33\tau\tau}_{LL}$ and $\cC^{23\tau\tau}_{LL}$ derived from other observables not directly related to the flavor anomalies. These will be analyzed in a more systematic way in the next section. However, we show in Figure~\ref{fig:EFTcomb} in a semi-quantitative way the three most relevant constraints. They arise from high-energy measurements of the $\tau^+\tau^-$ production cross section at the LHC, which can be affected in the presence of the four-fermion contact interactions in (\ref{eq:SMEFTLag}), from LFU tests in $\tau$ decays,
and from precision studies of the $B_s$--$\bar B_s$ mixing amplitude.

The bound from modifications of the high-$p_T$ tails in $pp\to\tau^+\tau^-+X$ processes, whose detailed derivation is postponed to Section~\ref{sect:U1}, is only weakly sensitive to the details of the UV completion of the EFT. At the energies accessible at the LHC, the effect of the heavy (multi-TeV scale) mediators is expected to be well described by the contact interactions. 
We include in our estimates the contributions from $b\bar b$-, $b\bar s$- and $s\bar b$-initiated scattering processes. In the benchmark scenario with $\cC_{LR}^{i3\tau\tau}=0$ (left panel) we also set $\cC_{RR}^{i3\tau\tau}=0$, while in the scenario with $\cC_{LR}^{i3\tau\tau}=-\cC_{LL}^{i3\tau\tau}$ (right panel) we take $\cC_{RR}^{33\tau\tau}=-\cC_{LR}^{33\tau\tau}=\cC_{LL}^{33\tau\tau}$, as expected from a UV completion with a Pati-Salam-like $U_1$ leptoquark.

\begin{figure}[t]
\centering
\includegraphics[scale=0.5]{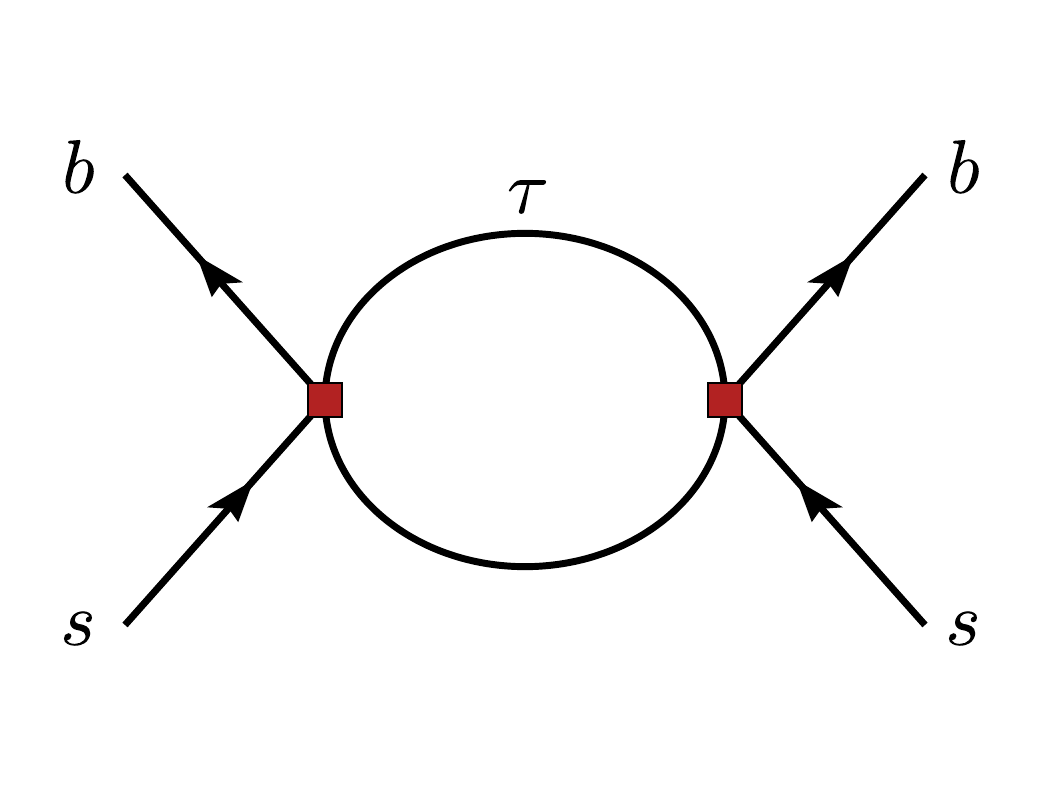}
\quad
\includegraphics[scale=0.5]{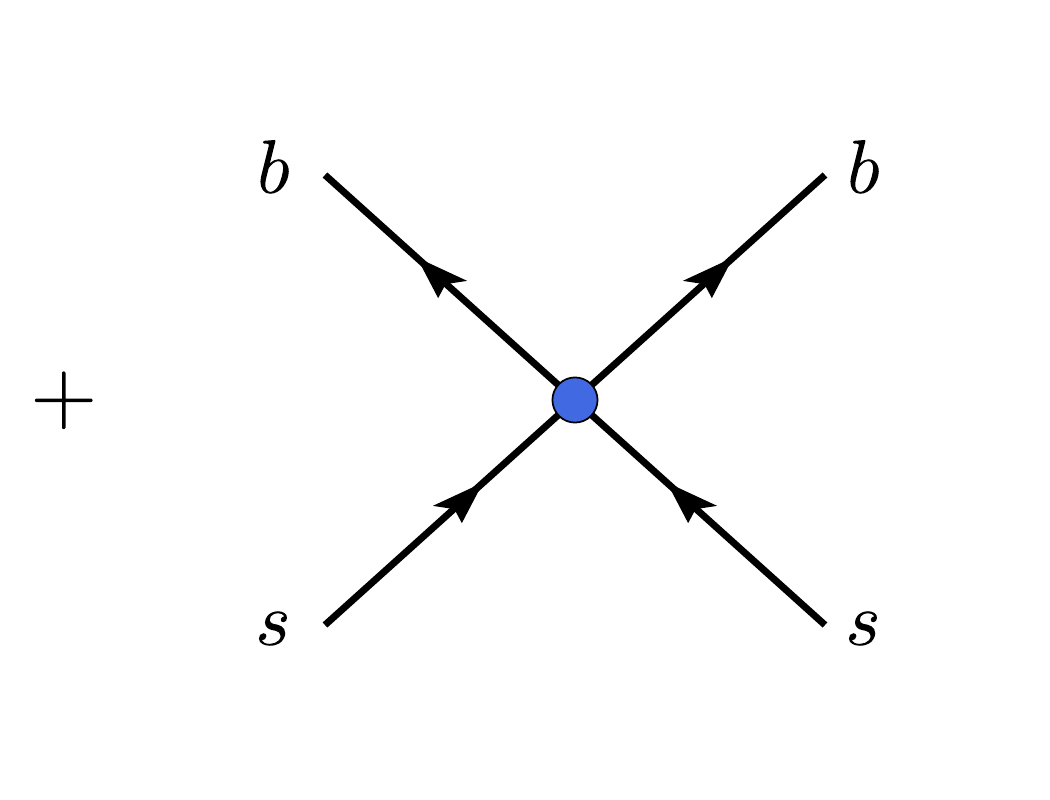}
\caption{\label{DF2loop} Different NP contributions to $B_s$--$\bar B_s$ mixing in the EFT. The red vertices denote the insertion of 
$\cO^{32\tau\tau}_{LL}$ from (\ref{eq:SMEFTLag}), while the blue vertex indicates generic short-distance contributions probing, in general, all heavy degrees of freedom present in the UV completion of the EFT. The first diagram contributes at dimension-8 order in SMEFT power counting, but the fact that it is quadratically divergent indicates that the blue vertex must contain a corresponding contribution with the same coupling structure.}
\end{figure}

The EFT treatment of $B_s$--$\bar B_s$ mixing is more model dependent. In general, we define the short-distance contributions to the effective Lagrangian for the $\Delta F=2$ mixing amplitudes as 
\be
   \cL^{\Delta F=2} 
   = - \cC_{bs} \left( \bar b_L \gamma_\mu s_L \right)^2
    - \cC_{bd} \left( \bar b_L \gamma_\mu d_L \right)^2
    - \cC_{uc} \left( \bar u_L \gamma_\mu c_L \right)^2
    + \text{h.c.} \,,
    \label{eq:DF2gen}
\ee
where for later convenience we also write the four-quark operators needed for the description of the $B_d$\,--$\bar B_d$ and $D$--$\bar D$ mixing amplitudes. The SM contribution to $B_s$--$\bar B_s$ mixing is given by 
\be
   \cC^{\rm SM}_{bs} 
   = \frac{G_F^2 m_W^2}{4\pi^2}\,\big( V_{tb}^* V_{ts} \big)^2\,S_0(x_t) \,,
\ee
with $x_t=m_t^2/m_W^2$ and $S_0(x_t)\approx 2.37$ \cite{Buchalla:1995vs}. 
Without reference to a concrete UV completion, there is no model-independent relation between the Wilson coefficient $\cC_{bs}$ and the Wilson coefficients entering the $\Delta F=1$ effective Lagrangian in (\ref{eq:SMEFTLag}). Note, in particular, that loop diagrams such as the first graph in Figure~\ref{DF2loop}, containing two insertions of  dimension-6 SMEFT operators, contribute only at dimension-8 in the SMEFT power counting. However, naive dimensional analysis shows that these loop graphs are quadratically UV-divergent, and therefore there must unavoidably exist a corresponding short-distance contribution to the Wilson coefficient of the dimension-6 operator $(\bar b_L\gamma_\mu s_L)^2$ with the same coupling structure, as shown by the second graph in Figure~\ref{DF2loop}. We thus expect that (with $i=s,d$)
\be\label{eq:MesonMixing}
   \frac{\cC_{b i}^{\rm NP}}{\cC_{b i}^{\rm SM}} 
   = \frac{\Lambda_{b i}^2}{2m_W^2\,S_0(x_t)}
    \left[ \frac{ \left(\cC^{i3\tau\tau}_{LL}\right)^*}{V_{tb}^* V_{t i}} \right]^2 
    + \dots \,, \qquad
   \cC_{uc}^{\rm NP} 
   = \frac{G_F^2 \Lambda_{uc}^2}{8\pi^2}\,\sum_{i,j=1}^3\,
    \big(V_{ui} V_{cj}^*\,\cC_{LL}^{ij\tau\tau} \big)^2 + \dots \,,
\ee
where $\Lambda_{ij}$ are combinations of mass parameters associated with new heavy particles in the UV theory, and the dots represent possible other UV contributions associated with different coupling parameters. Depending on the details of the model, the scale $\Lambda_{bs}$ can be lighter than the mass of the mediator responsible for the $\Delta F=1$ contact interactions shown in (\ref{eq:SMEFTLag}).

For instance, in the class of models analyzed in \cite{DiLuzio:2018zxy,Fuentes-Martin:2020hvc}, in which the $U_1$ leptoquark is a massive gauge boson, the scale $\Lambda_{bs}$ relevant to $B_s$--$\bar B_s$ mixing
is related to the mass $M_L$ of the vector-like leptons, which are responsible for the 2-3 flavor 
mixing in the $U_1$ couplings. More precisely, for $M_L\ll M_U$ one finds that $\Lambda_{bs}\approx\,M_L$ \cite{Fuentes-Martin:2020hvc}. The fact that in these models $\Lambda_{bs}$ is associated with the mass of a colorless particle (a vector-like lepton) is a highly non-trivial feature, which allows for a relatively low value $\Lambda_{bs}\lesssim 1$\,TeV without conflicting with current bounds from direct searches at the LHC (see Section~\ref{sect:UV} for more details).

In the case of $B_s$--$\bar B_s$ and $B_d$\,--$\bar B_d$ mixing, there exist stringent constraints on both the magnitude and the phase of the mixing amplitudes. It is then convenient to define (with $i=s,d$)
\be
   \delta (\Delta m_{B_i}) 
   = \frac{\Delta m_{B_i}-\Delta m^{\rm SM}_{B_i}}{\Delta m^{\rm SM}_{B_i}}
   = \left| 1 + \frac {\cC^{\rm NP}_{bi}}{\cC_{bi}^{\rm SM}} \right| - 1 \,,
    \qquad   
   \delta (\phi_{B_i}) 
   = \arg\left( 1 + \frac {\cC^{\rm NP}_{bi}}{\cC_{bi}^{\rm SM}} \right) .
\label{eq:MesonMixing2}
\ee
The horizontal red line in Figure~\ref{fig:EFTcomb} corresponds to $\delta (\Delta m_{B_s})=0.1$, i.e. to a 10\% correction to the magnitude of $\Delta m_{B_s}$, under the assumption that $\Lambda_{bs}=1$~TeV. 
Without entering into model-dependent considerations, we note that the absence of direct signals of new physics at the energy frontier implies that is very hard to conceive explicit models with $\Lambda_{bs}$ much below 1~TeV. This is why the region above the red line should be considered as disfavored by $B_s$--$\bar B_s$ mixing, barring cancellations between the contribution shown explicitly in (\ref{eq:MesonMixing}) and additional contributions related to different coupling parameters. 

\paragraph{Discussion.} From the two plots in Figure~\ref{fig:EFTcomb} 
we draw the following conclusions: 
\begin{itemize}
\item{} Overall there is a good consistency between the values of $\cC^{i3\tau\tau}_{LL,LR}$ 
necessary to fit $\bctnu$ data, the value of $\cC^{23\tau\tau}_{LL}$ indicated by
$\bsll$ data, and the flavor scaling assumed in Section~\ref{sect:OPbasis} (gray lines in Figure~\ref{fig:EFTcomb}).
\item{} The bound from $pp\to \tau^+\tau^-$ prevents a solution of the  $b \to c \tau \bar\nu$  anomalies with vanishing $\cC^{23\tau\tau}_{LL,LR}$, while $B_s$--$\bar B_s$ mixing tends to favor the smallest possible value of $|\cC^{23\tau\tau}_{LL}|$. 
\item{} While the benchmark scenario with $\cC^{i3\tau\tau}_{LR}=0$ appears to be favored by the $\bctnu$ data alone (see Figure~\ref{fig:bctn}), the $B_s$--$\bar B_s$ mixing bound is more stringent in this case. Hence, with present data it is still useful to consider both benchmark scenarios.  
\end{itemize}

\section{The simplified \texorpdfstring{$U_1$}{U1} model} \label{sect:U1}

In this section we analyze the case where the effective operators in (\ref{eq:SMEFTLag}) are generated 
by the tree-level exchange of a $U_{1}^{\mu} \sim (\mathbf{3},\mathbf{1})_{2/3} $ leptoquark.  
At the expense of introducing some model dependence,
this assumption allows to consider a wider class of observables in terms of a reduced number of free parameters (the effective leptoquark couplings), and it simplifies the interpretation of available data in the class of UV completions we are interested in.
Note that introducing a new heavy vector boson necessarily requires an additional new-physics sector, which gives mass to this particle. Here we consider a simplified model, focusing only on the $U_1$ as the dominant source of new flavor-changing interactions. In Section~\ref{sect:UV}, we will then explore a concrete example of a consistent UV completion.

The most general Lagrangian for a $U_1$ vector leptoquark coupling to SM particles is given by
\begin{align}\label{eq:LQLag}
\begin{aligned}
\mathcal{L}_U=&-\frac{1}{2}\,U_{\mu\nu}^\dagger\, U^{\mu\nu}+M_{U}^2\,U_\mu^\dagger\, U^\mu -ig_s\,(1-\kappa_c)\,U_\mu^\dagger\,T^a U_\nu\,G^{\mu\nu,a} \\
&- \frac{2i}{3}\,g_Y\, (1-\kappa_Y)\,U_\mu^\dagger\,U_\nu\,B^{\mu\nu}+\frac{g_U}{\sqrt{2}}\,(U^\mu J^U_\mu + \mathrm{h.c.})\,,
\end{aligned}
\end{align}
where $U_{\mu \nu} = D_{\mu} U_\nu - D_{\nu} U_\mu\,,$ with $D_\mu = \partial_\mu - i g_s\, G_\mu^aT^a - i\,\frac23 g_{Y}  B_{\mu}$. Here $G_\mu^a$ ($a=1,\dots,8$) and $B_\mu$ denote the $SU(3)_c$ and $U(1)_Y$ gauge bosons, $g_s$ and $g_Y$ are the corresponding gauge couplings, and $T^a$ are the generators of $SU(3)_c$. In models in which the vector leptoquark has a gauge origin, $\kappa_c=\kappa_Y=0$, while this is not necessarily the case in models in which the $U_1$ arises as a bound state from a strongly-coupled sector. The interaction of the $U_1$ with the SM fermions involves the currents 
\be
J^{U}_\mu=   \beta_{L}^{i\alpha}\,(\bar q_{L}^{\,i} \gamma_{\mu}  \ell_{L}^{\alpha})  +     \beta_{R}^{i \alpha }\,(\bar d_{R}^{\,i}\gamma_{\mu}   e_{R}^{\alpha}) \,,
\label{eq:JU0}
\ee
where the couplings $\beta_{L}$ and $\beta_{R}$ are complex $3 \times 3$ matrices in flavor space.
Following our hypothesis on the flavor structure of the theory, we can write 
\begin{equation}\label{eq:minCoup}
\begin{aligned}
\beta_{L} = 
\begin{pmatrix} 
0 & 0 & \beta_L^{d\tau} \\[5pt]
0 & \beta_{L}^{s \mu} &  \beta_{L}^{s \tau}\\[5pt]
0 & \beta_{L}^{b \mu} &  1
\end{pmatrix}
\,,\qquad\qquad 
\beta_{R} = 
\begin{pmatrix} 
0 & 0 & 0 \\[5pt]
0 & 0 & 0\\[5pt]
0 & 0 &  \beta_{R}^{b \tau} 
\end{pmatrix}
\,,
\end{aligned}
\end{equation}
with $|\beta_L^{d\tau,s\mu}|\ll|\beta_L^{s\tau,b\mu}|\ll 1$ and $\beta_R^{b\tau}=\mathcal{O}(1)$. The normalization of $g_U$ is chosen such that $\beta_{L}^{b\tau}=1$. The null entries in (\ref{eq:minCoup}) should be understood as small terms 
which have a negligible impact on the observables we analyze. It is worth stressing that this structure is a direct consequence of the hypothesis of a $U(2)^5$ flavor symmetry with sizable breaking 
only along the $U(2)_q$ direction.\footnote{As far as right-handed mixing is concerned,  this symmetry 
hypothesis alone  implies that the natural size of the largest off-diagonal entries is $\beta_{R}^{s \tau} \sim (m_s/m_b)\,\beta_{L}^{s \tau}$ and $\beta_{R}^{b \mu} \sim (m_\mu/m_\tau)\,\beta_{L}^{s \tau}$, well below the corresponding 
left-handed entries.} Under the stronger assumption of a single spurion transforming as doublet of $U(2)_q \in U(2)^5$,
we further expect $\beta_{L}^{d \tau}/\beta_{L}^{s \tau} = V^*_{td}/V^*_{ts}$ (see Appendix~\ref{app:U2}).

By integrating out the vector leptoquark at tree level, we obtain the following matching conditions for the 
effective operators introduced in Section~\ref{sect:OPbasis}: 
\be\label{eq:4Fmatching}
\cLL^{ij\alpha\beta} = C_U \beta_{L}^{i \alpha}  (\beta_{L}^{ j \beta})^*\,, \qquad 
\cLR^{ij\alpha\beta}  = C_U \beta_{L}^{i \alpha}  (\beta_{R}^{ j \beta})^*\,, \qquad 
\cRR^{ij\alpha\beta} = C_U \beta_{R}^{i \alpha}  (\beta_{R}^{ j \beta})^*\,, 
\ee
where $C_{U} \equiv g_{U}^{2} v^{2}/(4 M_{U}^{2})$.

\subsection{Low-energy fit in the simplified model}
In this section, we perform a fit to the $U_1$ model parameters described above with the value of $\betaR$ fixed to one of the two reference values $\betaR=0$ and $\betaR=-1$, corresponding 
to the two benchmark scenarios discussed in Section~\ref{sect:DF2}.
We recall that in models with third-family Pati-Salam unification, and in absence of a mixing of the SM fermions with exotic fermions, 
one expects $|\beta_{R}^{b \tau}|=1$ (see Section~\ref{sect:UV}). A value $|\beta_{R}^{b \tau}|\ll 1$ can 
be obtained, for instance, in models where the $U_1$ is a composite state. 
The condition $|\beta_{R}^{b \tau}|=1$ does not fix the phase of $\betaR$. In the large-$\betaR$ 
scenario we set $\betaR=-1$ in order to maximize the constructive interference of left-handed 
and right-handed contributions in the charged-current anomalies~\cite{Bordone:2017bld}. For similar reasons, we assume $\beta_{L}^{s \tau}$ to be real (in the down-quark mass basis). In the absence of observables providing stringent 
constraints on the corresponding phases, we also assume $\beta_{L}^{b \mu}$ and $\beta_{L}^{s \mu}$ to be real.
Note, however, that we treat $\beta_{L}^{d \tau}$ as a complex parameter, because its phase plays an 
important role in $B_d$\,--$\bar B_d$ mixing. The observables entering the fit (in addition to those discussed in Section~\ref{sect:EFT} and collected in Tables \ref{tab:bslldata} and \ref{tab:bclnudata}), together with their SM predictions and experimental values, are reported in Table~\ref{tab:fitobs}. These include LFU tests in $\tau$ decays, encoded in the ratios $\left(g_\tau/g_{e,\mu}\right)_{\ell, \pi, K}$ defined in \eqref{eq:deftauLFUV}, $B$ decays based on the $b \to s \tau^+ \tau^-$ and $b \to s \tau^{\pm} \mu^{\mp}$ transitions, the LFV tau decays $\tau \to \mu \gamma$ and $\tau \to \mu \phi$, and $\Delta F = 2$ amplitudes. The choice of these observables is motivated by their potential in constraining the fit parameters. 

\begin{table}[t]
\centering
\renewcommand{\arraystretch}{1.2} 
\begin{tabular}{cccc}
\toprule
{\bf Observable} & {\bf Experiment/constraint}  & {\bf SM prediction}  & {\bf Theory expr.}  \\
\specialrule{0.75pt}{1pt}{1pt}
$ ( g_{\tau}/g_{e,\mu} )_{\ell, \pi, K} $ & $1.0012 \pm 0.0012$ \cite{Amhis:2016xyh} & $1$ &\eqref{eq:tauLFUV} \\
$\mathcal{B}(B_{s} \to \tau^{+} \tau^{-})$ & $(-0.8\pm 3.5) \times 10^{-3}$ \cite{Aaij:2017xqt} &$(7.73\pm 0.49)\times 10^{-7}$ \cite{Bobeth:2013uxa} &\eqref{eq:B2ll}\\
$\mathcal{B}(B^+ \to K^+ \tau^{+} \tau^{-})$ & $(1.35\pm 0.70) \times 10^{-3}$ \cite{TheBaBar:2016xwe} &$(1.4\pm0.2)\times 10^{-7}$~\cite{Cornella:2019hct} &\eqref{eq:B2Ktautau}\\
$\mathcal{B}(B_s \to \tau^{\pm} \mu^{\mp}) $ & $<4.2\times 10^{-5} ~ (95\%\,\mathrm{CL})$ \cite{Aaij:2019okb}& $0$ & \eqref{eq:Bs2taumu} \\
$\mathcal{B}(B^{+} \to K^{+} \tau^{+} \mu^{-}) $ & $<3.3 \times 10^{-5}~ (95\%\,\mathrm{CL})$ \cite{Lees:2012zz} & $0$ & \eqref{eq:B2Ktaumu} \\
$\mathcal{B}(\tau \to \mu \gamma) $ & $<5.2 \times 10^{-8} ~ (95\%\,\mathrm{CL})$ \cite{Amhis:2016xyh} & $0$ &\eqref{eq:tau2mugamma}\\
$\mathcal{B}( \tau \to \mu \phi)$ & $< 1.0 \times 10^{-7} ~ (95\%\,\mathrm{CL})$ \cite{Miyazaki:2011xe} & $0$ &\eqref{eq:tau2muphi} \\ 
\midrule
$\delta (\Delta m_{B_s})$  &  $ 0.0 \pm0.1$  [*] & $0$ &   \\
$\delta (\Delta m_{B_d})$  &  $ 0.0 \pm 0.1$ [*] & $0$ &  \\
$\delta(\phi_d)\,[^o]$ &  $-1.0\pm0.9$~\cite{Bona:2007vi,Alpigiani:2017lpj}   & $0$ &\eqref{eq:MesonMixing}--\eqref{eq:MesonMixing2} \\
${\rm Im}(\cC_{uc}^{\rm NP})~[{\rm GeV}^{-2}]$ &  $(-0.03 \pm 0.46)\times10^{-14}$~\cite{Dmix,Bona:2007vi}  & $0$ &  \\
${\rm Re}(\cC_{uc}^{\rm NP})~[{\rm GeV}^{-2}]$ &  $ (0.3 \pm 1.4)\times10^{-13}$~\cite{Dmix,Bona:2007vi}   & $0$ &  \\
\bottomrule
\end{tabular}
\caption{\label{tab:fitobs} Low-energy observables included in the fit of the $U_1$ couplings (in addition to the observables in Tables \ref{tab:bslldata} and \ref{tab:bclnudata}). The entries marked with a [*] denote our constraint imposed on the magnitude of the 
$\Delta F=2$ amplitudes (see text for further explanation).}
\end{table}

\begin{figure}[p]
\centering
\includegraphics[width=0.49\textwidth]{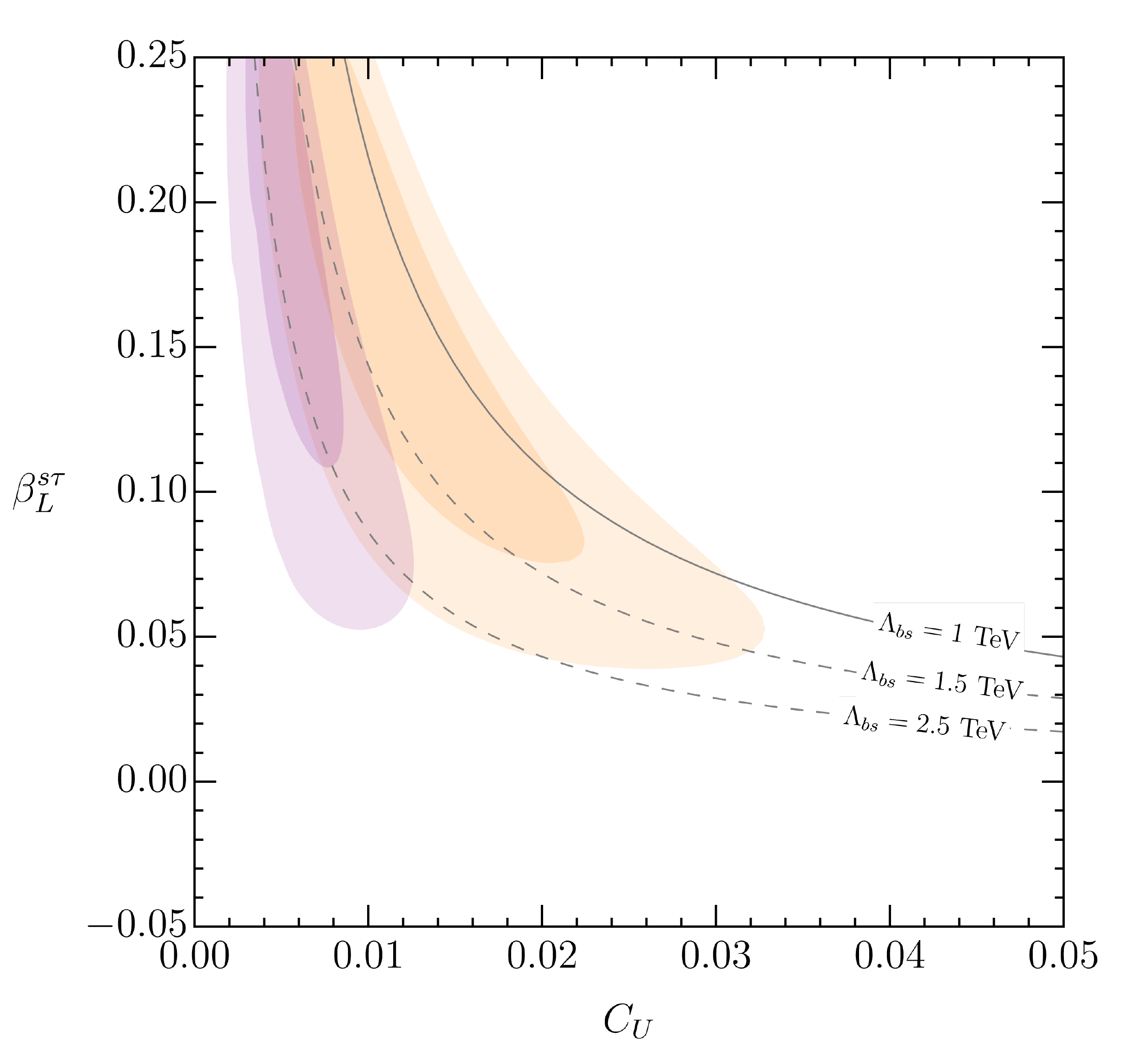} 
\includegraphics[width=0.49\textwidth]{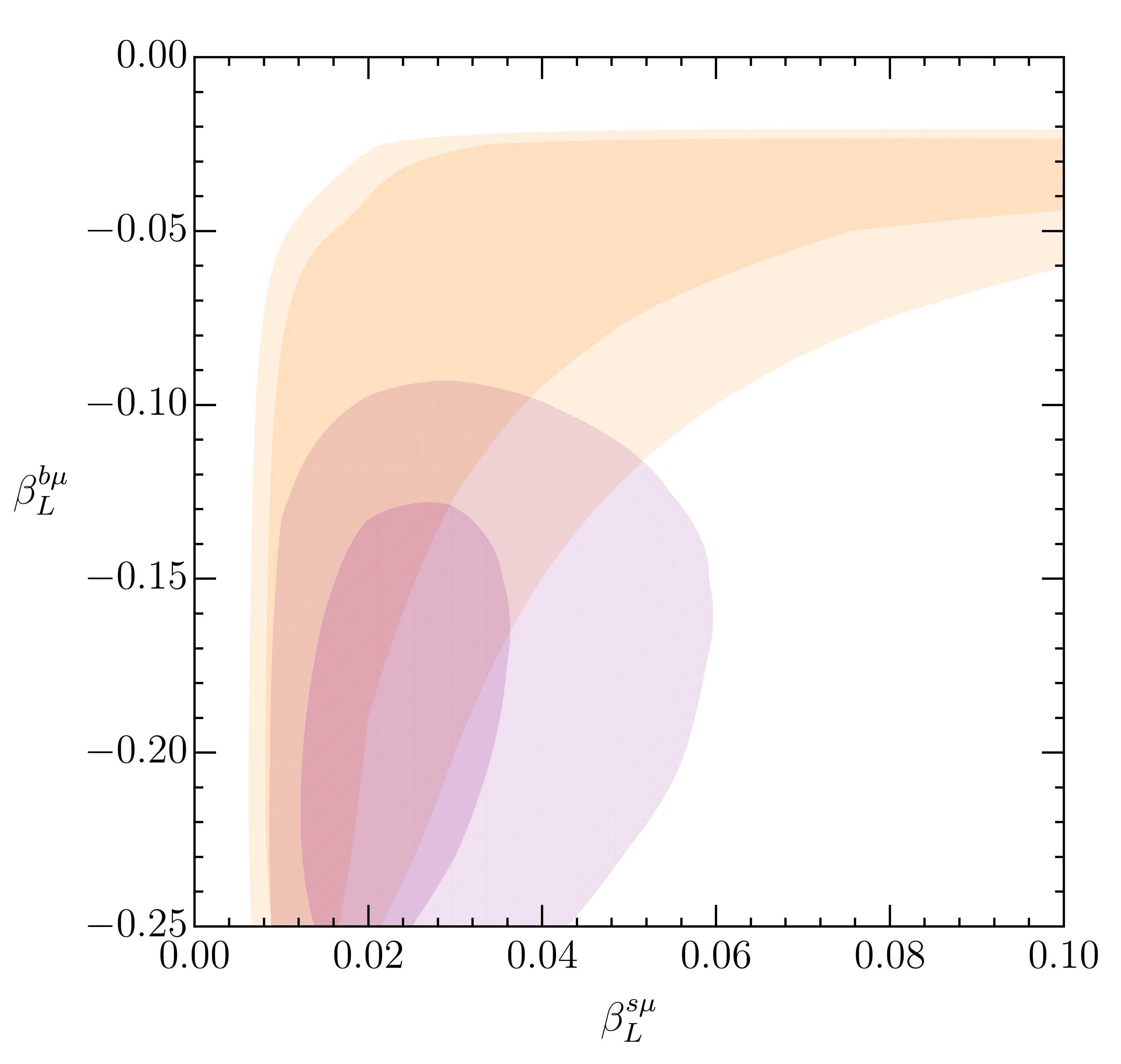} 
\includegraphics[width=0.49\textwidth]{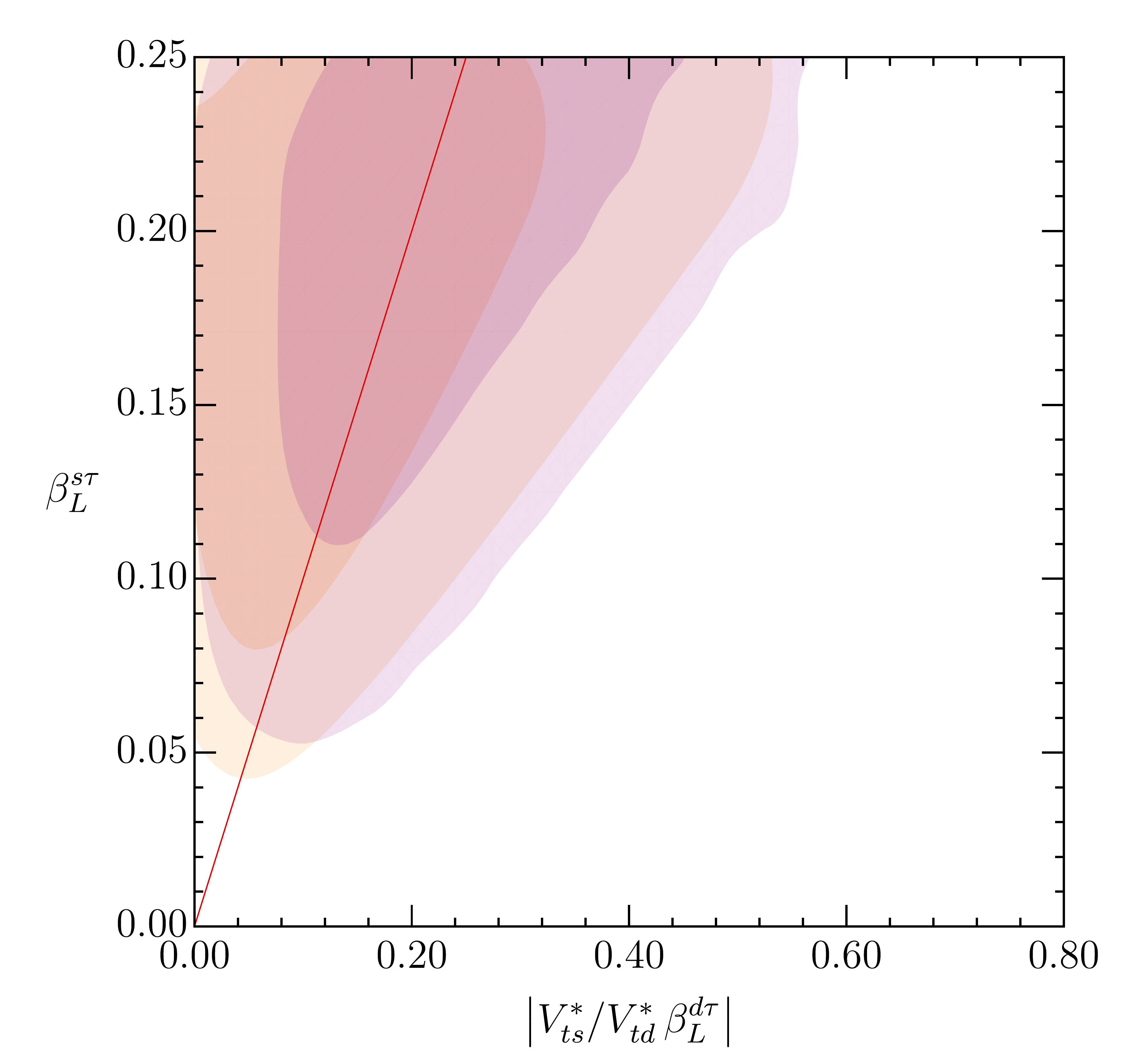} 
\includegraphics[width=0.49\textwidth]{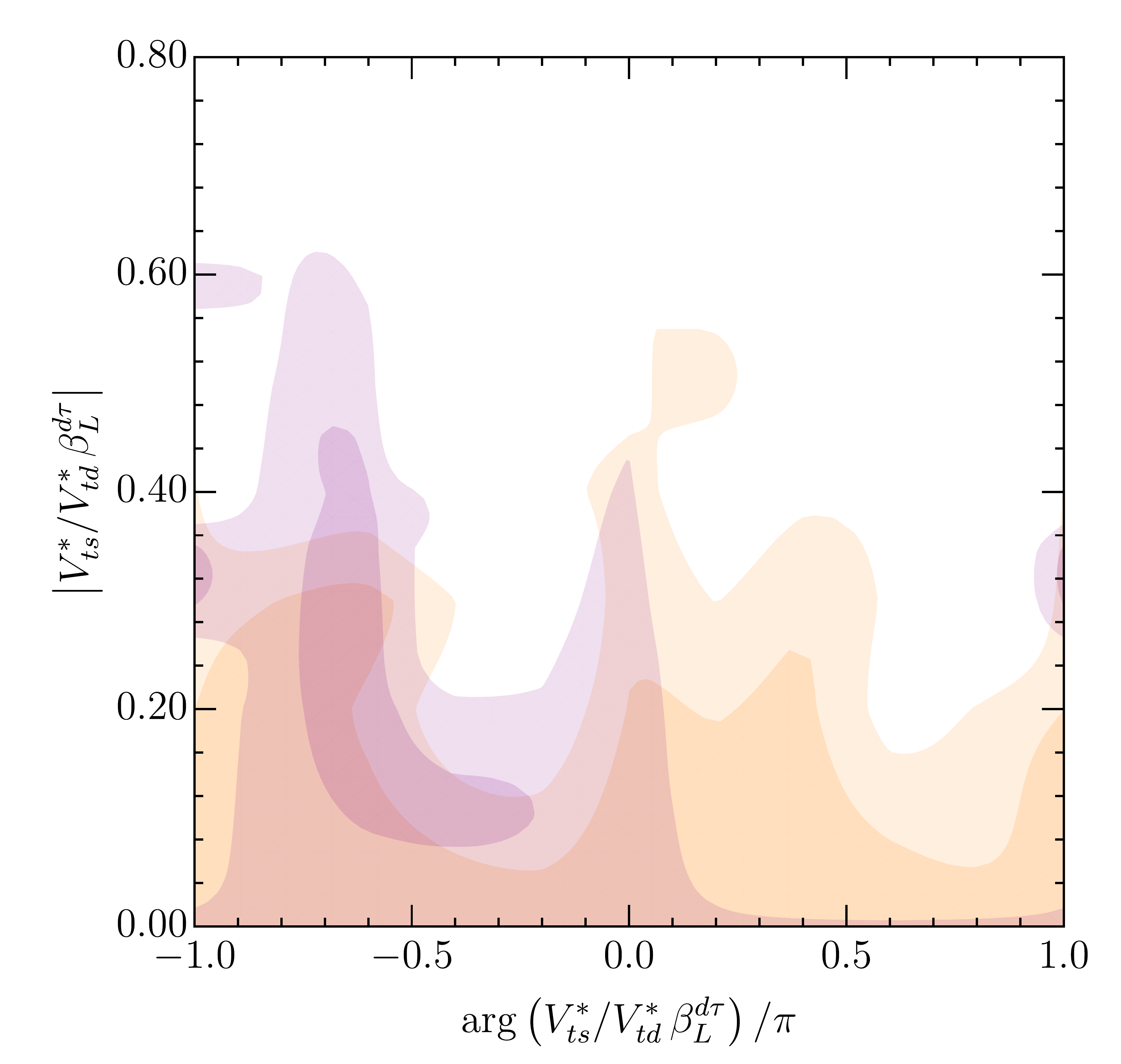} 
\caption{\label{fig:U1fit} Two-dimensional constraints on the $U_1$ couplings obtained by including the low-energy inputs in Tables \ref{tab:bslldata}, \ref{tab:bclnudata} and \ref{tab:fitobs}. The two colors correspond to the benchmarks $\betaR=0$  (orange) and $\betaR=-1$ (purple), with $\beta_L^{d \tau}$ unconstrained. For each benchmark, the darker and lighter regions denote $\Delta \chi^2 \leq 2.30$ ($1\sigma$) and $\Delta \chi^2 \leq 6.18$ ($2\sigma$), respectively.  In the $C_U$--$\beta_L^{s\tau}$ plot (upper left), we also show dashed contour lines corresponding to $\delta (\Delta m_{B_s})=0.1$ for different values of  $\Lambda_{bs}$ in (\ref{eq:MesonMixing}). In the $\beta_L^{s\tau}$--$\beta_L^{d\tau}$ plot (lower left) the red solid line corresponds to the $U(2)^5$ relation $\beta_{L}^{d \tau}/\beta_{L}^{s \tau} = V^*_{td}/V^*_{ts}$.}
\end{figure}

The fit results are shown in Figure~\ref{fig:U1fit} and summarized in  Table~\ref{tab:fitresults}. Such results represent a refined version 
of the analysis presented in \cite{Cornella:2019hct}, with updated inputs 
and a series of relevant differences, as listed below.
\begin{itemize}
\item{} Contrary to \cite{Cornella:2019hct}, we analyze two scenarios with $\betaR=0$ or $\betaR=-1$.
Moreover, in both cases we analyze the impact of relaxing the minimally-broken $U(2)^5$ relation $\beta_{L}^{d \tau}/  \beta_{L}^{s \tau} = {V_{td}^\ast}/{ V_{ts}^\ast}$
on the subleading coupling $\beta_{L}^{d \tau}$.
\item{} We impose a smooth cutoff on large values of $|\beta_L^{s\tau}|$ and $|\beta_L^{b\mu}|$ via a Gaussian suppression factor 
with $\sigma =0.05$ for $|\beta_L^{s\tau,b\mu}|>0.2$.
\item{} We include constraints from the $\Delta F=2$ mixing amplitudes as reported in Table~\ref{fig:U1fit}. 
Since these amplitudes depend on the UV completion of the model, we implement the corresponding constraints in a ``mild'' way in order to minimize the model-dependence. 
In practice, we use the estimates derived in (\ref{eq:MesonMixing}) and evaluate them setting 
$\Lambda_{bs}=\Lambda_{bd}=\Lambda_{uc}=1~{\rm TeV}$.
Moreover, in order to avoid a possible bias from UV-sensitive observables, we require the contributions to 
$|\delta(\Delta m_{B_{s,d}})|$ thus estimated 
not to exceed $10\%$ (at $1\sigma$),
which is in line with the present error on $|\delta(\Delta m_{B_{s,d}})|$ from global CKM fits~\cite{Bona:2007vi}.
\item{} We implement the one-loop contribution to $\tau\to \mu\gamma$ 
according to the complete result presented in~\cite{Fuentes-Martin:2020hvc}. As explained in \cite{Cornella:2019hct},  
this observable remains largely insensitive to the UV completion of the model (see the discussion in Section~\ref{sect:UV}).
\end{itemize} 

\begin{table}[t]
\centering
\renewcommand{\arraystretch}{1.3} 
\begin{tabular}{cccccc}
\toprule
\multicolumn{2}{c}{\bf{Scenario}}  &  $\boldsymbol{\Delta \chi^2}$ & {\bf Parameter} & {\bf best fit} & \bf $\boldsymbol{1 \sigma}$  \\
\specialrule{0.75pt}{1pt}{1pt}
\multirow{10}{90pt}{\centering no RH currents ($\betaR = 0$)} &  \multirow{4}{110pt}{\centering min. $U(2)^5 $ breaking ($\beta_L^{d \tau} = V_{td}^{\ast}/ V_{ts}^\ast  \, \beta_L^{s \tau}$)}  &  \multirow{4}{*}{$58.6$} 
& $C_U$ & $ 0.010 $ &  $  \left[0.007,  0.017 \right] $  \\
&&& $\beta_L^{b \mu}$ & ${-0.15}$ & $ {\left[-0.26,  -0.02 \right]}$ \\
&&& $\beta_L^{s \tau}$ & $ 0.19 $ &  $\left[0.10,  0.25 \right]$  \\
&&& $\beta_L^{s \mu}$ & $ 0.014 $ &  $ \left[0.004,  0.14 \right]$   \\
 \cmidrule(lr{1em}){2-6}
  &  \multirow{6}{*}{\centering $\beta_{L}^{d \tau}$ free}  &  \multirow{6}{*}{$59.5$}   &$C_U$ & $ 0.011 $ &  $  \left[0.007,  0.018 \right] $  \\
&&& $\beta_L^{b \mu}$ & $-0.14$  & $\left[-0.25,  -0.02 \right]$  \\
&&& $\beta_L^{s \tau}$ & $ 0.19 $ &  $ \left[0.11,  0.24 \right] $  \\
&&& $\beta_L^{s \mu}$ & $ 0.013 $  &  $\left[0.005,  0.125 \right]$  \\
&&& $ \abs{ V_{ts}^\ast/V_{td}^\ast \, \beta_L^{d \tau}} $ & $ 0.11 $ & $\left[0.04, 0.24  \right]$  \\
&&& $ \mathrm{arg}\left( V_{ts}^\ast/V_{td}^\ast \, \beta_L^{d \tau} \right)$ & $ 0.4\, \pi $ &  $ \left[0.0, 0.5 \right] \pi$  \\
\midrule
\midrule
 \multirow{10}{90pt}{\centering max. RH currents ($\betaR = -1$)} &  \multirow{4}{110pt}{\centering min. $U(2)^5 $ breaking ($\beta_L^{d \tau} = V_{td}^{\ast}/ V_{ts}^\ast  \, \beta_L^{s \tau}$)}  &  \multirow{4}{*}{$54.0$}  
&$C_U$ & $  0.004 $ &  $ \left[0.002,  0.006 \right]$ \\
&&& $\beta_L^{b \mu}$ & $ -0.21 $ & $ \left[-0.26,  -0.16 \right]$  \\
&&& $\beta_L^{s \tau}$ & $ 0.21 $ &  $ \left[0.12,  0.26 \right] $  \\
&&& $\beta_L^{s \mu}$ & $ 0.03 $ &  $ \left[0.01,  0.04 \right]$  \\
 \cmidrule(lr{1em}){2-6}
 &  \multirow{6}{*}{\centering $\beta_{L}^{d \tau}$ free}  &  \multirow{6}{*}{$56.7$}   
&$C_U$ & $ 0.005  $ &  $\left[0.004,  0.007  \right]$ \\
&&& $\beta_L^{b \mu}$ & $ - 0.21 $ & $ \left[-0.25,  -0.14 \right] $  \\
&&& $\beta_L^{s \tau}$ & $ 0.21  $ &  $ \left[0.15,  0.26 \right]$  \\
&&& $\beta_L^{s \mu}$ & $0.02$ &  $ \left[0.01,  0.07 \right]$  \\
&&& $  \abs{V_{ts}^\ast/V_{td}^\ast \, \beta_L^{d \tau}} $ & $0.24$ & $\left[0.13,0.34\right]$  \\
&&&  $ \mathrm{arg}\left( V_{ts}^\ast/V_{td}^\ast \, \beta_L^{d \tau} \right) $ & $ -0.6\, \pi $ &  $ \left[-0.7,-0.5 \right] \pi $ \\
\bottomrule
\end{tabular}
\caption{\label{tab:fitresults}
Fit results for $\beta_{R}^{b \tau} = 0$ and $\beta_{R}^{b \tau} = -1$, with and without the additional assumption of minimal $U(2)^5$ breaking. For each scenario we report the $\Delta \chi^2 = \chi^2 - \chi^2_{\rm SM}$, the best-fit values and the $1 \sigma$ confidence intervals (marginalizing over the other parameters) for all fit parameters.}
\end{table} 

\begin{figure}[t]
\centering
\includegraphics[width=0.48\textwidth]{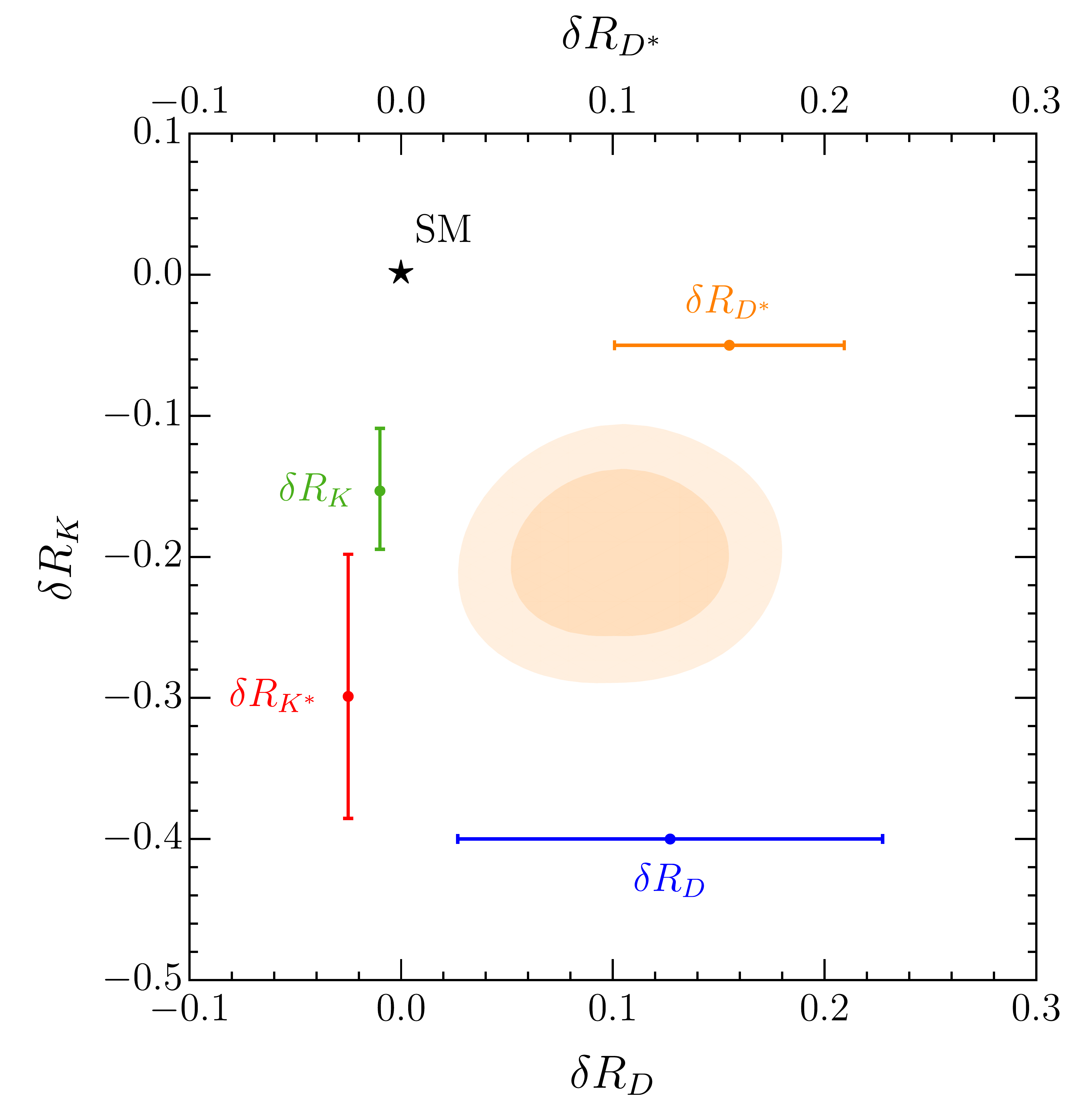}   \quad
\includegraphics[width=0.48\textwidth]{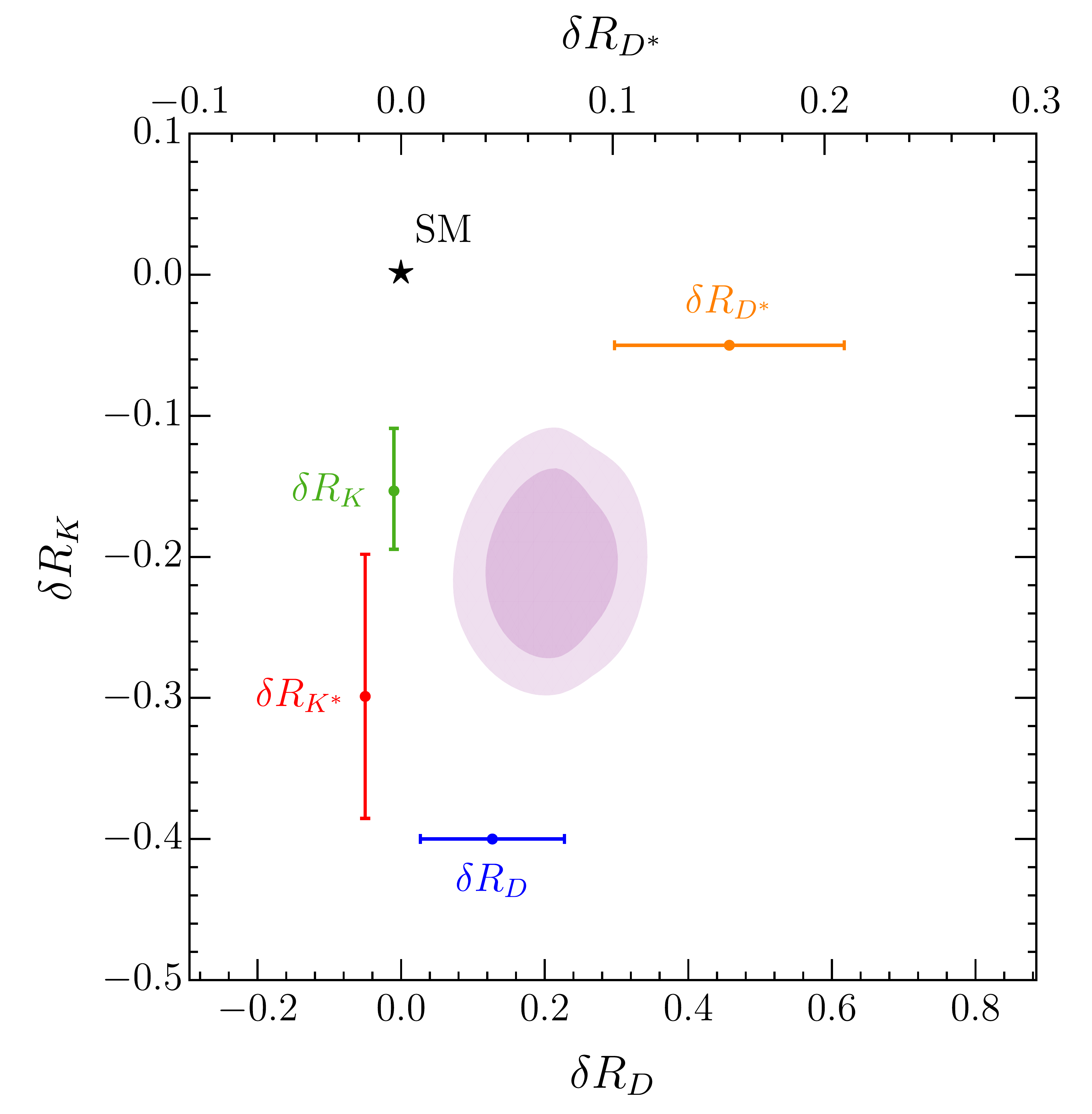}  
\caption{  \label{fig:RKrD} 
Preferred $1\sigma$ and $2\sigma$ regions for the ratios $\delta R_{D^{(\ast)}}$ 
and $\delta R_{K}^{[1.1,6]}$ resulting from the low-energy fit for $\betaR=0$ (orange) and $\betaR=-1$ (purple). Note that in both cases $\delta R_K^{\ast} \approx \delta R_K$ to a very good approximation. 
The colored error bars show the current experimental measurements at $1 \sigma$.}
\end{figure}

\paragraph{Discussion.} From Table~\ref{tab:fitresults} and the plots in Figure~\ref{fig:U1fit} 
we draw the following conclusions: 
\begin{itemize}
\item{} All four scenarios provide a very good fit to the data.  Taking into account that the NP fits have 4 or 6 parameters, 
 depending on whether we leave  $\beta_{L}^{d \tau}$ free to vary in magnitude and phase,
 the $\Delta \chi^2$ values reported in Table~\ref{tab:fitresults} indicate a strong 
 significance of the NP hypothesis   (well above the $5\sigma$ level) compared to the SM in all four cases.
\item{} A different way to illustrate the previous statement is offered by Figure~\ref{fig:RKrD}, where we show the 
$1\sigma$ and $2\sigma$ preferred regions for the LFU-violating observables
 \be
\delta R_{K^{(*)}} =  \frac{ R_{K^{(*)}}-  R^{\rm SM}_{K^{(*)}} }{
R^{\rm SM}_{K^{(*)}} }\,, \qquad  \delta R_{D^{(*)}} =  \frac{ R_{D^{(*)}} -  R^{\rm SM}_{D^{(*)}} }{ 
 R^{\rm SM}_{D^{(*)}} }\,, \qquad 
 \ee
obtained for the two reference options $\betaR=0$ with minimal $U(2)^5$ breaking, and $\betaR=-1$ with generic $\beta_{L}^{d \tau}$, corresponding to the two benchmark scenarios analyzed in Section~\ref{sect:EFT}.
 As can be seen, the overall agreement with data is very good in both cases. The 
$\betaR=-1$ case is slightly disfavored (as indicated by the lower $\Delta \chi^2$ values in Table~\ref{tab:fitresults}), because of the lower prediction of $R_{D^*}$ compared to data.
\item{}  As found in previous studies~\cite{Buttazzo:2017ixm,Cornella:2019hct}, there is a strong correlation between $C_U$ and 
$\beta_{L}^{s \tau}$ (upper left plot in Figure~\ref{fig:U1fit}). This is a consequence of  $\delta R_{D^{(*)}}$, which  
fixes the product of $C_U$ and $\beta_L^{s\tau}$. 
\item{} The value of $\delta R_{K^{(*)}}$ fixes the size and the sign of the product of $\beta_L^{b\mu}$ and $\beta_L^{s\mu}$. The individual signs of the two couplings are not determined, and we choose $\beta_L^{b\mu}<0$ and $\beta_L^{s\mu}>0$ when displaying the fit results. The magnitudes of the $\beta_{L}^{i\ell}$ parameters are consistent with the expected scaling rules $|\beta_L^{s\tau}|\sim\epsilon_q\sim 0.1$, $|\beta_L^{b\mu}|\sim\epsilon_\ell\sim 0.1$ and $|\beta_L^{s\mu}|\sim\epsilon_q\hspace{0.3mm}\epsilon_\ell\sim 10^{-2}$.  
\item{}  As already found in the EFT analysis, the $\Delta F=2$ constraint is particularly relevant, especially in the scenario with only left-handed couplings. 
Depending on the value of the effective mass parameter $\Lambda_{bs}$ in (\ref{eq:MesonMixing}), it cuts out a significant region parameter space above the dashed contour lines shown in the upper left plot in Figure~\ref{fig:U1fit}. For the relatively low value $\Lambda_{bs}=1$~TeV adopted in our analysis, the effect of imposing this cut is still relatively mild, but larger values would give rise to much tighter cuts. As mentioned earlier, the value of 1~TeV is motivated by the fact that in the specific UV completion we have in mind the parameter $\Lambda_{bs}$ is associated with the mass of a heavy vector-like lepton 
(see Section~\ref{sect:UV} for more details). 
\item{}  The minimal $U(2)^5$-breaking relation $\beta_{L}^{d \tau}/\beta_{L}^{s \tau} = V^*_{td}/V^*_{ts}$
is well supported by data. This becomes evident when one compares the $\Delta \chi^2$ values in Table~\ref{tab:fitresults} obtained with and without imposing this hypothesis. However, as shown by the lower plots in Figure~\ref{fig:U1fit}, $\cO(1)$ deviations in both magnitude and phase
are still allowed (and slightly favored in the $\betaR=-1$ case).
 \end{itemize}

\begin{figure}[t]
\centering
\includegraphics[width=0.48\textwidth]{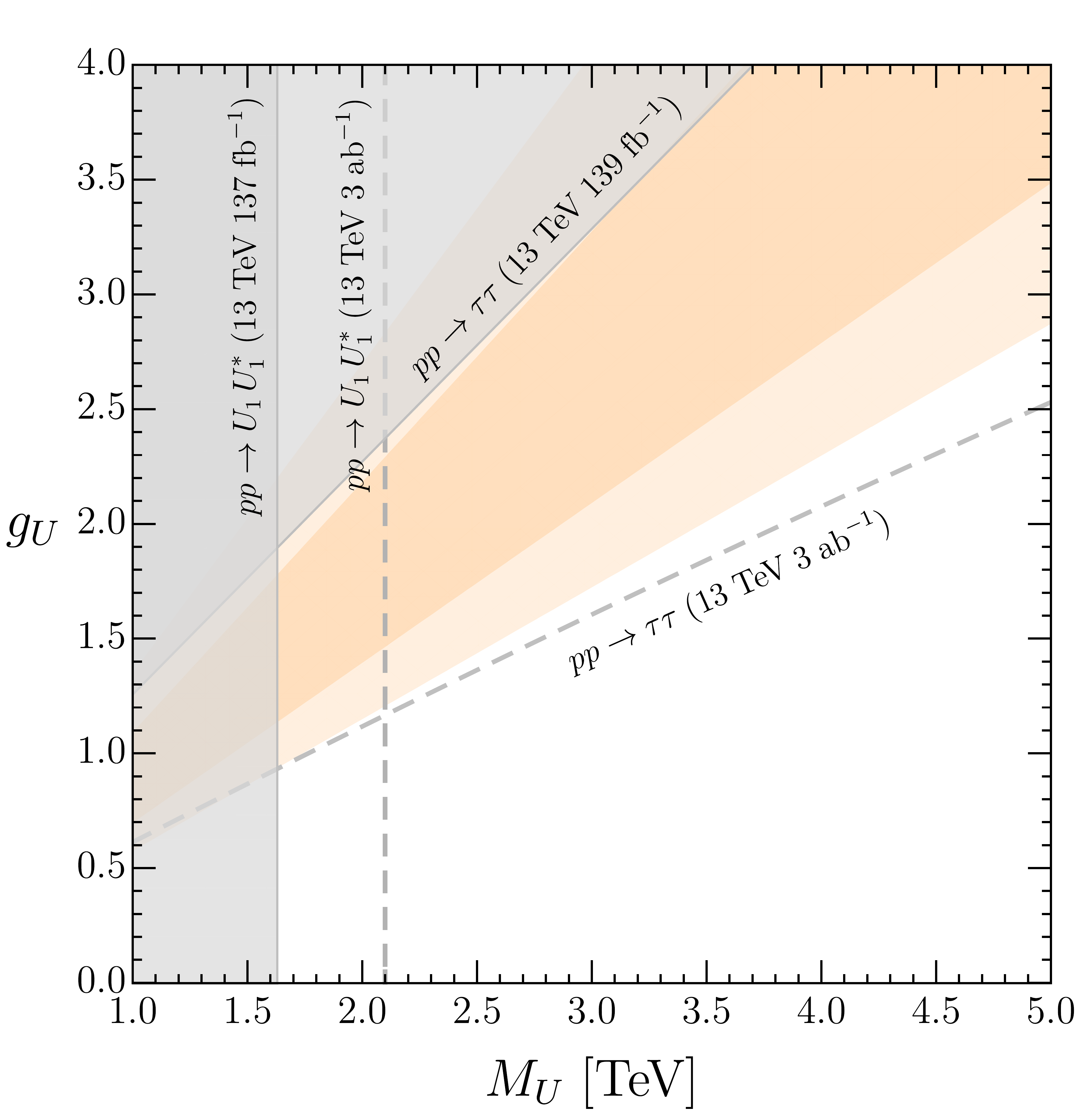}   \quad
\includegraphics[width=0.48\textwidth]{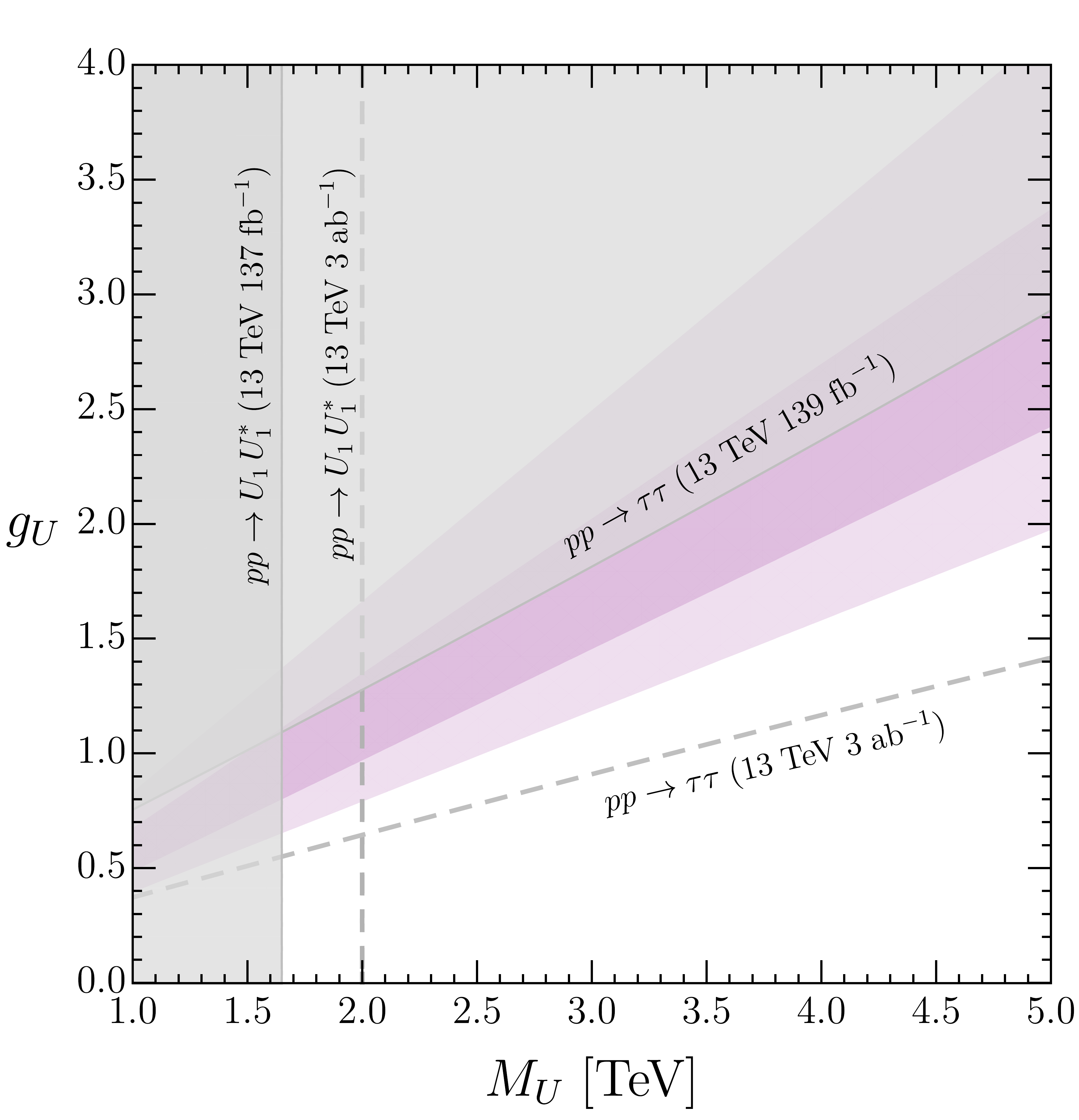}  
\caption{LHC constraints for the $U_1$ vector leptoquark for the benchmark scenarios with  $\beta_R^{b\tau}=0$ (left) and $\beta_R^{b\tau}=-1$ (right). The $1\sigma$ and $2\sigma$ regions obtained from the fit to low-energy data are also shown.}\label{fig:U1highpT}
\end{figure}

\subsection{Constraints from high-\texorpdfstring{$p_T$}{pT} observables}

Having discussed the most relevant low-energy constraints, we now turn our attention to the bounds from collider (``high-$p_T$'') physics. We focus here on the constraints on the $U_1$ leptoquark that can be derived within the simplified model defined by the Lagrangian in~\eqref{eq:LQLag}, and postpone the discussion of effects of possible additional TeV-scale states to Section~\ref{sect:HighpTAdditional}. As we did for the low-energy fit, we consider the two reference benchmark scenarios $\beta_R^{b\tau}=0$ and $\beta_R^{b\tau}=-1$, and we assume the same $U(2)$-inspired scaling rules for the $\beta_L^{i\alpha}$ couplings discussed at the beginning of this section and supported by the low-energy fit. 

\begin{figure}[t!]
\centering
\includegraphics[width=0.32\textwidth]{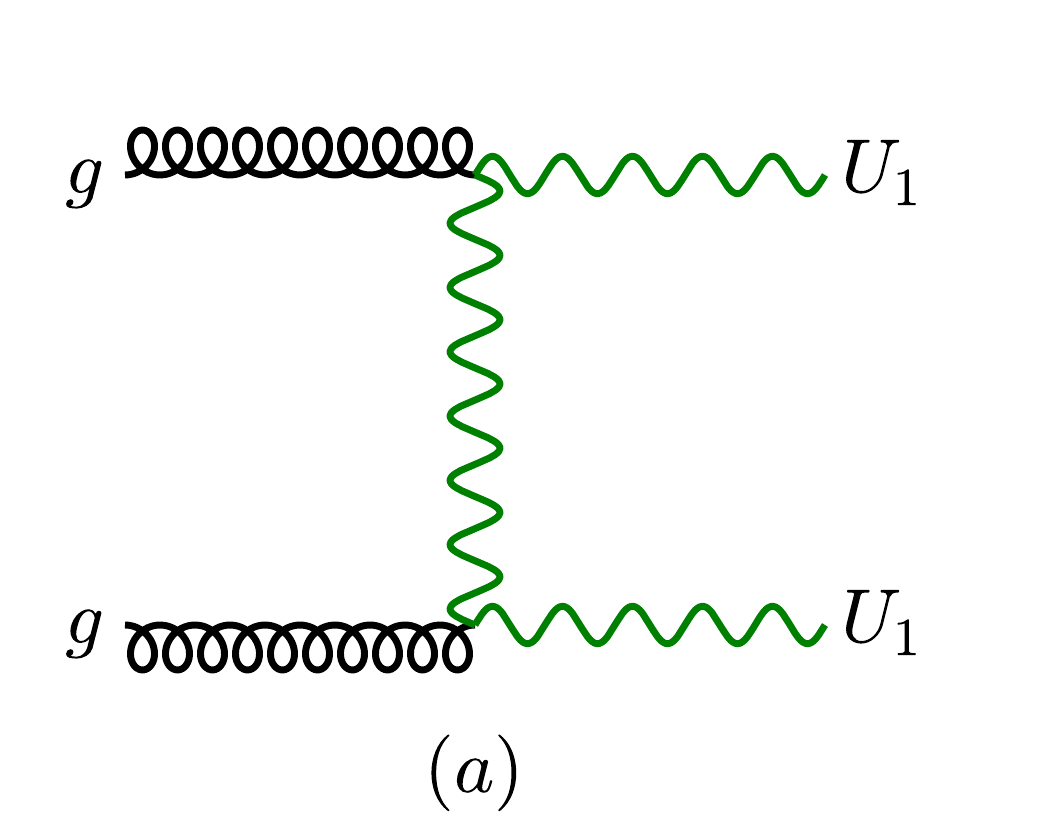}~~ 
\includegraphics[width=0.32\textwidth]{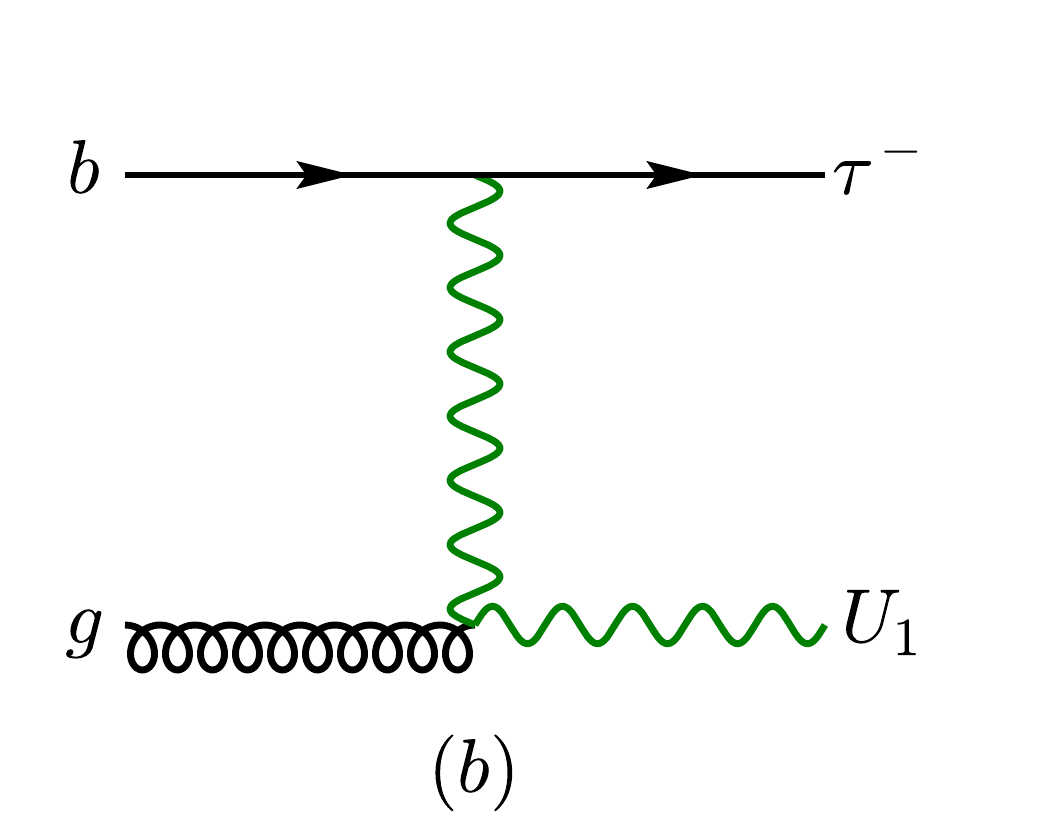}~~  
\includegraphics[width=0.32\textwidth]{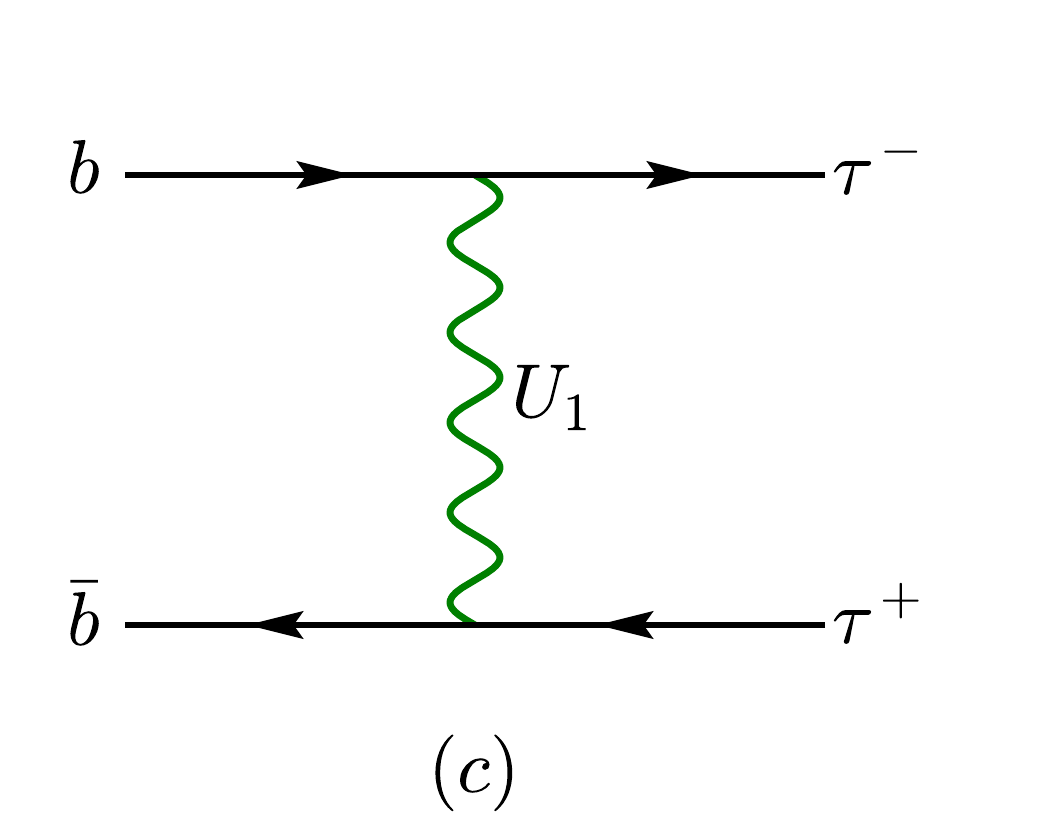}  
\caption{  \label{fig:DiagramsU1} Representative Feynman diagrams for vector leptoquark pair production (a), single-leptoquark production (b), and $t$-channel Drell-Yan production (c).}
\end{figure}

Leptoquark pair-production cross sections at the LHC are dominated by QCD dynamics~\cite{Blumlein:1996qp,Dorsner:2018ynv,Diaz:2017lit,Baker:2019sli} (figure \ref{fig:DiagramsU1} a) and thus are largely independent of the leptoquark couplings to fermions. Nevertheless, a certain model dependence is still retained in the form of non-minimal couplings to gluons, parameterized in the Lagrangian in~\eqref{eq:LQLag} by the quantity $\kappa_c$. In models where the vector leptoquark has a gauge origin, this non-minimal coupling is absent ($\kappa_c=0$), allowing for robust theory predictions for the pair-production cross section. As a consequence, the largest model dependence for this type of searches arises through the leptoquark branching fractions to its different decay channels~\cite{Mecaj:2020opd}. The flavor structure emerging from our analysis of the $B$-meson anomalies suggests that the dominant decay channels are those involving pairs of third-generation fermions, namely $U_1\to b\tau^+$ and $U_1\to t\bar\nu$, with branching fractions that depend on the value of $\beta_R^{b\tau}$. For the benchmark scenarios considered here, the largest cross section is obtained for $pp\to U_1^* U_1\to b\tau t\nu$. The CMS collaboration has performed a dedicated search for this channel using $137~\mathrm{fb}^{-1}$ of $13~\mathrm{TeV}$ data~\cite{Sirunyan:2020zbk}. The corresponding exclusion regions (obtained for $\kappa_c=0$) are shown in Figure~\ref{fig:U1highpT} for both benchmark scenarios, together with the $1\sigma$ and $2\sigma$ regions obtained from the low-energy fit. We also show the projected limits for the high-luminosity phase of the LHC (HL-LHC with $3\,\mathrm{ab}^{-1}$ of integrated luminosity) under the assumption that no NP signal is detected and that statistical and systematic uncertainties scale with the square root of the luminosity. As can be seen, these searches offer only a relatively small coverage of the parameter space favored by the low-energy fit. Other direct searches, such as single-leptoquark production from quark-gluon scattering~\cite{Alves:2002tj,Hammett:2015sea,Mandal:2015vfa,Dorsner:2018ynv} (see figure \ref{fig:DiagramsU1} b) or resonant production via lepton-quark fusion~\cite{Haisch:2020xjd,Buonocore:2020erb,Greljo:2020tgv} (exploiting the recently determined lepton PDFs from photon splitting~\cite{Buonocore:2020nai}) will play a crucial role in the event of a discovery, but are currently not competitive with other high-$p_T$ searches.

Another interesting collider constraint is obtained by searching for modifications of the high-$p_T$ tail in the dilepton invariant mass distribution in the Drell-Yan process $pp\to\tau^+\tau^- +X$ induced by $t$-channel $U_1$ exchange~\cite{Faroughy:2016osc,Schmaltz:2018nls,Baker:2019sli,Angelescu:2021lln} (see figure \ref{fig:DiagramsU1} c).\footnote{Analogous limits from $pp\to\mu\tau$~\cite{Baker:2019sli,Angelescu:2020uug} and $pp\to\mu^+\mu^-$~\cite{Greljo:2017vvb} do not provide competitive bounds because of the flavor suppression of the light-lepton couplings, though they might play a relevant role in the future in the event of discovery. Similarly, limits derived from $pp\to\tau\bar\nu$~\cite{Greljo:2018tzh} are found to be weaker due to the smallness of $V_{cs}^*\,\beta_L^{s\tau}$ and $V_{cb}^*$ compared to the dominant third-generation couplings.} 
The dominant production mechanism for this channel is via a $b\bar b$ initial state, while contributions from $b\bar s$- and $s\bar s$-initiated processes are subdominant due to the underlying flavor structure of the leptoquark couplings. Stringent limits from $pp\to\tau^+\tau^- +X$ data can be obtained by recasting the ATLAS analysis in~\cite{Aad:2020zxo} with $139~\mathrm{fb}^{-1}$ of $13$~TeV data, following the same recasting procedure described in~\cite{Baker:2019sli}. As shown in Figure~\ref{fig:U1highpT}, high-$p_T$ lepton tails provide important constraints on the parameter space preferred by the low-energy fit, especially for $\beta_R^{b\tau}=-1$, where the limit is about two times stronger than in the $\beta_R^{b\tau}=0$ scenario. However, for both benchmark scenarios a large region of the parameter space still remains viable. Together with the present bounds, we also show the projected limits for the HL-LHC, again assuming a naive luminosity scaling of the uncertainties. Interestingly, we find that the preferred $1\sigma$ and $2\sigma$ regions for both benchmarks are completely within the reach of the HL-LHC. We stress that this sensitivity projections do not consider possible improvements in these searches, e.g.\ due to a finer and more extended binning of the transverse mass, which will be available when more events are collected, or by searching for $b$-tagged jets in the final state (see e.g.~\cite{Afik:2018nlr,Marzocca:2020ueu}).

\section{UV completion and further predictions} 
\label{sect:UV}

Any UV completion for a TeV-scale $U_1$ field coupled to SM fermions via the 
simplified Lagrangian in~(\ref{eq:LQLag}) requires additional TeV-scale heavy vectors and vector-like fermions. New scalar degrees of freedom (radial modes associated to the symmetry breaking sector) are also expected to arise, but their masses can be pushed in the $\cO(10~{\rm TeV})$ domain, yielding negligible impact on collider phenomenology and low-energy observables.
In all models based on the 4321 gauge group~\cite{DiLuzio:2017vat}, 
\be
\mathcal{G}_{4321}\equiv SU(4)\times SU(3)^\prime\times SU(2)_L\times U(1)_X\,, 
\ee
the additional heavy vectors feature a color octet, $G^\prime\sim(\boldsymbol{8},\boldsymbol{1},0)$, and 
a SM-singlet, $Z^\prime\sim(\boldsymbol{1},\boldsymbol{1},0)$. As discussed in~\cite{Baker:2019sli}, this choice is unavoidable in any UV completion (including composite models) by the closure of the $SU(4)$ algebra hosting  the $U_1$ generators and by the requirement of flavor non-universality.  The vector-like fermions, transforming as doublets of $SU(2)_L$,  are a key ingredient to generate a non-trivial flavor structure for the $U_1$ couplings to left-handed fermions, yielding flavor off-diagonal couplings such as $\beta_{L}^{s \tau}$ and $\beta_{L}^{b \mu}$.  As shown in~\cite{DiLuzio:2018zxy}, the vector-like fermion mixing can be realized in such a way that sizable off-diagonal entries are generated in the $U_1$ couplings while preserving a flavor-diagonal structure in the $G^\prime$ and $Z^\prime$ (tree-level) couplings. The underlying mechanism resembles the up-down (or $SU(2)_L$) CKM flavor misalignment but now in quark-lepton (or $SU(4)$) space.

The flavor non-universal but flavor-diagonal nature of $G^\prime$ and $Z^\prime$ couplings holds in a specific $SU(2)_L$ basis for quarks and leptons. The strong constraints from down-type $\Delta F=2$ observables and LFV charged lepton decays, which would receive tree-level contributions if off-diagonal couplings of the $G^\prime$ and $Z^\prime$ were present, are the main reason why we adopt the down-quark and charged-lepton mass eigenstate basis in \eqref{eq:DownBasis} as interaction basis, i.e the basis defining the different 4321 charges of the SM fermions. In such basis, all fields couple with the same strength, up to group factors, to third-generation fermions. The $U_1$ has sizable off-diagonal 
couplings to the light families controlled by the mixing with vector-like fermions, while $G^\prime$ and $Z^\prime$ couple (almost) universally 
to light fermions via small couplings suppressed by $g_s^2/g_U^2$ and $g^2_Y/g_U^2$ compared to the third-generation case~\cite{Cornella:2019hct}.

In this section, we analyze the predictions of this set up for a series of low- and high-energy observables.
We can group them in three categories: i)~low-energy observables insensitive to the details of the UV dynamics;
ii)~UV-sensitive low-energy observables; iii)~high-$p_T$ observables related to the 
additional heavy states predicted by the UV completion.

\subsection{UV-insensitive low-energy observables} \label{sect:UVinsensitive} 

\begin{figure}[t]
\centering
\includegraphics[width=0.48\textwidth]{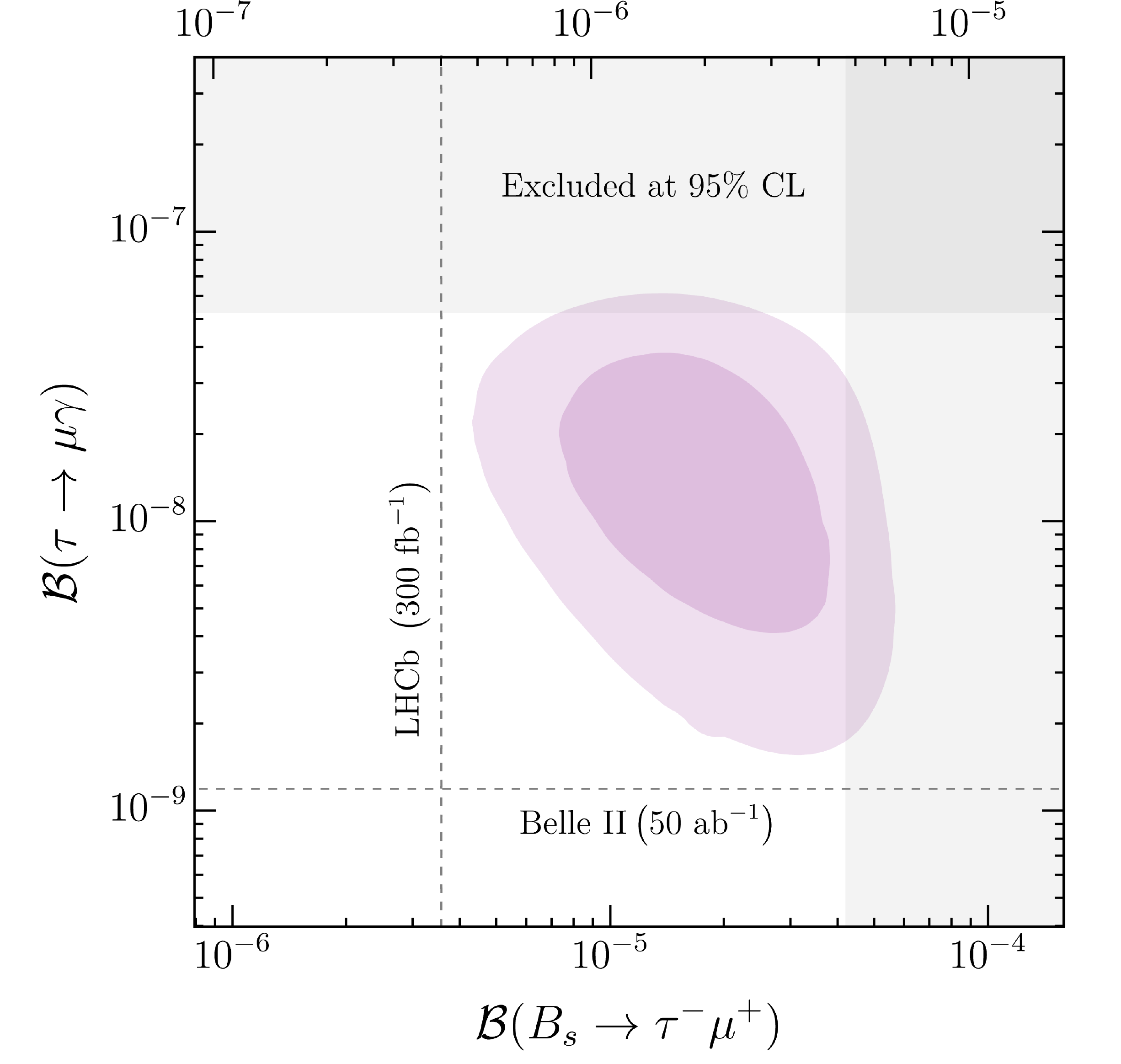}   \quad
\includegraphics[width=0.48\textwidth]{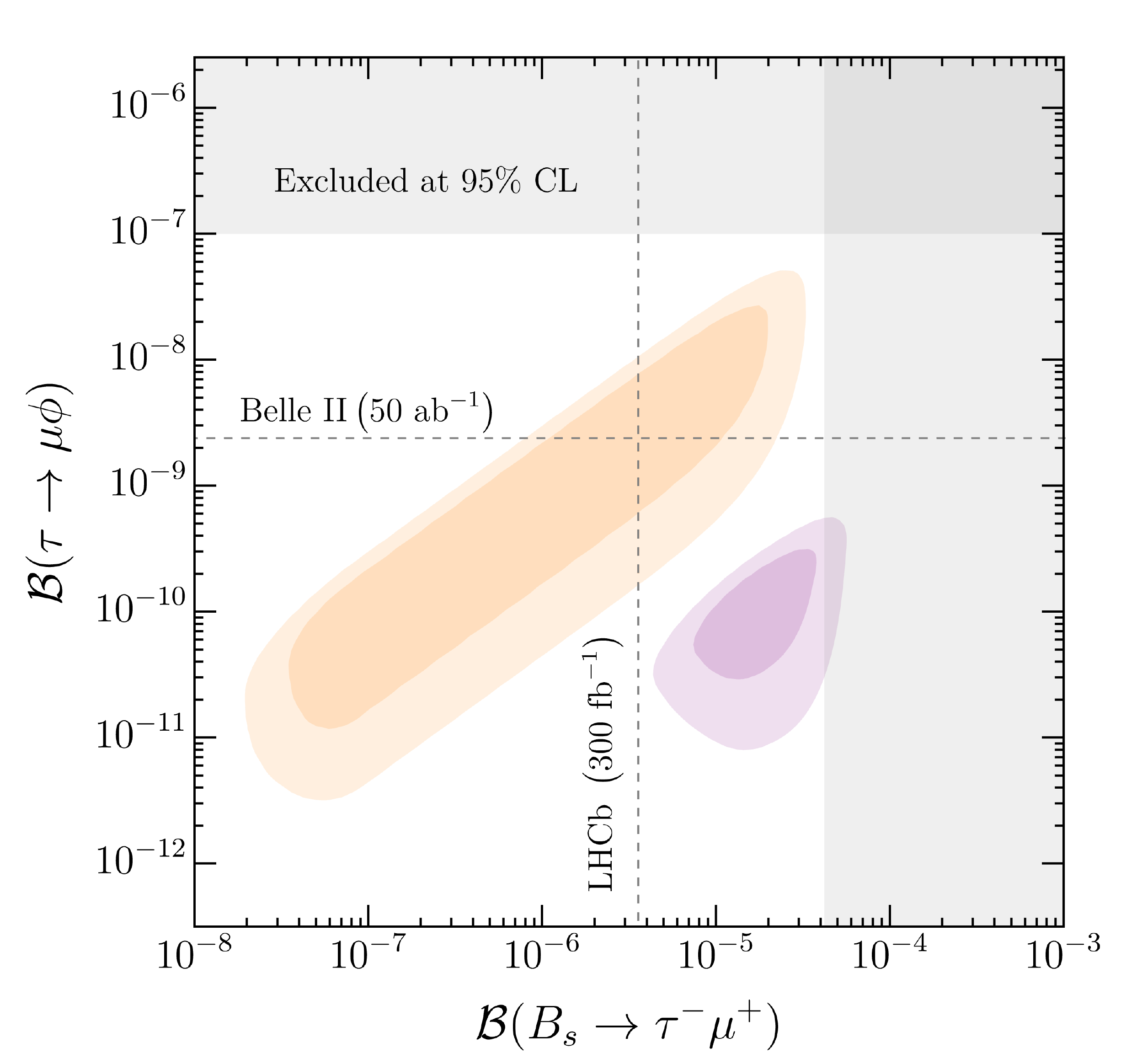}  
\caption{ \label{fig:LFV} Preferred $1\sigma$ and $2 \sigma$ regions for LFV processes resulting from the low-energy fit in the $U_1$ simplified model for $\betaR=0$ (orange) and $\betaR=-1$ (purple). The gray bands show the $95 \%$ CL experimental exclusion limits. Future projections (rescaled to $95 \%$ CL) are denoted by dashed lines. In the $x-$axis of the left plot only the 
more stringent exclusion band from $B_s \to \tau^- \mu^+$ is shown.}
\end{figure}

\begin{figure}[p]
\centering
\includegraphics[width=0.45\textwidth]{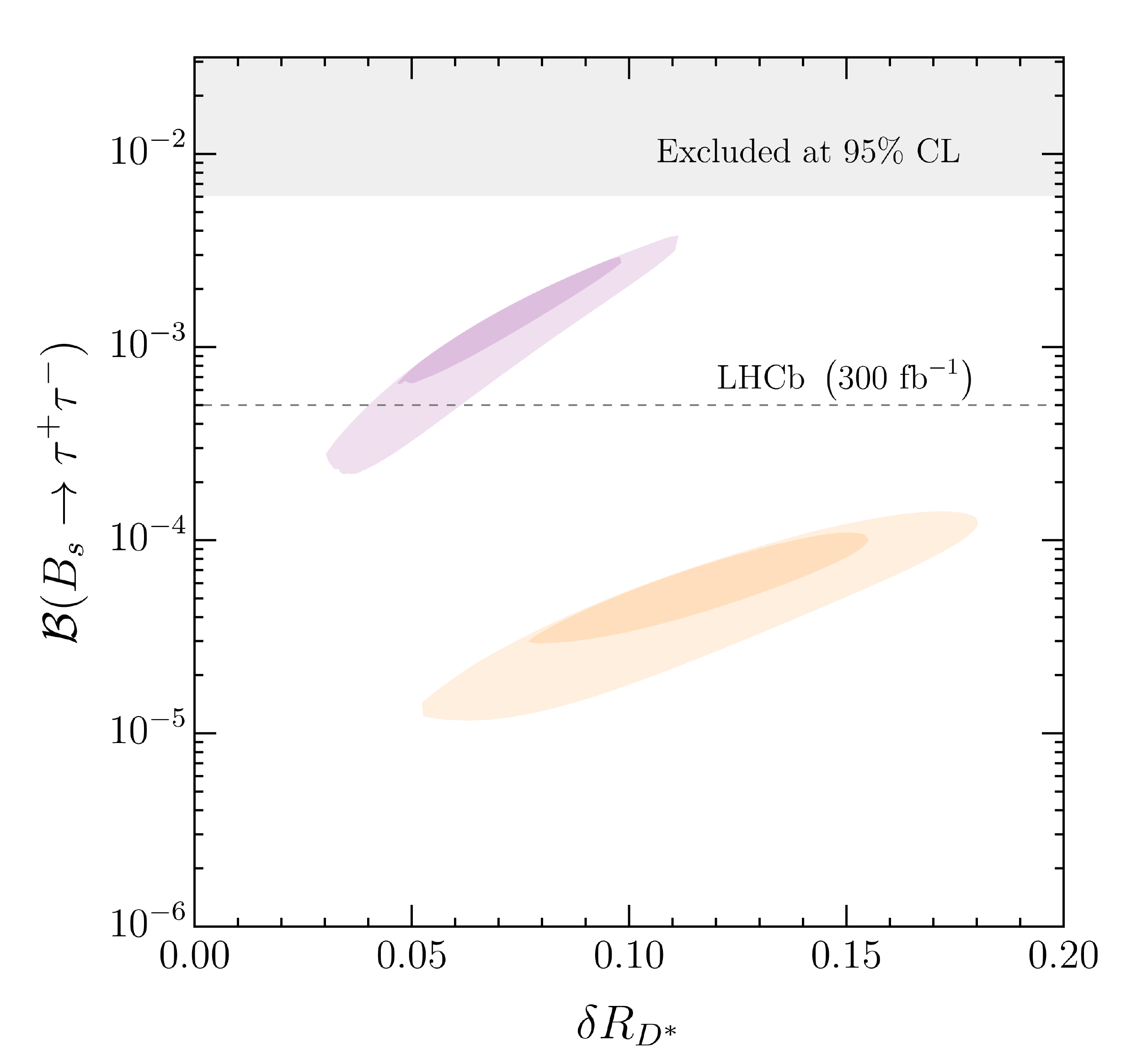} \quad
\includegraphics[width=0.45\textwidth]{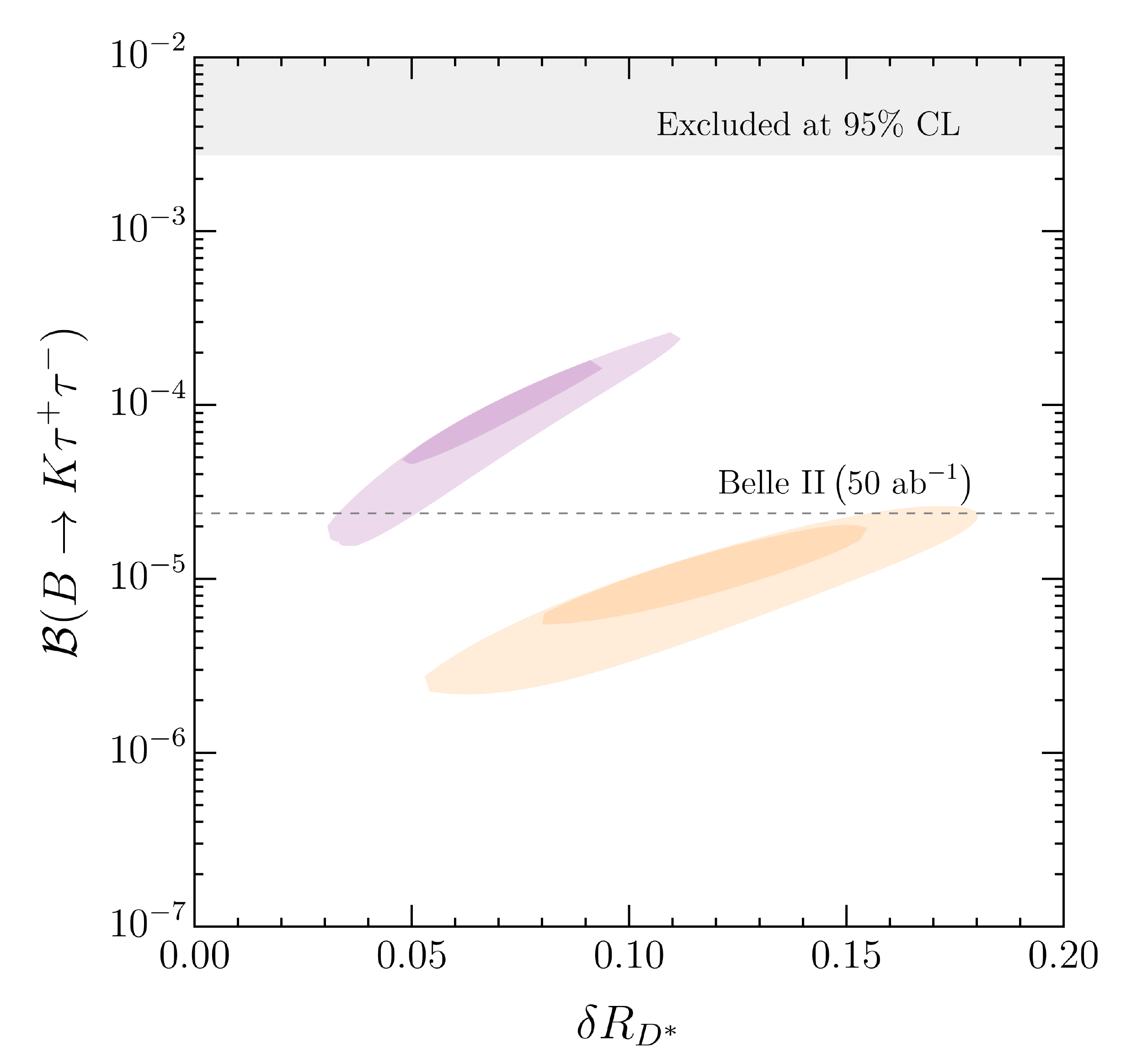} 
\includegraphics[width=0.45\textwidth]{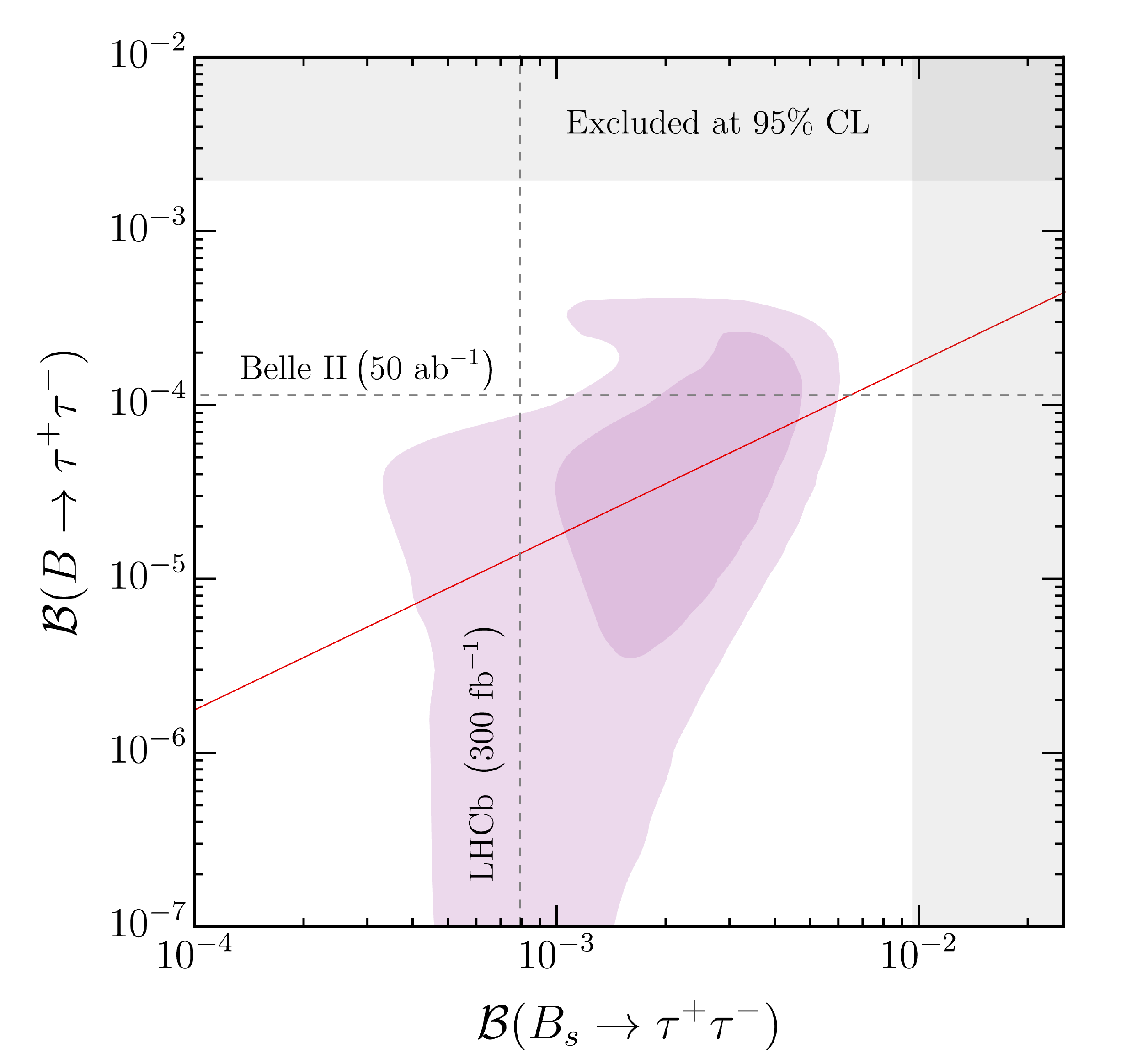}   
\caption{  \label{fig:ditau} Preferred $1\sigma$ and $2 \sigma$ regions for di-tau final states resulting from the low-energy fit in the $U_1$ simplified model for $\betaR=0$ (orange) and $\betaR=-1$ (purple). The gray bands show the $95 \%$~CL experimental exclusion limits. Future projections (rescaled to $95 \%$ CL) are denoted by dashed lines. The red line in the lower plot shows the expected relation between $\mathcal{B}(B_s \to \tau^+ \tau^-)$ and $\mathcal{B}(B \to \tau^+ \tau^-)$ for $\beta_{L}^{d \tau}/\beta_{L}^{s \tau} = V^*_{td}/V^*_{ts}$ (minimally broken $U(2)^5$).}
\end{figure}

Several low-energy observables can be predicted directly using the results of the $U_1$ simplified model fit.
The most interesting ones are the Lepton Flavor Violating (LFV) decays $B^+ \to K^+ \tau^+ \mu^-$, $B_s \to \tau^- \mu^+$, $\tau \to  \mu \gamma$ and $\tau \to  \mu \phi$ (Figure~\ref{fig:LFV}), and the rare $B_{(s)}$ decays into di-tau final states $B_{(s)} \to \tau^+ \tau^-$ and $B \to K \tau^+ \tau^-$ (see Figure~\ref{fig:ditau}).
With the exception of $\tau\to\mu \gamma$, all these observables are dominated by the tree-level contribution from the $U_1$ 
exchange. The radiative decay $\tau\to\mu \gamma$ is generated at the loop level, but for $\beta_R^{b \tau} =-1$ it is completely dominated by the contribution due to $b$-quark running inside the loop (see Section~\ref{sect:UVsensitive}), and is therefore insensitive to the UV completion of the model. The explicit expressions of the amplitudes in  terms of the $U_1$ couplings are reported in the Appendix~\ref{app:Obs}.
It is worth recalling that we used the present bounds on all these observables but $B \to \tau^+ \tau^-$ (which does not yield significant constraints on the parameter space) to determine the best fit points. As a consequence,
Figure~\ref{fig:LFV} and  Figure~\ref{fig:ditau} do not show unbiased predictions, but rather expected ranges for the observables, 
taking into account all the available information on the allowed parameter space of the model. As can be seen from these plots, the case with large $|\betaR|$ leads to predictions for  
$B_s \to \tau^+\tau^-$, $B \to K \tau^+\tau^-$, $B_s \to \tau^- \mu^+$, and $\tau\to\mu \gamma$ close to present bounds.
In the dilepton modes, this is because of the chiral-enhancement of the corresponding hadronic matrix element while for $\tau\to\mu \gamma$, this occurs because of the large loop-induced amplitude. This observable is not shown for
$\betaR=0$ since in this case it is well below present bounds and sensitive to the details of the UV completion, see Section~\ref{sect:UVsensitive}. 

We recall that the two benchmarks we have chosen for $\betaR$ are representative of two extreme options. 
In gauge models, the na\"ive expectation is $|\betaR|=1$, but with a suitable mixing with right-handed vector-like fermions 
it is possible to  achieve smaller values. A possible detection of  $B_s \to \tau^- \mu^+$ would therefore represent a key 
ingredient to determine the size of $|\betaR|$. An important role is also played by $\tau \to \mu \phi$, which is a very promising LFV observable for the LHC and Belle II experiments, 
and which does not lie far from present bounds in the pure left-handed case~\cite{Angelescu:2018tyl,Kumar:2018kmr}. In the right panel of Figure~\ref{fig:ditau} we show the predictions of $\cB(B \to \tau^+ \tau^-)$ vs.~$\cB(B_s \to \tau^+ \tau^-)$. 
The ratio of these two rates is controlled by $\beta_L^{d\tau}/\beta_L^{s\tau}$, and would therefore be an ideal 
probe to test the assumption of minimal breaking of the $U(2)^5$ flavor symmetry in the 1--3 sector, in the scenario with sizable right-handed couplings. In the pure left-handed case, subleading $U(2)^5$ breaking 
terms could be tested via the LFU ratio $R_\pi =\cB(B^+\to \pi^+\mu^+\mu^-)/\cB(B^+\to \pi^+e^+e^-)$~\cite{Fuentes-Martin:2019mun,Bordone:2021olx},
which would allow us to probe $\beta_L^{d\mu}/\beta_L^{s\mu}$.

Significant improvements in the experimental searches for all the processes shown in Figure~\ref{fig:LFV} and Figure~\ref{fig:ditau} are expected from future Belle~II and LHCb data~\cite{Kou:2018nap,Bediaga:2018lhg}. In particular, with $5\,\mathrm{ab}^{-1}$ Belle~II expects to set the bounds $\mathcal{B}(B^+\to K^+\tau^+\tau^-)< 6.5 \times 10^{-5}$ ($ 90 \%$ CL) and $\mathcal{B}(B_s\to\tau^+\tau^-)<8.1 \times 10^{-4}$ ($ 90 \%$ CL) in the absence of any signals\footnote{Access to $B_s$ at Belle~II requires dedicated runs at the $\Upsilon(5S)$ resonance. The limit we report assumes an integrated luminosity of $5\,\mathrm{ab}^{-1}$ \cite{Kou:2018nap}.},
while LHCb should reach $\mathcal{B}(B_s\to\tau^+\tau^-)<5\times 10^{-4}$ ($95 \%$ CL) by the end of the Upgrade~II, thus probing the preferred parameter space for the $|\betaR|=1$ case almost entirely. For the LFV processes, Belle~II is expected to reach a sensitivity of $1-2\times 10^{-9}$ ($90 \%$ CL) for both $\mathcal{B}(\tau\to\mu\gamma)$ and $\mathcal{B}(\tau\to\mu\phi)$ with $50\,\mathrm{ab}^{-1}$, and LHCb should achieve $\mathcal{B}(B_s\to\tau\mu)<3\times 10^{-6}$ ($90 \%$ CL) at the end of Upgrade~II.

\subsection{UV-sensitive low-energy observables} 
\label{sect:UVsensitive}

We denote as ``UV-sensitive'' the observables receiving contribution from dimension-six operators other than those introduced 
in Section~\ref{sect:OPbasis}. In order to correlate them to the observables discussed so far 
(i.e.~semileptonic processes involving charged leptons) we need a complete model. 
Within this category, we concentrate here on three classes of rare processes which are 
particularly interesting given the strong suppression within the SM and the 
stringent experimental constraints: meson-antimeson mixing, $\cB(B \to K^{(*)} \nu\bar\nu)$, and $\tau \to \mu \gamma$. 

In all these observables we can identify a $U_1$-induced loop contribution, which is 
unavoidable in our setup,
and additional contributions due to the other mediators. 
The latter are controlled by additional parameters, unrelated to those introduced so far, 
such as the mixing angles controlling the 
misalignment of the interaction basis from the down-quark and charged-lepton mass eigenstate basis in \eqref{eq:DownBasis}.
As discussed in Refs.~\cite{Cornella:2019hct,Fuentes-Martin:2020pww}, such misalignment is strongly constrained, especially in the quark sector.
In the following, we concentrate mainly on the  
$U_1$ loop-induced amplitudes, 
which can be considered the irreducible
contributions to these rare processes 
barring  fine-tuned cancellations
with tree-level mediated $G^\prime$ 
and/or $Z^\prime$ amplitudes. 
We will discuss in more detail this point 
in the case of $\Delta F=2$ observables.

A systematic analysis of the loop-induced amplitudes in the non-universal 4321 model has been presented in \cite{Fuentes-Martin:2020hvc}.
A key role in evaluating the $U_1$ loop-induced contributions is provided by the additional 
vector-like fermions. To take their effect into account, we modify 
the $J^{U}_\mu$ current in (\ref{eq:JU0}) as follows
\be
J^{U}_\mu  \to  J^{U}_\mu  +  \beta_{L}^{Q \alpha}\,(\bar Q \gamma_{\mu}  \ell_{L}^{\alpha})  +   \beta_{L}^{i L}\,(\bar q_L^i \gamma_{\mu}  L)~, 
\label{eq:JUfull}
\ee
where $Q$ ($L$) denote vector-like fermions with left-handed quark (lepton) quantum numbers that mix with the third and second generation  left-handed  chiral quarks (leptons). For the sake of simplicity, and given we focus mainly on $3\to2$ transitions in the quark sector, 
we assume a single family of vector-like fermions.
In order to take advantage of the calculations presented in \cite{Fuentes-Martin:2020hvc}, 
we match the notation for the $U_1$ couplings to fermions used therein to the one adopted in the present paper via
\begin{align}\label{eq:betaDict}
\begin{aligned}
\beta_L^{b \tau} &= W_{11}\,,&
\beta_L^{s \tau} &= -s_Q W_{21}\,,&
\beta_L^{b \mu}  &= -s_L W_{12}\,,&
\beta_L^{s \mu}  &= s_Q s_L W_{22}\,,\\
\beta_L^{Q \tau} &= c_Q W_{21}\,,&
\beta_L^{b L}    &= c_L W_{12}\,,&
\beta_L^{Q \mu}  &= -c_Q s_L W_{22}\,,&
\beta_L^{s L}    &= -s_Q c_L W_{22}\,.
\end{aligned}
\end{align}
Here $s_{L(Q)}$ and $c_{L(Q)}$ denote sine and cosine of the mixing angles of the vector-like fermions with the light (second-generation) 
chiral fermions, whereas $W_{ij}$ are the elements of the complex $2\times 2$ unitary 
matrix describing the mixing among vector-like fermions and third-generation chiral fermions. The natural expectation 
is $|s_{L(Q)}| \ll1$, $|c_{L(Q)}|\approx 1$, and $W_{ij} =\cO(1)$.\footnote{Note that in (\ref{eq:betaDict}) we have not  redefined the value of $g_U$
to absorb the difference of $\beta_L^{b \tau}$ from unity.} 
The mixing matrix $W_{ij}$ is real in the limit where the 
mass matrix of the vector-like fermions and the 
mixing between chiral and vector-like fermions are aligned 
in phase. We work under this hypothesis, which 
justifies having treated the $\beta_L^{i\ell}$ 
corresponding to $3 \leftrightarrow  2$ mixing as real couplings.

In the following we discuss the three classes of UV sensitive low-energy observables using the results obtained in \cite{Fuentes-Martin:2020hvc}. 
While these results have been obtained in the non-universal 4321 model, 
they can easily be extended to the other realizations of 4321 models
(i.e.~4321 models with different charge assignments for the SM fermions~\cite{DiLuzio:2017vat}) 
provided the only relevant flavor mixing terms are those defined in
(\ref{eq:JUfull}).

\subsubsection*{$\Delta F=2$}\label{sect:DF2UV}

\begin{figure}[p]
\centering
\includegraphics[width=0.47\textwidth]{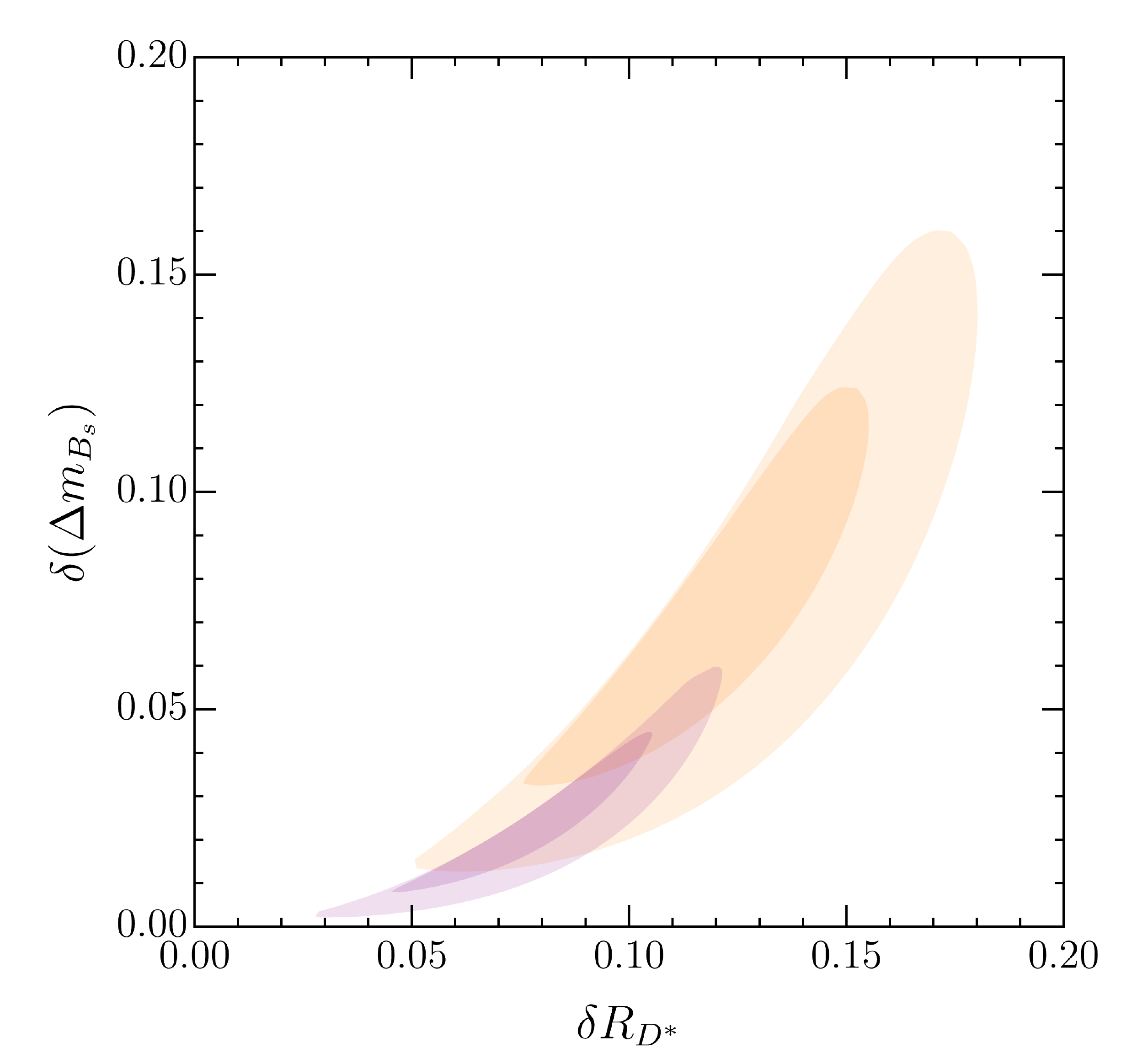}   \quad
\includegraphics[width=0.47\textwidth]{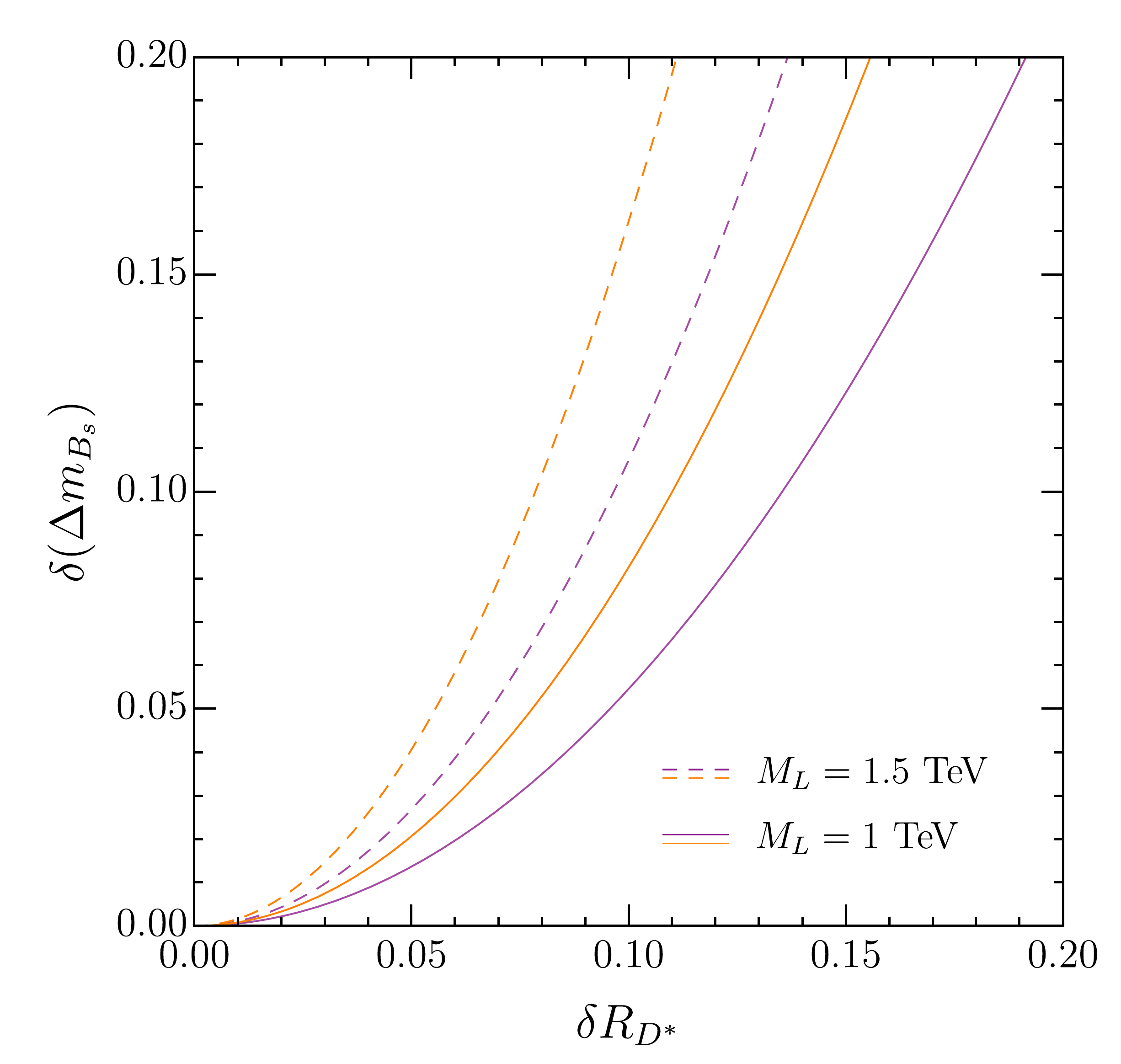}  
\includegraphics[width=0.47\textwidth]{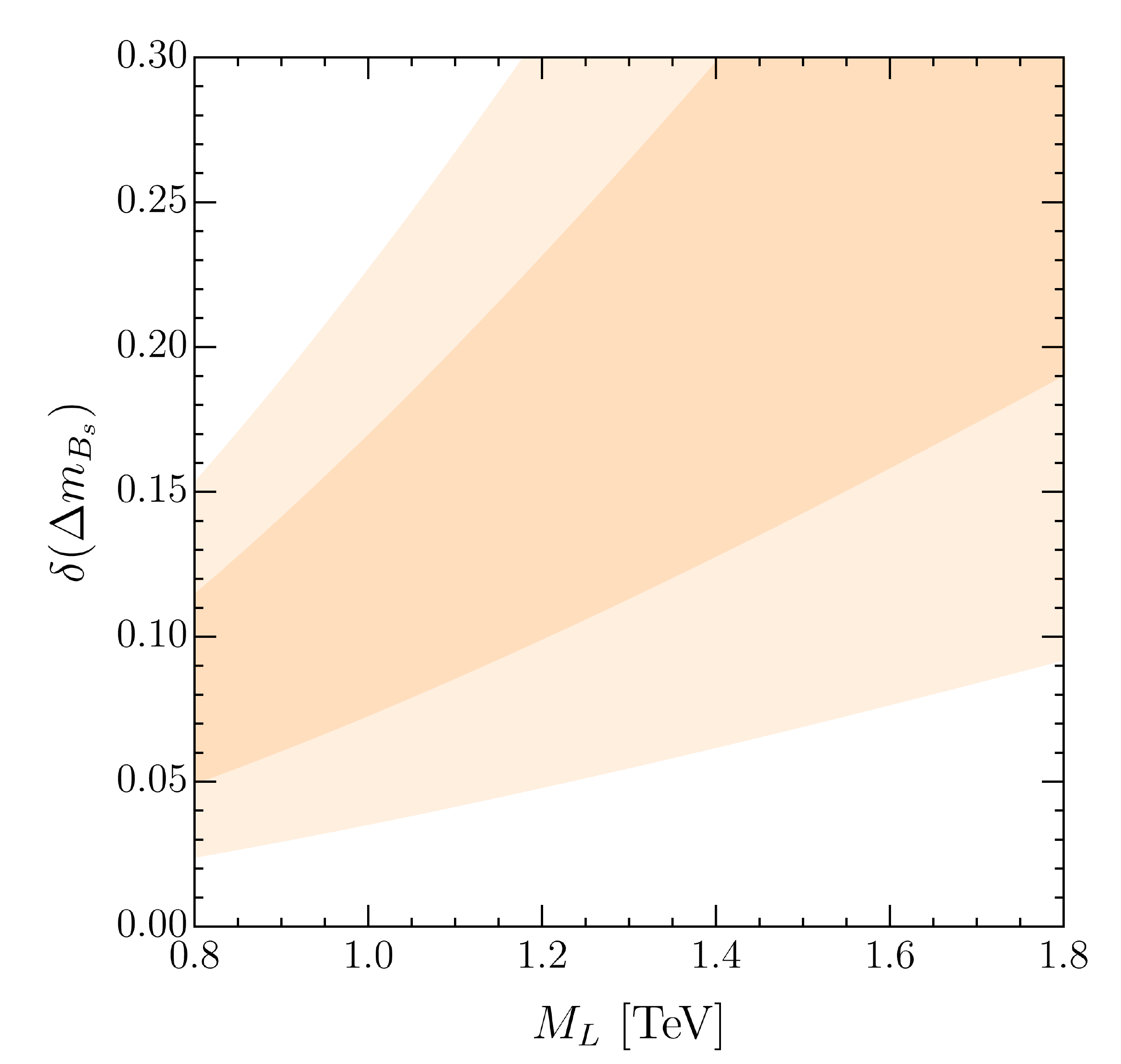}   \quad
\includegraphics[width=0.47\textwidth]{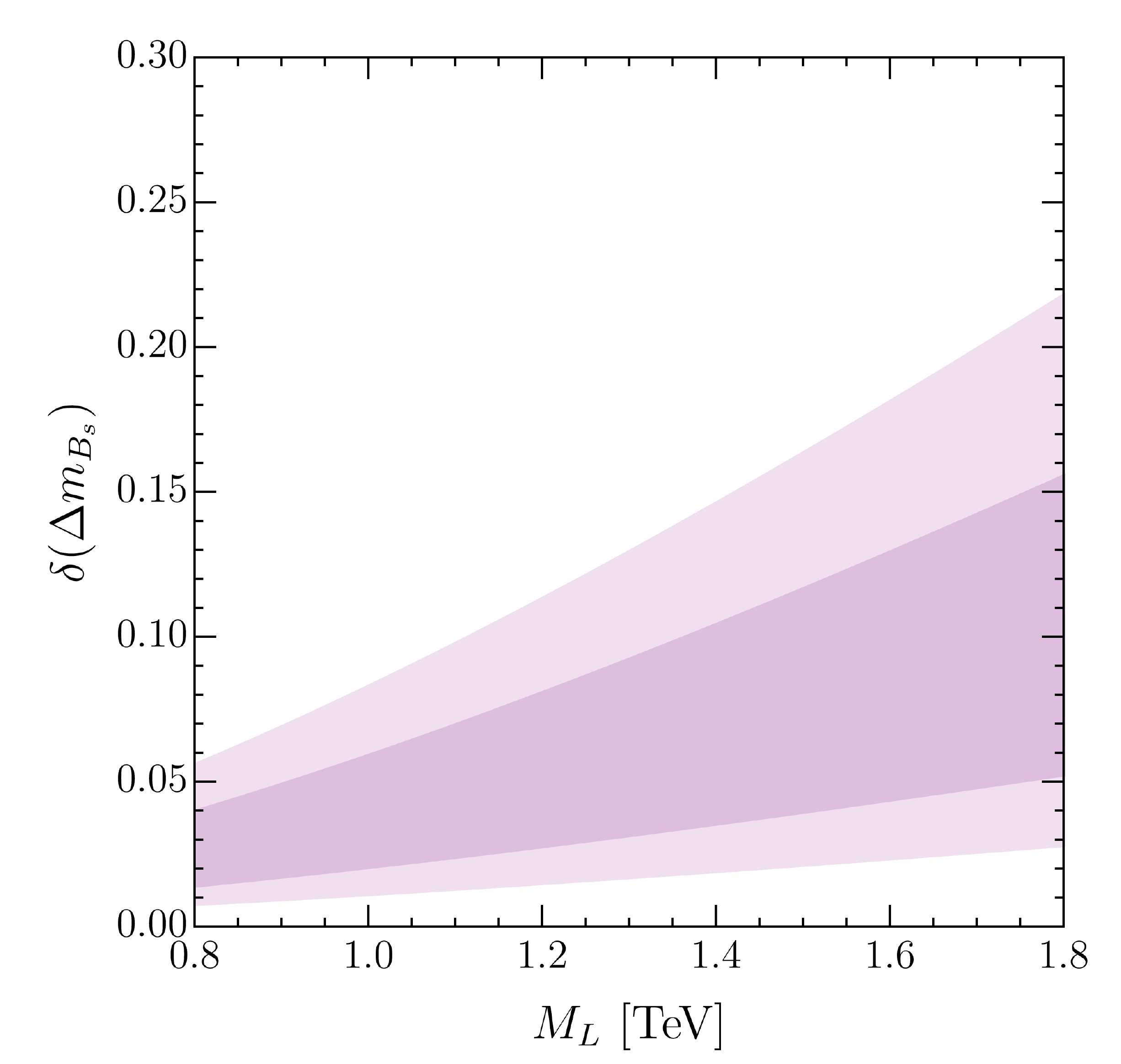}  
\caption{  \label{fig:DF2ML} 
{\em Upper left:} $1\sigma$ and $2\sigma$ regions in the ($\delta R_{D^*}$, $\delta(\Delta m_{B_s})$) plane preferred by the low-energy fit, with $\Lambda_{bs} = 1 \, \mathrm{TeV}$.  {\em Upper right:} $\delta(\Delta m_{B_s})$ as a function of $\delta R_{D^*}$ in the UV complete model for different values of $M_L$, fixing $M_U = 4$~TeV and $\beta_L^{s \tau} = 0.15$. 
{\em Lower plots:} Preferred $1\sigma$ and $2\sigma$ regions for $\delta(\Delta m_{B_s})$ as a function of the vector-like lepton mass for $M_U = 4 \, \mathrm{TeV}$. As in the other plots, orange and purple correspond to the benchmarks $\betaR=0$ and  $\betaR=-1$.}
\end{figure}

We start the discussion of the $\Delta F=2$ amplitudes by addressing the possible tree-level contributions from the additional massive gauge bosons.
In the limit where the small and flavor-universal couplings to light families are neglected, the tree-level exchange of 
$G^\prime$ and $Z^\prime$ fields in non-universal 4321 models leads to four-quark contact interactions among third-generation fermions. The relevant Lagrangian for these interactions reads
\begin{align}\label{eq:4q33}
\cL^{(4q)}_{\rm  eff}  &= - \frac{2}{v^2}\,  \left[  C_{G^\prime} \left( \bar q_L^3 \gamma^\mu\,T^a q^3_L 
+ \bar q_R^3 \gamma_\mu\,T^a q_R^3 \right)^2 
+ \frac{C_{Z^\prime}}{24} \left( 
\bar q_L^3\gamma^\mu q^3_L + \bar q_R^3 \gamma_\mu q_R^3 \right)^2 \right] ,
\end{align}
where we have defined $q_R^3\equiv(u_R^3\;d_R^3)^\intercal$ and $C_{G^\prime,Z^\prime}= g_4^2\,v^2/(4\,M_{G^\prime,Z^\prime}^2)+\mathcal{O}(g_{s,Y}^2/g_4^2)$, with $g_4=g_U$ denoting the $SU(4)$ coupling. Note that the right-handed currents could have a different overall weight (or even be absent) in other 4321 models.

In the limit where we assume perfect alignment of the right-handed quarks to their mass basis and of the left-handed quarks to the down-quark mass basis, there are no contributions to down-type flavor-changing amplitudes. However, contributions to $D$--$\bar D$ mixing are still present. Using the notation introduced in \eqref{eq:DF2gen} we get 
\be
   \cC_{uc}^{\rm NP-tree}  =    \frac{2}{v^2}\, \left(\frac{C_{G^\prime}}{3}+\frac{C_{Z^\prime}}{24}\right)  \left(V_{ub}^* V_{cb}\right)^2 \approx  1.2 \times 10^{-15}\, {\rm GeV}^{-2} 
   \times 
    \left( \frac{C_U}{0.01} \right)   \left(   \frac{  V_{ub}^* V_{cb} }{ 1.0 \times 10^{-4}  }  \right)^2~,
\ee
where, in the second equality, we have taken the limit of degenerate heavy-vector masses $M_{G^\prime}=M_{Z^\prime}=M_U$, for which $C_{G^\prime}\approx C_{Z^\prime}\approx C_U$ up to terms of $\cO(g_{s,Y}^2/g_U^2)$.  Given the limits reported in Tables~\ref{tab:fitobs}, this contribution falls below the current experimental sensitivity on $D$--$\bar D$ mixing for both the real and the imaginary part of the amplitude.

On the other hand, if we allow for a misalignment from the right-handed-quark or down-quark mass bases, a potentially sizable contribution appears in $B_{(s)}$--$\bar B_{(s)}$ mixing.  Considering a misalignment only in the left-handed sector, and denoting by
$L_d$ the unitary matrix connecting the 4321 interaction basis to the down-quark mass basis 
($q^i_L = (L_d)_{id_i}\, q^{d_i}_L$, see Appendix~\ref{app:U2}), in the case of $B_s$ mixing we get 
\be
   \frac{\cC_{b s}^{\rm NP-tree}}{\cC_{b s}^{\rm SM}} =
    \frac{ 16\pi^2 v^2 }{ m_W^2\,S_0(x_t)}   \left(\frac{C_{G^\prime}}{3}+\frac{C_{Z^\prime}}{24}\right)  
    \left[ \frac{  \left(L_d\right)^*_{3b} (L_d)_{3s}   }{V_{tb}^* V_{t s}} \right]^2  \approx 2.3 \times  \left( \frac{C_U}{0.01} \right) 
     \left[ \frac{  \left(L_d\right)^*_{3b} (L_d)_{3s}   }{V_{tb}^* V_{t s}} \right]^2~,
     \label{eq:Bsmixingtree}
\ee
where, once more, we took the limit of degenerate heavy-vector masses in the second equality. For this contribution to give $\delta (m_{B_s}) < 0.1$, the left-handed down-quark rotation needs to satisfy $|(L_d)_{3s}|  \lsim 0.2 |V_{ts}|$. This does not require a strong tuning, since the natural size of the possible 
misalignment is $\cO(|V_{ts}|)$. In models with minimal breaking of the $U(2)^5$ flavor symmetry, 
the term between square brackets in (\ref{eq:Bsmixingtree})
is real~\cite{Fuentes-Martin:2019mun} and yields a positive contribution to  $B_s$ mixing.
Contributions of arbitrary sign can be obtained 
only by relaxing the assumption of minimal $U(2)^5$ breaking (either in the left-handed sector or in the right-handed sector).
However, conceiving models where this tree-level contribution 
cancels significantly against the 
$U_1$-induced loop-induced contribution 
(which is also positive, as we show below) 
requires a tuning of magnitude and phase of the mixing terms for both the
 $B_s$ and the $B_d$ amplitudes, which is rather unnatural.  This is why we prefer to adopt the assumption of small misalignment 
 from the down-type basis, i.e. $|(L_d)_{3s}| \ll |V_{ts}|$, 
 which is a more motivated choice from the model-building point of view.

Having analyzed the tree-level contribution to $\Delta F=2$ amplitudes, we now discuss the $U_1$ loop-induced amplitudes.
For simplicity, we focus on $B_s-\bar B_s$ mixing. 
Following the analysis of \cite{Fuentes-Martin:2020hvc}, the (finite) loop contribution to  $\cC_{b s}$, taking into account both light and heavy leptons, is
\be
   \cC_{b s}^{\rm NP-loop} = \frac{2 C_U}{v^2}\, \frac{g_4^2}{16\pi^2}\,  \left[\left(\beta_L^{sL}\right)^* \beta_L^{bL} \right]^2\,  F_{\Delta F=2}(x_L)~,
\ee
where $x_L =M_L^2/M_U^2$, with $M_L$ being the vector-like lepton mass, 
and the expression for the loop function reads~\cite{Fuentes-Martin:2020hvc}
\begin{align}
F_{\Delta F=2}(x)&=\frac{x\,(x+4)(x+1)}{8(x-1)^2}\left(\frac{1}{2}+\frac{x\ln x}{1-x^2}\right)\,.
\end{align}
For the sake of completeness
we kept a complex notation for $\beta^{i\ell}_L$ 
although, as anticipated,  they are real in the limit where we introduce a single family of vector-like leptons.

Normalizing to the SM contribution, and taking into account the relations among the 
different $\beta^{i\ell}_L$ in the limit $|s_{L(Q)}| \ll1$, 
we can write
\be
   \frac{\cC_{b s}^{\rm NP-tree}}{\cC_{b s}^{\rm SM}} \approx  \frac{ 4 M_U^2}{  m_W^2\,S_0(x_t) } C_U^2 \left[ \frac{ \left(\beta_L^{s\tau}\right)^* }{ 
   V_{tb}^* V_{t s} }\right]^2\, F_{\Delta F=2}(x_L)~.
   \label{eq:DFUV}
\ee
Using the expansion $F_{\Delta F=2}(x) =  x/4 + \mathcal{O}(x^2)$, the effective mass combination $\Lambda_{bs}$ introduced in (\ref{eq:MesonMixing}) can be identified with $\Lambda_{bs} = \sqrt{2}\, M_L$ in the small $x_L$ limit. 
We stress that, rather than the value of $M_L$, what controls the overall size of the $\Delta F=2$ amplitude
is the component of the vector-like mass that breaks the 
$SU(4)$ custodial symmetry and is responsible for the 
flavor mixing of the $SU(4)$-charged 
fermions, namely 
$\Delta M_L \propto  W_{12} M_L$~\cite{Fuentes-Martin:2020hvc}. 
In other words,  the $\Delta F=2$ amplitude is finite in the  
limit of heavy vector-like masses. However, for $M_L\to \infty$ 
the effective mixing $\beta_L^{s\tau}$ vanishes.

In Figure~\ref{fig:DF2ML}, we show the expected value of $\delta(\Delta m_{B_s})$ as a function of $\delta R_{D^*}$ and of the vector-like mass using the results of the low-energy fit. In both cases, we observe an approximate quadratic dependence, which can be easily understood from \eqref{eq:DFUV}. The quadratic dependence from $\delta R_{D^*}$ is closely connected to the quadratic dependence from $|\beta_L^{s\tau}|$ that, to a large extent, controls the size of  $\delta R_{D^*}$. The quadratic dependence from $M_L$ is due  to the behavior of $F_{\Delta F=2}(x)$ at small $x$.  From these plots it is clear that, in absence of tuned scenarios allowing a partial cancellation between tree and loop amplitudes, the $\Delta F=2$ bounds require a vector-like lepton with mass not far from 1~TeV.

\subsubsection*{$B\to K^{(*)} \nu\bar\nu$}

\begin{figure}[t]
\centering
\includegraphics[width=0.5\textwidth]{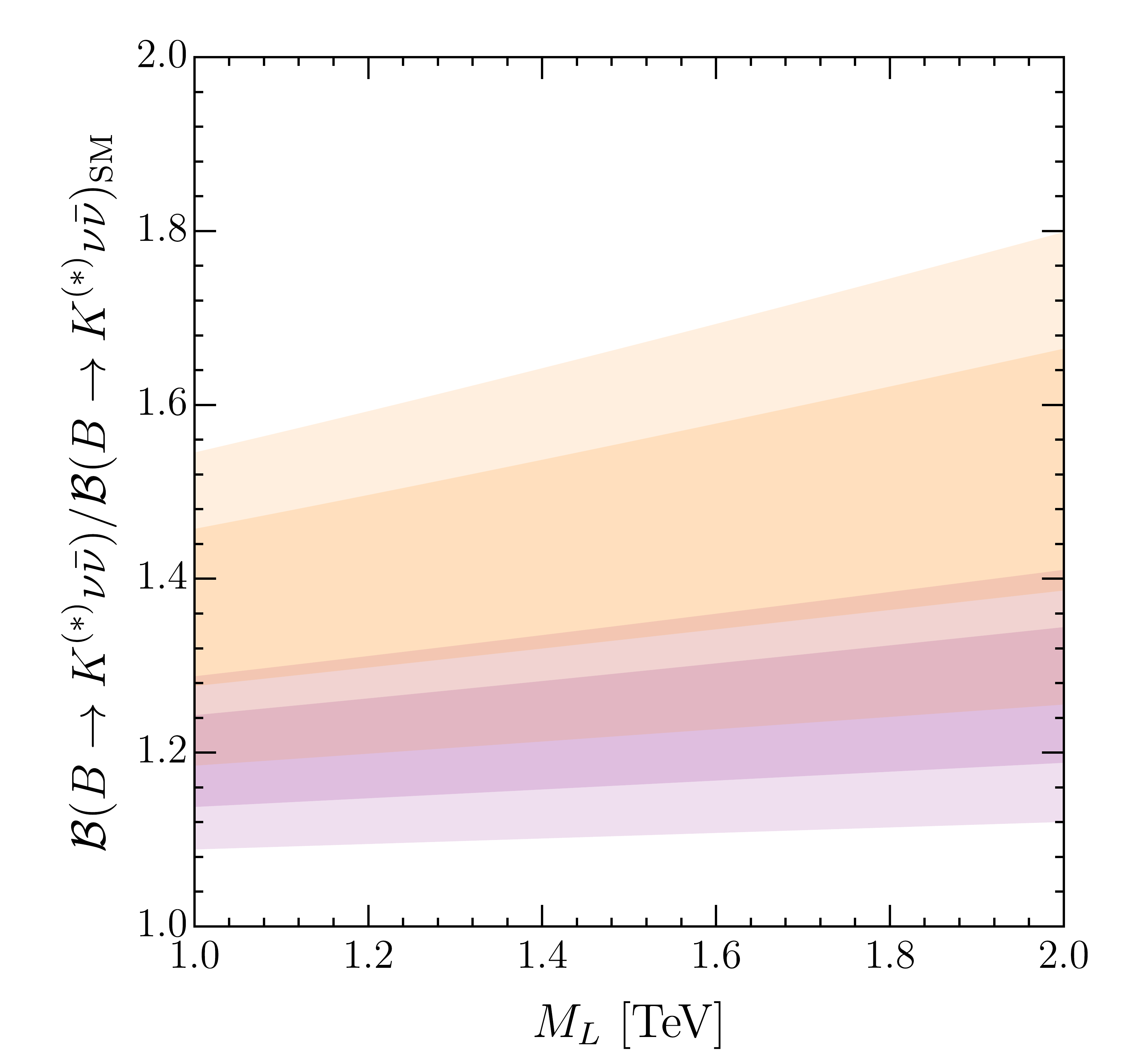}  
\caption{Model predictions (best $1\sigma$ and $2\sigma$ fit regions) for the $B\to K^{(*)}\nu\bar\nu$ branching ratio as a function of $M_L$, setting $(L_d)_{3s}^*=0$, $M_U=4$~TeV and $g_4=3$, and varying the scale of the scalar degrees of freedom and the $Z^\prime$ mass in the $M_R=[1,2\pi]\,M_U$ and $M_{Z^\prime}=[0.5,1]\,M_U$ ranges. As in the other plots, orange and purple correspond to the benchmarks $\betaR=0$ and  $\betaR=-1$.} \label{fig:B2Knn}
\end{figure}

We now analyze $b\to s\nu\bar\nu$ transitions. As already mentioned in Section~\ref{sect:EFT}, the $U_1$ leptoquark does not contribute to these transitions at the tree-level. These are however generated once we include loop corrections. Moreover, in the complete UV model, we also have contributions from tree-level exchange of the $Z^\prime$, making $b\to s\nu\bar\nu$ transitions a clear example of  UV-sensitive low-energy observable. \\
We define the relevant Lagrangian as
\begin{align}\label{eq:bsnn_LEET}
\cL_{b\to s\nu\bar\nu} = -\frac{4 G_F}{\sqrt{2}}\,V_{ts}^*V_{tb}\,  \cC_\nu^{\alpha\beta}\, 
(\bar s_L \gamma_\mu b_L) (\bar\nu_L^\alpha\gamma^\mu \nu_L^\beta)\,.
\end{align}
Neglecting effects suppressed by light quark masses, the lepton-flavor-conserving and -universal SM contribution reads
\begin{align}\label{eq:bsnn_CSM}
\cC_{\nu,\,{\rm SM}} = \frac{\alpha_W}{2\pi}\,X_t\,,
\end{align}
where $X_t = 1.48\pm 0.01$~\cite{Buchalla:1998ba}, and $\alpha_W = g_L^2/(4\pi)$ with $g_L$ being the $SU(2)_L$ coupling. Due to the underlying $U(2)^5$ flavor structure, NP effects are dominant in the Wilson coefficient involving the third family, while the other flavor combinations receive negligible contributions. Namely, we have ($\ell, \ell^\prime=e,\,\mu$)
\begin{align}
\cC_\nu^{\tau\tau}  &=  \cC_{\nu,\,{\rm SM}}  + \cC_{\nu,\,{\rm NP}}^{\tau\tau}\,,&
\cC_\nu^{\ell\ell}  &\approx \cC_{\nu,\,{\rm SM}}\,, &
\cC_\nu^{\tau\ell}  &\approx \cC_\nu^{\ell\tau}\approx \cC_\nu^{\ell\ell^\prime} \approx 0\,.
\end{align}
With these definitions, the NP correction to the $B\to K^{(*)}\nu\bar\nu$ branching ratio reads
\begin{align}\label{eq:DeltaKnn}
\frac{ \cB(B\to K^{(*)} \nu\bar\nu ) }{    
\cB(B\to K^{(*)} \nu\bar\nu )_{\rm SM}}\approx \frac{2}{3}+\frac{1}{3}\,\left|\frac{\cC_{\nu,{\rm NP}}^{\tau\tau}+\cC_{\nu,{\rm SM}}}{\cC_{\nu,{\rm SM}}}\right|^2\,.
\end{align}
We further split the NP effects into $Z^\prime$-mediated and $U_1$ loop-induced contributions as follows:
\begin{align}
\cC_{\nu,\,{\rm NP}}^{\tau\tau}&=\cC_{\nu,\,Z^\prime}^{\tau\tau} + \cC_{\nu,\,U}^{\tau\tau}\,.
\end{align}
At NLO accuracy, we have~\cite{Fuentes-Martin:2019ign,Fuentes-Martin:2020hvc}
\begin{align}
\begin{aligned}
\mathcal{C}_{\nu,\,U}^{\tau\tau} &\approx\mathcal{C}_{\nu,\,U}^{\rm RGE}+\frac{\alpha_4}{4\pi}\,C_U\left[-2\,\frac{(L_d)^*_{3s}\,(L_d)_{3b}}{V_{ts}^*V_{tb}}- \frac{\beta_L^{sL}\beta_L^{bL\,*}}{V_{ts}^*V_{tb}}\,F_{\Delta Q=1}(x_L)\right]\,,\\
\mathcal{C}_{\nu,\,Z^\prime}^{\tau\tau} &\approx  -\,\frac{C_U}{4 x_{Z^\prime}}\left[ \frac{(L_d)^*_{3s}\,(L_d)_{3b}}{V_{ts}^*V_{tb}}\left(1 +  \frac{3}{8}\,\frac{\alpha_4}{4\pi}\right) -\frac{\alpha_4}{4\pi}\,\frac{\beta_L^{sL}\beta_L^{bL\,*}}{V_{ts}^*V_{tb}}\,G_{\Delta Q=1}(x_L,x_{Z^\prime},x_R)\right]\,,
\end{aligned}
\end{align}
where the loop functions are given by,\footnote{For concreteness we choose the model I scenario in~\cite{Fuentes-Martin:2020hvc} when writing $G_{\Delta Q=1}$. As shown in this reference, similar predictions are obtained in the model II scenario.} 
\begin{align}
F_{\Delta Q=1}(x)&\approx\frac{2x}{1-x}+\frac{2x\ln x}{(x-1)^2}\,,&
G_{\Delta Q=1}(x_1,x_2,x_3)&\approx\frac{5}{4}\,x_1+\frac{x_1}{2}\left(x_2-\frac{3}{2}\right)\left(\ln x_3-\frac{5}{2}\right)\,,
\end{align}
and we defined $x_{Z^\prime}\equiv M_{Z^\prime}^2/M_U^2$, $x_L\equiv M_L^2/M_U^2$ and $x_R\equiv M_R^2/M_U^2$, with $M_R$ being a scale associated to new scalar degrees of freedom. The coefficient $\mathcal{C}_{\nu,\,U}^{\rm RGE}$ encodes the RGE-induced contribution from the tree-level leptoquark-mediated operator $\mathcal{O}_{LL}^{23\tau\tau}$. Using DsixTools~\cite{Celis:2017hod} and setting $\Lambda=2~\mathrm{TeV}$, we find
\begin{align}\label{eq:CnuURGE}
\mathcal{C}_{\nu,\,U}^{\rm RGE}&=-0.047 \,\frac{\cC_{LL}^{23\tau\tau}(\Lambda)}{V_{ts}^*V_{tb}}=-0.047\,C_U\,\frac{\beta_{s \tau}}{V_{ts}^*V_{tb}}\,.
\end{align}
Out of the different contributions, those proportional to $(L_d)_{3s}^*$ are tightly constrained by $B_s$-mixing (see discussion above) and can therefore induce at most a $\pm 3\%$ correction to $\cB(B\to K^{(*)} \nu\bar\nu )$. The other contributions, on the other hand, can be sizable, yielding up to $\mathcal{O}(1)$ corrections to the SM value. Moreover, the sign of this correction is unambiguously connected to the NP effect in $\cB(B\to K^{(*)} \nu\bar\nu)$. More precisely, noting that $\beta_L^{sL}\beta_L^{bL\,*}\approx-\beta_L^{s\tau}$, an enhancement in the $R_{D^{(*)}}$ ratio unavoidably yields an enhancement also in $\cB(B\to K^{(*)} \nu\bar\nu)$. 
Note that this enhancement does not depend directly on $\beta_R^{b\tau}$; 
however, the value of $\beta_R^{b\tau}$ indirectly influences the effect via the extraction of $C_U$ from the fit of $R_{D^{(*)}}$
(see Figure~\ref{fig:U1fit}). This is why the NP impact in 
$\cB(B\to K^{(*)} \nu\bar\nu)$ shown in Figure~\ref{fig:B2Knn} is smaller
for $\beta_R^{b\tau}=-1$.  

In Figure~\ref{fig:B2Knn}, we show the model predictions for $\cB(B\to K^{(*)}\nu\bar\nu)$ as function of $M_L$ for a fixed value of $g_4=3$ and $M_U=4$~TeV. Current experimental limits $\mathcal{B}(B\to K\nu\bar\nu)/\mathcal{B}(B\to K\nu\bar\nu)_{\rm SM}=2.4\pm0.9$~\cite{BelleIIB2Knn,Grygier:2017tzo,Lutz:2013ftz,Lees:2013kla} and $\mathcal{B}(B\to K^*\nu\bar\nu)/\mathcal{B}(B\to K^*\nu\bar\nu)_{\rm SM}<3.2\;(95\%\,\mathrm{CL})$~\cite{Grygier:2017tzo} are still far from our model expectations. However, the Belle II Collaboration will measure $\cB(B\to K^{(*)} \nu\bar\nu)$ to an accuracy of 10\% of its SM value~\cite{Kou:2018nap}, thus probing the entire relevant parameter space of the model.

\subsubsection*{$\tau\to\mu\gamma$}
The matching contribution to the dipole operator is given by
\begin{align}\label{eq:ta2mugaMatching}
\begin{aligned}
\cC_{e \gamma}^{\mu \tau} (\Lambda) &=   - \frac{C_U }{16 \pi^2 }   \left[   \frac{y_\tau(\Lambda)}{2} \beta_L^{Q \tau} \beta_L^{Q \mu  \ast}\, \big(  G_1(x_Q)  - 2 G_2(x_Q)   \big) + \frac{3  \, y_b(\Lambda)}{2} \beta_R^{b \tau} \beta_L^{ b \mu \ast}   \right]\,,\\
\cC_{e \gamma}^{\tau \mu \ast} (\Lambda) &=  -  \frac{C_U }{16 \pi^2 } \, \frac{y_\tau(\Lambda)}{2} \beta_L^{Q \mu \ast} \beta_L^{Q \tau }\,\big(  G_1(x_Q)  - 2 G_2(x_Q)   \big)  \,,
\end{aligned}
\end{align}
where $x_Q = M_Q^2/M_U^2$ and the functions $G_{1,2}(x)$ are
given by~\cite{Fuentes-Martin:2020hvc} 
\begin{align}
G_1(x)&=x\left[\frac{2 - 5 x}{2 (x-1)^4}\,\ln x-\frac{4 - 13 x + 3 x^2}{4 (x-1)^3}\right]\,,&
G_2(x)&=x\left[\frac{4 x-1}{2 (x-1)^4}\,x\ln x+\frac{2 - 5 x - 3 x^2}{4 (x-1)^3}\right]\,.
\end{align}
Using these results, and given the size of the various  $\beta^{i\ell}_L$ couplings, it is easy to verify that  the leading contribution
is the one proportional to $y_b$,  which is independent from the mass of the vector-like fermions. An even larger contribution to the  decay 
rate is the one generated by the RGE contribution of  $\cO^{33\mu \tau}_{LR} (m_\tau)$ in \eqref{eq:taumugRGE}.
This is why we considered $\cB(\tau\to\mu\gamma)$ among the UV-insensitive observables in Section~\ref{sect:UVinsensitive},
at least in the case of large $\betaR$ (which is the only case where it can reach  values close to present bounds).

\subsection{Collider signatures from the additional TeV-scale states} 
\label{sect:HighpTAdditional}
\begin{figure}[t]
\centering
\includegraphics[width=0.5\textwidth]{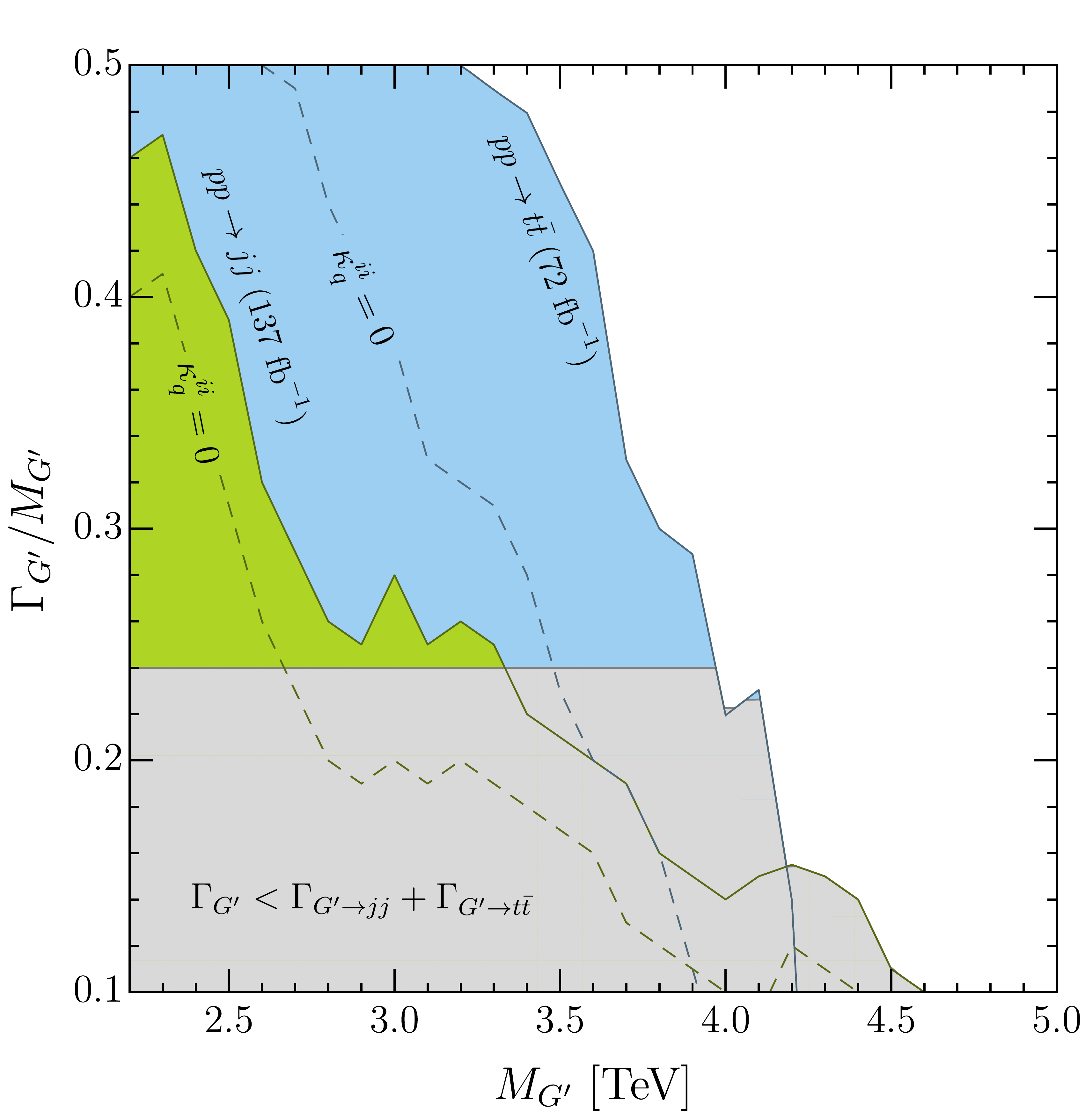}  
\caption{Collider constraints from resonant $G^\prime$ production with dijet (green) and $t\bar t$ (blue) final states for the benchmark couplings $\kappa_q^{ii}=-g_s(M_G^\prime)^2/g_{G^\prime}^2$ (solid) and $\kappa_q^{ii}=0$ (dashed), and $\kappa_R=1$ in both cases (see text for details). We mark in gray those values of the width which are below the sum of $t\bar t$ and dijet partial widths for $g_{G^\prime}=3$.} \label{fig:coloron}
\end{figure}

\paragraph{Coloron searches.} \!\!\!\!  
The most general Lagrangian for a coloron coupling to SM particles is given by
\begin{align}
\mathcal{L}_{G^\prime}&=-\frac{1}{4}\,G_{\mu\nu}^{\prime\,a}\, G^{\prime\,a\,\mu\nu}+\frac{1}{2}\,M_{G^\prime}^2\,G_\mu^{\prime\,a} \,G^{\prime\,a\,\mu}+\frac{1}{2}\,\kappa_{G^\prime}\,G_{\mu\nu}^{\prime\,a}\, G^{a\,\mu\nu}+g_s\,\tilde\kappa_{G^\prime} f_{abc}\,G_{\mu}^{\prime\,a}\,G_{\nu}^{\prime\,b}\, G^{c\,\mu\nu}\nonumber\\
&\quad+g_{G^\prime}\,G^{\prime\,a\,\mu}\,(\kappa_q^{ij}\,\bar q^i_L\,T^a\,\gamma_\mu\, q_L^j+\kappa_u^{ij}\,\bar u^i_R\,T^a\,\gamma_\mu\,u_R^j+\kappa_d^{ij}\,\bar d^i_R\,T^a\,\gamma_\mu\, d_R^j)\,.
\end{align}
In models where the coloron has a gauge origin, such as the ones discussed here, we have $\kappa_G=\tilde\kappa_G=0$. This is an important feature, since it implies that coloron couplings to two gluons are absent at the tree-level, thus effectively reducing the coloron production cross-section at the LHC. Concerning the coloron couplings to fermions, as already discussed in Section~\ref{sect:DF2UV}, off-diagonal couplings are strongly constrained by meson-antimeson mixing and thus do not play a relevant role at high-$p_T$. Furthermore, in all UV completions aimed at the explanation of the $B$-anomalies, the following relations are satisfied
\begin{align}\label{eq:GpLag}
g_{G^\prime}&\approx K_{\rm NLO}\,g_U\,,&
\kappa_q^{33}&=1\,,&
\kappa_{u,d}^{33}&=\kappa_R\,,&
\kappa_q^{ii}&=s_Q^2-\frac{g_s^2}{g_{G^\prime}^2}\,,&
\kappa_{u,d}^{ii}&=-\frac{g_s^2}{g_{G^\prime}^2}\,,
\end{align}
with $i=1,2$ and $g_s$ evaluated at the coloron mass scale 
($g_s(M_{G^\prime})\approx1$). In this expression, $s_Q$ parametrizes possible contributions from vector-like fermion mixing, assuming the latter enter in an $U(2)$-invariant way, namely that they are the same (up to small corrections) for first and second families. This condition is necessary to avoid the strong constraints from $D$--$\bar D$ mixing~\cite{DiLuzio:2018zxy}. Since $s_Q$ is connected to the value of $\beta_L^{s\tau}$ (c.f.~\eqref{eq:betaDict}),  the constraint from the low-energy fit naturally leads to a significant suppression of the $\kappa_q^{ii}$ couplings. To illustrate this effect, we consider two benchmark values for these couplings in our analysis: $\kappa_q^{ii}=0$ and $\kappa_q^{ii}=-g_s^2/g_U^2$. The value of $\kappa_R$ can vary depending on the UV completion. For concreteness, here we take $\kappa_R=1$, which is common to all models with third-family quark-lepton unification. Finally, in UV models where $U_1$ and $G^
\prime$ have a common origin, we have $g_{G^\prime}\approx g_U$. In order to account for NLO corrections to coloron on-shell production, in \eqref{eq:GpLag} we modified this relation by adding the $K$-factor~\cite{Fuentes-Martin:2019ign,Fuentes-Martin:2020luw}
\begin{align}
K_{\rm NLO}\approx\left(1+2.65\,\frac{\alpha_4}{4\pi}+8.92\,\frac{\alpha_s}{4\pi}\right)^{-1/2}\,,
\end{align}
which amounts to an $\mathcal{O}(10\%)$ reduction of $g_{G^\prime}$ for $g_U=3$.

\begin{figure}[t!]
\centering
\includegraphics[width=0.32\textwidth]{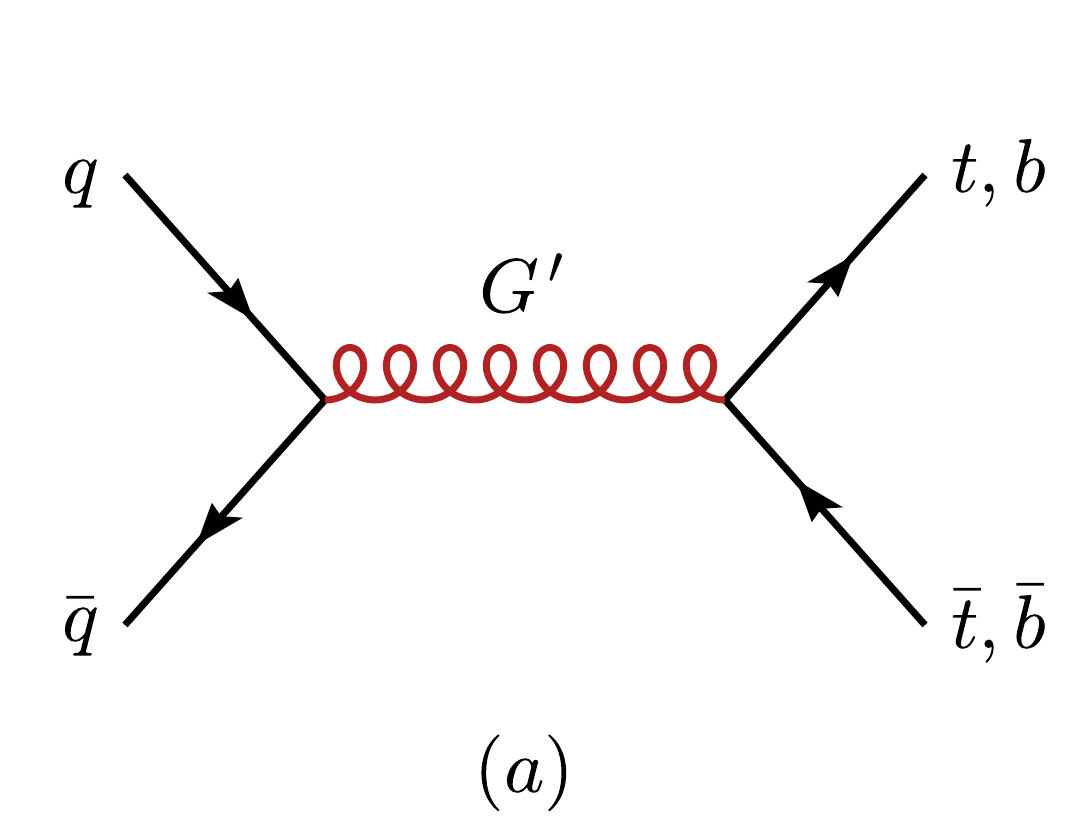}~  
\includegraphics[width=0.32\textwidth]{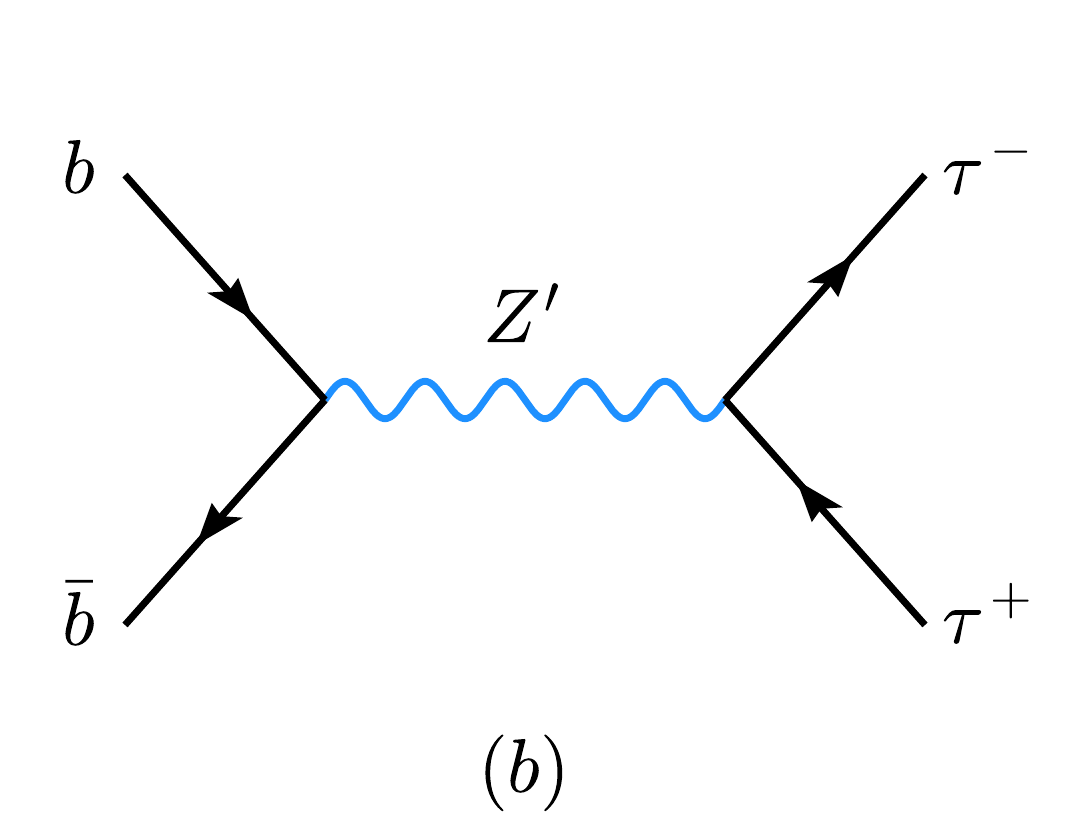}  
\includegraphics[width=0.32\textwidth]{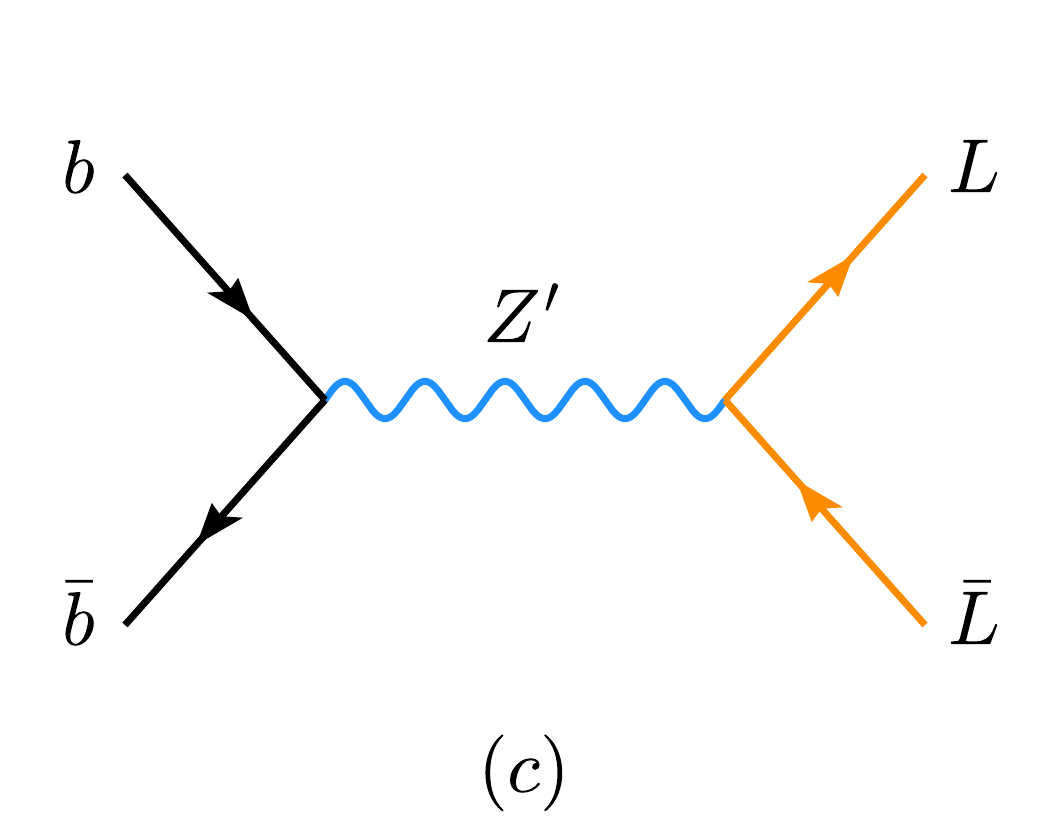}  
\caption{  \label{fig:DiagramsAdditional} Feynman diagrams describing the main processes for $G^\prime$, $Z^\prime$, and vector-like lepton production at the LHC. }
\end{figure}

Given the coupling structure in~\eqref{eq:GpLag}, the most effective coloron searches consist in resonant production with $t\bar t$ and $b\bar b$ final states. Since the coloron couples to two gluons only at the loop level, the dominant production channel is via quark fusion, i.e. $q\bar q\to G^\prime$ (see Figure~\ref{fig:DiagramsAdditional}). Even though the couplings to light quarks are suppressed by $g_s^2/g_{G^\prime}^2$, the PDF enhancement of valence quarks relative to third generation quarks ensures that production via valence quarks is nevertheless dominant. As a result, for fixed coloron width, the cross section for $pp\to G^\prime\to t\bar t/ b\bar b$ has only a mild dependence on $g_{G^\prime}$, except for extreme values of this coupling, where the $b\bar b$ initiated process starts to dominate. For concreteness, in what follows we fix $g_{G^\prime}=3$, noting that the limits we derive will not be significantly affected for lower values of $g_{G^\prime}$ (up to a trivial rescaling of the width). The partial width of the coloron to SM particles for $g_{G^\prime}=3$ is $\Gamma_{G^\prime\to q\bar q}/M_{G^\prime}\approx 0.24$. In realistic UV models, the total width could be larger if the decay channel to vector-like quarks becomes kinematically available. Therefore, we leave the total width of the coloron as a free parameter in our analysis.

For the inclusive dijet final state, we use the recent CMS analysis in~\cite{Sirunyan:2019vgj} with $137~\mathrm{fb}^{-1}$ of $13$~TeV data. This analysis targets both narrow and broad resonances with widths up to $55\%$ of the resonance mass, which makes it particularly suitable for the model discussed here. We calculate the model predictions for the dijet process with {\tt Madgraph5\_aMC@NLO}~\cite{Alwall:2014hca} using the default {\tt NNPDF23LO} PDF set and the coloron UFO model presented in~\cite{Baker:2019sli}, publicly available in the {\tt Feynrules}~\cite{Alloul:2013bka} model database (\url{https://feynrules.irmp.ucl.ac.be/wiki/LeptoQuark}). In order to set limits, we compute for different values of the total width the $\sigma(pp\to G^\prime \to jj)$ cross-section subject to the same rapidity cuts $|\eta(j_1)|,|\eta(j_2)|<2.5$ and $|\Delta\eta(j_1,j_2)|<1.1$ used in the experimental analysis. The resulting cross-sections are then confronted directly with the $95\%$ CL limits for a $q\bar q$-initiated  spin-1 resonance provided by CMS in Ref.\cite{Sirunyan:2019vgj}, Figure~10. 

The extracted limits are shown in Figure~\ref{fig:coloron} (green region). As can be seen, despite the large-width coverage of the search, the limits significantly weaken for larger values of the coloron width. Taking as a reference $\Gamma_{G^\prime}/M_{G^\prime}\approx0.24$, an exclusion limit is found at $M_{G^\prime}\approx 3.3$~TeV for $\kappa_q^{ii}=-g_s^2/g_U^2$ and $M_{G^\prime}\approx 2.7$~TeV for $\kappa_q^{ii}=0$. Since our dijet signal is made almost entirely of $b\bar b$ pairs, improved limits could be obtained if at least one of the final-state jets is $b$-tagged. In this work we have not extracted limits from $b$-tagged dijet searches, because these usually focus on narrow resonances and the improvement on the limits over the inclusive dijet searches is not substantial \cite{DiLuzio:2018zxy}. Moreover, the traditional $b$-taggers used to identify very heavy resonances decaying to $b$-jets suffer from a fast decline in the $b$-tagging efficiency with increasing jet transverse momentum. Recently, new $b$-taggers based on deep neural networks, like {\tt DL1r} \cite{ATL-PHYS-PUB-2015-039}, {\tt DeepCSV} \cite{Sirunyan:2017ezt} and {\tt DeepJet} \cite{Bols:2020bkb}, have shown very promising results in identifying boosted heavy flavor jets. These novel techniques may be useful in the near future to improve searches for broad colorons in the $b$-tagged dijet invariant mass spectrum.

Much stronger coloron limits are found in the $pp\to G^\prime\to t\bar t$ channel. The precise measurements of the differential cross-sections for boosted top-quarks from $t\bar t $ production are powerful probes for heavy resonances, the invariant mass distribution $m_{t\bar t}$ in particular. Following the same procedure as in~\cite{Baker:2019sli,DiLuzio:2018zxy}, we use the unfolded parton-level data of the normalized $m_{t\bar t}$ distribution provided by the experimental collaborations to set limits on the coloron mass and width. A novelty with respect to these previous studies is that in addition to the ATLAS search with $36.1~\mathrm{fb}^{-1}$~\cite{Aaboud:2018eqg}, a new search by CMS with $35.9~\mathrm{fb}^{-1}$~\cite{CMS:2020kdq} has recently been released. Below we present results combining the two searches for a total luminosity of $72$~fb$^{-1}$, under the assumption that the measurements from each experiment (and the invariant mass bins within) are completely uncorrelated. While both experimental searches are very similar, one key difference between the two unfolded data sets is their energy reach: the data provided by ATLAS extends up to 3~TeV in the invariant mass distribution, while the data provided by CMS reaches 4~TeV. The latter is therefore more relevant to set limits on the coloron. As discussed in \cite{Baker:2019sli}, in order to obtain good agreement between the SM leading-order predictions from MC event generators and the experimental data, it is necessary to exclude the lower invariant mass bins from our analysis. We find very good agreement with the MC predictions by CMS (ATLAS) for bins above $m_{t\bar t}>1250$~GeV ($m_{t\bar t}>1200$~GeV). After computing the normalized $m_{t\bar t}$ distributions subject to partonic cuts\footnote{$m_{t\bar t}>800$~GeV, $p_T(t)>400$~GeV, $|\eta(t)|<2.4$ for CMS, and $p_T(t_1)>500$~GeV, $p_T(t_2)>350$~GeV for ATLAS.} for a range of coloron masses and widths, we performed a combined fit to the ATLAS and CMS data sets. Excluded regions in the mass-width plane are shown in Figure~\ref{fig:coloron} (light blue region) together with the dijet bounds discussed above. As shown there, in spite of the lower statistics, $t\bar t$ provides much stronger limits in the entire relevant region in parameter space, making this a key channel to test the $U_1$ solution of the anomalies. The largest energy reach of the CMS search compared to the one of ATLAS gives rise to limits that are stronger than those presented in~\cite{Baker:2019sli,DiLuzio:2018zxy} by about $500$~GeV. As a result, we find coloron exclusion limits of $M_{G^\prime}\approx4.0$~TeV for $\kappa_q^{ii}=-g_s^2/g_U^2$ and $M_{G^\prime}\approx 3.5$~TeV for $\kappa_q^{ii}=0$, for a nominal width of $\Gamma_{G^\prime}/M_{G^\prime}\approx0.24$.

\paragraph{$Z^\prime$ searches.} \!\!\!\!
The dominant search channel for the $Z^\prime$ at hadron colliders is via the Drell-Yan process $pp\to Z^\prime\to\tau^+\tau^-$, see Figure~\ref{fig:DiagramsAdditional}~b.\footnote{Because of the coupling structure of the $Z^\prime$, decay channels into dijets and $t\bar t$ pairs are suppressed by about a factor of 3 compared to $pp\to Z^\prime\to\tau^+\tau^-$ and are therefore less competitive~\cite{Baker:2019sli}.} Contributions to the production of the $Z^\prime$ from initial light quarks are suppressed by the coupling ratio $(g_Y/g_4)^2$ and can therefore be neglected in a first approximation. On the other hand, due to the underlying gauge structure of the model, the contributions from the $b\bar b$--initiated process are about 4 times smaller than those from the $U_1$ leptoquark when the masses of both mediators are equal and outside the region of resonant $Z^\prime$ production. As a result and as shown in~\cite{Baker:2019sli}, limits on the 4321 model from $Z^\prime$ searches are less competitive than those obtained from the $U_1$ and $G^\prime$ gauge bosons. We checked this explicitly by matching the $Z^\prime$ to the effective operators in \eqref{eq:opbasis} for which limits on the Wilson coefficients from $pp\to\tau^+\tau^-$ are extracted. For example, if we assume equal left-handed and right-handed couplings, a 100\% branching ratio into di-taus and a coupling satisfying $g_{Z^\prime}\approx g_U\ge3$, the limits from the effective operators translate to an exclusion limit for the mass around $M_{Z^\prime}\gtrsim3$~TeV. Note that using the EFT limits is already a good approximation as long as the $Z^\prime$ mass ranges around $\sim3$~TeV. The resulting bounds for the $Z^\prime$ are generically weaker than the $U_1$ limits displayed in Figure.~\ref{fig:U1highpT}. Hence, the presence of the $Z^\prime$ is not expected to modify the exclusion limits discussed there.

\paragraph{Searches for vector-like fermions.} \!\!\!\!
The phenomenological consequences of heavy vector-like quarks and leptons are of fundamental importance in characterizing the UV completion. Besides their indirect effects at low-energies, these heavy states lead to a rich array of signatures at hadron colliders \cite{DiLuzio:2018zxy}. Of particular importance are the vector-like leptons $L$. In order to comply with the stringent limits from $\Delta F=2$ transitions at low energies, the mass of $L$ must be relatively light, around $M_L\sim1$ TeV, well within the energy reach of the LHC. The main production mechanisms at the LHC are the resonant and electroweak pair production modes $q\bar q\to \gamma/Z^*/Z^\prime\to L\bar L$ and the associated single production mode $q^i g\to L^i U_1$. The resonant production of  $L\bar L$ from the decay of an on-shell $Z^\prime$ (Figure~\ref{fig:DiagramsAdditional} c) is a promising search channel for the vector-like lepton, provided such channel is kinematically available. The heavy leptons can decay either into a Higgs boson via the 2-body process $L^i\to h\ell^i$, or via the 3-body process $L^i\to q^i \ell^j\bar q^k$ induced by the vector leptoquark mediator, depending on the UV completion. In either case, one expects high-multiplicity final states composed of $\tau$-leptons, $b$-jets, top-quarks and missing energy from neutrinos. In a similar way, vector-like quarks $Q$ can be searched at the LHC. These states are pair produced purely from QCD interactions via gluon fusion, or through the decay of the coloron $pp\to G^\prime\to Q\bar Q$ (if kinematically allowed). Once produced, these heavy fermions decay via $Q^i\to hq^i$, $Q^i\to \ell^iq^j\bar\ell^k$, $Q^i\to \ell^iq^j\bar L^k$, leading to a large amount of third-generation SM fermions in the final state. The vector-like quarks, however, are in principle allowed to have larger masses than the vector-like leptons, and can lie beyond the current reach of heavy fermion searches at the LHC. To this date, no dedicated search by the LHC experimental collaborations for vector-like quarks or leptons with the characteristics described above is available. ATLAS and CMS have recently released searches for `excited' leptons $\ell^\star$ \cite{Aaboud:2019zpc,Sirunyan:2020awe} in the 3-body decay channel $\ell^\star\to \ell q\bar q$ through a contact interaction. Unfortunately, these typically target first and second generation decay products, and the excited leptons are singly produced $pp\to \ell^\star \bar\ell$ (single vector-like lepton production $pp\to L \bar\ell$ in the 4321 models is suppressed by small mixing angles). Existing searches for SUSY by ATLAS \cite{Aaboud:2017dmy,Aad:2020sgw,Aad:2021oos} and CMS \cite{Sirunyan:2020tyy,Sirunyan:2019kgv}, in particular stop searches, can be reinterpreted to set limits on the heavy vector-like leptons and quarks. A naive comparison of $\sigma(pp\to L\bar L)$ with the limits on the signal cross-sections from \cite{Aaboud:2017dmy} indicates that search strategies with multi-lepton and multi-jet final states are starting to become sensitive to vector-like leptons with masses around $M_L\approx 0.8$~TeV \cite{DiLuzio:2018zxy}. These experimental analyses are however not completely optimized for our vector-like lepton signatures and are challenging to recast. We therefore leave for future work a dedicated collider analysis of vector-like fermions in the 4321 model.
 
\section{Conclusions}

The recent  LHCb result on $R_K$~\cite{Aaij:2021vac} marks an important milestone in the study of the $B$ anomalies, putting on a firmer statistical basis the evidence of  LFU violation in $\bsll$ decays. As we have shown, a conservative combination of $\bsll$ observables leads to a 4.6$\sigma$ significance for the motivated NP hypothesis of a purely left-handed LFU-violating contact interaction. 

The evidence of LFU in $\bsll$ observables, taken alone, does not provide a clear indication of the scale of NP. As shown in \cite{DiLuzio:2017chi}, this could be well above the TeV scale and, in practice, not accessible via direct searches. The situation is different if we combine the evidence of LFU violation in $\bsll$ transitions with the less significant, but still
tantalizing hints of LFU violation in $\bctnu$ decays. This is the path we have followed in this work. 

Besides the phenomenological interest of dealing with a TeV-scale model of NP, the attempt of combining the two anomalies has two strong theoretical motivations: a possible connection to the flavor problem, for instance along the lines proposed in \cite{Bordone:2017bld,Fuentes-Martin:2020pww,Barbieri:2021wrc}, 
and to the Higgs hierarchy problem, as discussed in \cite{Barbieri:2016las,Barbieri:2017tuq,Fuentes-Martin:2020bnh,Barbieri:2021wrc}.
  
In this work we have presented a detailed phenomenological analysis of a combined solution of the $B$ anomalies based on the hypothesis of a TeV-scale vector leptoquark $U_1$ as leading mediator, which previous studies have singled out as the most viable option. We have analyzed the problem on three levels: i)~a pure EFT approach, where the $U_1$ hypothesis enters only in the selection of the relevant set of semileptonic operators;  ii)~a simplified model, which has allowed us to analyze in general terms the leading signatures of the $U_1$ boson at high energies; iii)~a complete UV scenario based on the non-universal 4321 gauge group, which has allowed us to address some important UV-sensitive observables at both low and high energies. 

The main conclusions of our analysis can be summarized as follows:
\begin{itemize}
\item{} 
A combined explanation of both  $\bsll$ and $\bctnu$  anomalies in terms of a massive $U_1$ is perfectly consistent with all available data. Interestingly, the low-energy data imply a preferred parameter region for mass and coupling of the $U_1$ that can be probed entirely within the high-luminosity phase of the LHC, as shown in Figure~\ref{fig:U1highpT}.
\item{} 
The above conclusion is driven by the $\bctnu$  anomalies and holds irrespective of various unknown parameters, such as the strength of right-handed leptoquark couplings or the size 
of subleading $U(2)^5$-breaking terms. As we have pointed out, the latter could be determined by a series of measurements of low-energy observables. As already found in previous studies, a general expectation is a large enhancement over the SM of the $B$-meson decay rates into di-tau final states. Moreover, lepton flavor-violating $B$ decays into $\tau^\pm\mu^\mp$ final states are expected to be within the experimental reach. However, contrary to previous studies, we have shown that the predictions for both these classes of 
rare processes are subject to large uncertainties due to model parameters which are still poorly constrained. The predictions presented in Section~\ref{sect:UVinsensitive} could allow one to determine these parameters using future data.
\item{}
$B_s$--$\bar B_s$ mixing poses a severe constraint on the UV completion of the simplified model. In the 4321 framework, this constraint can be satisfied in a natural way only if the 
vector-like leptons, which are a necessary ingredient of the model, have masses below 1.5--2~TeV (depending on the other model parameters).
\item{}
In the 4321 framework, the $B\to K^{(*)}\nu\bar\nu$ decay rates are expected to be significantly enhanced over their SM values, as shown in Figure~\ref{fig:B2Knn}. 
\item{}
As shown in Figure~\ref{fig:coloron}, in the 4321 framework the most severe constraint on the overall scale of the model  is represented by the bound on the coloron mass from $pp\to t\bar t$. Present data indicate $M_{G^\prime}\gsim 4$~TeV, 
close to the bounds of the perturbative regime. 
We therefore await with great interest the improvement of the experimental analyses of $pp\to t\bar t$ 
with the full LHC run-II data set.
\end{itemize}

In conclusion, following the path of a combined explanation of the two sets of flavor anomalies in neutral-current and charged-current decays of $B$ mesons, besides being a very interesting possibility from the theoretical point of view,
has the virtue of defining a concrete NP framework, which can be confirmed or excluded in the next few years thanks to a 
series of well-defined signatures both at low and high energies.

\subsection*{Note Added}

While this project was under completion, a few papers containing updated EFT analysis of the $\bsll$ anomalies \cite{Geng:2021nhg,Altmannshofer:2021qrr,Carvunis:2021jga} and of the simplified $U_1$ model \cite{Angelescu:2021lln,Hiller:2021pul} appeared. Our results in Section~\ref{sect:bsll} are compatible with those obtained by these authors.

\subsection*{Acknowledgements}

We thank Riccardo Barbieri for useful discussions. 
This project has received funding from the European Research Council (ERC) under the European Union's Horizon 2020 research and innovation programme under grant agreement 833280 (FLAY), and by the Swiss National Science Foundation (SNF) under contract 200021-175940. 
The work of J.F.\ and M.N.\ was supported by the Cluster of Excellence {\em Precision Physics, Fundamental Interactions, and Structure of Matter} (PRISMA$^+$, EXC 2118/1) funded by the German Research Foundation (DFG) within the German Excellence Strategy (Project ID 39083149) and by 
the BMBF-Project 05H2018 - Belle II. M.N.\ thanks the particle theory group at Zurich University and the Pauli Center for hospitality during a sabbatical stay. 

\appendix 

\section{The \texorpdfstring{$U(2)^5$}{U(2)} symmetry}
\label{app:U2}

The largest global flavor symmetry commuting with the gauge symmetry of the SM Lagrangian is $U(3)^5$~\cite{Chivukula:1987py}.
The $U(2)^5$ flavor symmetry is the subgroup of the $U(3)^5$ global symmetry   
that, by construction, distinguishes the first two generations of fermions 
from the third one~\cite{Barbieri:2011ci,Barbieri:2012uh,Blankenburg:2012nx}.
For each set of SM fermions with the same gauge quantum numbers, the first two generations form a doublet of a given $U(2)$ subgroup, whereas the third one transforms as a singlet. 
The five independent flavor doublets are denoted $Q,L,U,D,E$ and the flavor symmetry decomposes as 
\begin{align}
	U(2)^5 = U(2)_Q \otimes U(2)_L \otimes U(2)_U \otimes U(2)_D \otimes U(2)_E~.
\end{align}
In the limit of unbroken $U(2)^5$, only third-generation 
fermions can have non-vanishing Yukawa couplings, 
which is an excellent first-order approximation for the SM Lagrangian. 

A set of symmetry-breaking terms sufficient to reproduce 
the complete structure of the SM Yukawa couplings is
\begin{align}
\begin{aligned}
V_q&\sim\left(\boldsymbol{2},\boldsymbol{1},\boldsymbol{1},\boldsymbol{1},\boldsymbol{1}\right)\,,& V_\ell&\sim\left(\boldsymbol{1},\boldsymbol{2},\boldsymbol{1},\boldsymbol{1},\boldsymbol{1}\right)\,,\\ 
\Delta_{u(d)}&\sim\left(\boldsymbol{2},\boldsymbol{1},\boldsymbol{\bar 2}(\boldsymbol{1}),\boldsymbol{1}(\boldsymbol{\bar 2}),\boldsymbol{1}\right)\,,&\Delta_e&\sim\left(\boldsymbol{1},\boldsymbol{2},\boldsymbol{1},\boldsymbol{1},\boldsymbol{\bar 2}\right) \,.
\label{eq:U2spur}
\end{aligned}
\end{align}
In the quark sector, this is the minimal set of terms 
with different $U(2)^5$ transformation properties 
necessary to describe all quark masses and the off-diagonal entries of the CKM matrix~\cite{Barbieri:2011ci}.
In the lepton sector, the absence of mixing (for vanishing neutrino masses) does not allow us to conclude that 
$V_\ell$ is strictly necessary; however, it is natural to introduce such a breaking term to maintain a quark--lepton 
symmetric structure~\cite{Blankenburg:2012nx}. Given that the  $U(2)^5$ symmetric limit is a good approximation to the 
SM Lagrangian, all the $U(2)^5$-breaking terms can be chosen to be small in size.

By construction, $V_{q,\ell}$ are complex two-vectors and $\Delta_{e,u,d}$ are complex $2\times 2$~matrices. In terms of these spurions, we can express the Yukawa couplings as
\begin{align}
	Y_e = y_\tau\left(\begin{matrix}
		\Delta_e	 & x_\tau V_\ell \\
		0			 & 1
	\end{matrix}\right), && Y_u = y_t\left(\begin{matrix}
		\Delta_u	 & x_t V_q \\
		0			 & 1
	\end{matrix}\right), && Y_d = y_b\left(\begin{matrix}
		\Delta_d	 & x_b V_q \\
		0			 & 1
	\end{matrix}\right),
	\label{eq:YU2_5}
\end{align}
where $y_{\tau,t,b}$ and $x_{\tau,t,b}$ are free complex parameters expected to be of order~$\mathcal{O}(1)$. 
The observed mass hierarchies and the smallness of the CKM mixing angles are attributed to the size (and structure)
of the $U(2)^5$-breaking  spurions. The latter cannot be determined completely, but are constrained 
requiring no tuning in the  $\cO(1)$ parameters. In particular, from the 2--3 mixing in the CKM matrix we deduce 
$|V_q| =O(|V_{cb}|)$, while light quark and lepton masses imply $|\Delta_{u,d,e}|_{ij} \ll |V_{q}|$.
The assumption that $V_q$ is the only 
leading $U(2)^5$-breaking term in the quark sector ensures a suppression of 
flavor-violating terms in higher-dimensional operators as effective as the one implied by the hypothesis 
of Minimal Flavor Violation~\cite{Chivukula:1987py,DAmbrosio:2002vsn}.

An EFT constructed in terms of SM fields and arbitrary powers of the spurions in (\ref{eq:U2spur}), formally invariant under $U(2)^5$, 
is what we denote as the minimally-broken $U(2)^5$ hypothesis. The flavor scaling discussed in Section~\ref{sect:OPbasis}  for the leading semileptonic operators
addressing the anomalies  is obtained for 
\begin{align}\label{eq:leadingV}
|V_q|,~|V_\ell| = O(10^{-1})~.
\end{align}
This is  perfectly consistent with the estimate $|V_q| =O(|V_{cb}|)$ from the quark Yukawa sector 
and is compatible with a possible common origin for the two leading $U(2)^5$-breaking terms. 

\subsection*{Flavor bases and diagonalization of the Yukawa couplings}

We define as {\em interaction basis\/} the flavor basis 
in $U(2)^5$ space where $ V_{q,\ell}=  |V_{q,\ell}| \times \vec n$,  with $\vec n =(0,1)^\intercal$, and $\Delta_{u,d,e}^\dagger \Delta_{u,d,e}$ is diagonal.
 Without loss of generality, the Yukawa matrices can be brought to this basis via appropriate field redefinitions, yielding
\bea
& Y_u =|y_t|
\begin{pmatrix}
U_q^\dagger O_u^\intercal\, \hat\Delta_u & |V_q|\,|x_t|\,e^{i\phi_q}\,\vec n\\
0 & 1
\end{pmatrix}
\,, \qquad 
Y_d =|y_b|
\begin{pmatrix}
\;\;\;U_q^\dagger \hat\Delta_d& |V_q|\,|x_b|\,e^{i\phi_q}\,\vec n\\
\;\;\;0 & 1
\end{pmatrix}
\,, \no\\
& Y_e =|y_\tau|
\begin{pmatrix}
\;\;O_e^\intercal\,\hat\Delta_e\;\;& |V_\ell|\,|x_\tau|\,\vec n\\
\;\;0 & 1
\end{pmatrix}
\,,
\eea
where $\hat\Delta_{u,d,e}$ are $2\times 2$ diagonal positive matrices, $O_{u,e}$ are $2\times2$ orthogonal matrices, and $U_q$ is of the form
\begin{align}\label{eq:Uq}
U_q=
\begin{pmatrix}
c_d & s_d\,e^{i\alpha_d}\\
-s_d\,e^{-i\alpha_d} & c_d
\end{pmatrix}\,, \qquad 
s_d\equiv\sin\theta_d\,, \quad  c_d\equiv\cos\theta_d\,.
\end{align}
The unitary matrices that diagonalize the Yukawa matrices above, defined as $L_f^\dagger Y_f R_f  = {\rm diag}(Y_f)$ for $f=u,d,e$, assume the form~\cite{Fuentes-Martin:2019mun} 
\begin{align}\label{eq:YukRot}
\begin{aligned}
 L_d  &\approx 
\begin{pmatrix}
 c_d   &  -s_d\,e^{i\alpha_d}  & 0  \\
 s_d\,e^{-i\alpha_d}   &  c_d &  s_b   \\
-s_d\,s_b\,e^{-i(\alpha_d+\phi_q)}   & -c_d\,s_b\, e^{-i\phi_q} & e^{-i\phi_q}
\end{pmatrix}\,,
\qquad\qquad
L_e \approx    
\begin{pmatrix}
c_e       & -s_e          & 0  \\
s_e &  c_e           & s_\tau \\
-s_e s_\tau   &  -c_es_\tau & 1
\end{pmatrix}
\,,\\[5pt]
 R_d  &\approx  
\begin{pmatrix}
1   &  0  & 0  \\
0   &  1  & \frac{m_s}{m_b}\,s_b   \\
0  & -\frac{m_s}{m_b}\,s_b\,e^{-i\phi_q} & e^{-i\phi_q}
\end{pmatrix}
,\;\;
 R_u  \approx
\begin{pmatrix}
1   &  0  & 0  \\
0   &  1  & \frac{m_c}{m_t}\,s_t   \\
0  & -\frac{m_c}{m_t}\,s_t\,e^{-i\phi_q} & e^{-i\phi_q}
\end{pmatrix}
,\;\;
 R_e \approx  
\begin{pmatrix}
 1   & 0 & 0  \\
0 &  1& \frac{m_\mu}{m_\tau}\, s_\tau \\
0   &  -\frac{m_\mu}{m_\tau}\, s_\tau & 1
\end{pmatrix}\,,
\end{aligned}
\end{align}
with $L_u=L_d\,V_{\rm CKM}^\dagger$ and $s_t=s_b-V_{cb}$. 
The parameters $s_d$ and $\alpha_d$ are not free and can be expressed in terms of CKM parameters:
\begin{align}
s_d/c_d&=|V_{td}/V_{ts}|\,,&
\alpha_d&=\arg(V_{td}^*/V_{ts}^*)\,.
\end{align}
The fact that the $2\times 2$ upper block of $L_d$ is entirely determined in terms of CKM elements is the origin of the relation in \eqref{eq:U2min} and the condition $\beta_{L}^{d \tau}/\beta_{L}^{s \tau} = V^*_{td}/V^*_{ts}$ among the $U_1$ couplings. 
 
 The flavor-mixing parameters (spurion structures) 
which are left  unconstrained 
(i.e.~cannot be determined in terms of SM parameters) are:
 \begin{itemize}
 \item{} {\em quark sector}:  2--3 mixing angle $s_b/c_b=|x_b|\,|V_q|$ and CP-violating phase $\phi_q$;
 \item{} {\em lepton sector}: 2--3 mixing angle $s_\tau/c_\tau=|x_\tau|\,|V_\ell|$ and 1--2 mixing angle $s_e$ (encoded in $O_e$).
 \end{itemize}
 The approximate down-alignment in the 2--3 sector, necessary to satisfy the tight bounds from $B_s$ mixing when addressing the $B$ anomalies, is achieved for $|x_b| \ll 1$. In the exact down-alignment limit $|x_b| \to 0$ (or equivalently $s_b\to 0$), the phase $\phi_q$ becomes unphysical.

\section{Theoretical expressions for flavor observables} \label{app:Obs}

Here we collect the expressions for the relevant flavor observables used in our analysis in terms of Wilson coefficients and quark masses defined at the low scale $\mu_b=m_b$.

\subsection{\texorpdfstring{$b\to c(u)\tau\bar\nu$ transitions}{b->c(u) tau nu}}
\label{sec:appendix_bctaunu}

The low-energy effective Lagrangian describing $b\to u_i\tau \bar\nu$ transitions (with $i=1,2$) can be written as
\begin{align}\label{eq:LagCC}
\mathcal{L}_{b\to u_i\tau\bar\nu}=-\frac{4G_F}{\sqrt{2}}\,V_{ib}\,
\Big[\big(1+C_{V_L}^i\big)\,\mathcal{O}_{V_L}^i+C_{V_R}^i\,\mathcal{O}_{V_R}^i+C_{S_L}^i\,\mathcal{O}_{S_L}^i+C_{S_R}^i\,\mathcal{O}_{S_R}^i+C_T^i\,\mathcal{O}_T^i\Big] \,,
\end{align}
with operators
\begin{align}\label{eq:OpsCC}
\begin{aligned}
\mathcal{O}_{V_L}^i&=(\bar \tau_L \gamma_\mu {\nu}_{ L})(\bar u^i_L\gamma^\mu b_L)\,, \qquad\qquad&
\mathcal{O}_{V_R}^i&=(\bar \tau_L \gamma_\mu \nu_{ L})(\bar u^i_R\gamma^\mu b_R)\,, \\
\mathcal{O}_{S_L}^i&=(\bar \tau_R \,\nu_{ L})(\bar u^i_R\, b_L)\,, & 
\mathcal{O}_{S_R}^i&=(\bar \tau_R \,\nu_{ L})(\bar u^i_L\, b_R)\,, \\
\mathcal{O}_T^i~ &=(\bar \tau_R \sigma_{\mu\nu}\nu_{ L})(\bar u^i_R\sigma^{\mu\nu}b_L)\,.
\end{aligned}
\end{align}
In our case, we only generate NP contributions to the operators $\mathcal{O}_{V_L}$ and $\mathcal{O}_{S_R}$, with coefficients given by
\begin{align}
C_{V_L}^i \approx C_{LL}^{u_i}\,, \qquad
C_{S_R}^i \approx  - 2\,\eta_S\,C_{LR}^{u_i}\,,
\end{align}
where the coefficients $C_{LL}^{u_i}$ and $C_{LR}^{u_i}$ have been defined in (\ref{eq:CVSc}), and the evolution factor for the scalar current evaluates to $\eta_S\approx1.7$ for $\Lambda=2$~TeV.

The relevant observables in this category are the LFU ratios $R_{D}$ and $R_{D^\ast}$, defined as 
\begin{align}
\label{eq:RDRDs_def}
R_{D^{(\ast)}} &=\frac{\mathcal{B}\left(\bar B \to D^{(\ast)} \tau \bar \nu \right)}{\mathcal{B}\left(\bar B \to D^{(\ast)} \ell \bar \nu \right)}\,, \qquad  \ell = e,\mu\,,
\end{align}
and the leptonic decays $B^- \to \tau \bar \nu$ and $B_c^- \to \tau \bar \nu$. The expressions for these observables in terms of the Wilson coefficients in \eqref{eq:LagCC} read
\be
\begin{aligned}
\label{eq:RDRDs_EFT} 
R_{D} &= R_{D}^{\mathrm{SM}} \left[ \abs{1 + C_{V_L}^c}^2 + 1.5\, \mathrm{Re}\left \{ \left( 1 + C_{V_L}^c \right)  C_{S_R}^{c \ast} \right \}  + 1.03 \abs{ C_S^{c}}^2 \right] ,\\[10pt]
R_{D^\ast} &= R_{D^\ast}^{\mathrm{SM}} \left[ \abs{1 + C_{V_L}^c}^2 + 0.12\, \mathrm{Re}\left \{ \left( 1 + C_{V_L}^c \right)  C_{S_R}^{c \ast} \right \}  + 0.04 \abs{ C_{S_R}^{c}}^2 \right] , 
\end{aligned}
\ee
and
\begin{align}
\label{eq:Bitaunu_EFT} 
\mathcal{B}(B_i^- \to \tau \bar \nu ) & = \mathcal{B}(B_i^- \to \tau \bar \nu)_{\mathrm{SM}} \left| 1 + C_{V_L}^i+ C_{S_R}^i \, \frac{m_{B_i}^2}{m_\tau (m_b + m_i)} \right|^2 ,
\end{align}
where $B^-\equiv B_u^-$ is understood. 

\subsection{\texorpdfstring{$b\to s(d)\ell^-\ell^+$ transitions}{b->s (d) l l}}

The low-energy effective Lagrangian describing $b\to d_i  \ell_\alpha^-\ell_\beta^+$ transitions (with $i=1,2$) is given by
\begin{align}
\label{eq:LWET}
\mathcal{L}_{b \to d_i \ell_\alpha^-  \ell_\beta^+} &\supset  \frac{4G_F}{\sqrt2}\,V_{ti}^* V_{tb} \sum_k \, \mathcal{C}_k^{i \alpha \beta}\,\mathcal{O}_k^{i \alpha \beta} \,,
\end{align}
with operators
\begin{align}
\begin{aligned}
\label{eq:LWEToperators}
\mathcal{O}_9^{i \alpha \beta}&= \frac{\alpha}{4 \pi} \left(\bar d_L^i \gamma_\mu b_L \right) \left(\bar{\ell}_\alpha \gamma^\mu \ell_\beta \right) ,  & \mathcal{O}_{S}^{ i \alpha \beta}&=\frac{\alpha}{4 \pi}  (\bar d_L^i b_R )( \bar \ell_\alpha \ell_\beta) \,,  \\[5pt] 
\mathcal{O}_{10}^{i \alpha \beta}&=\frac{\alpha}{4 \pi} \left(\bar d_L^i \gamma_\mu b_L \right)
\left(\bar{\ell}_\alpha \gamma^\mu\gamma_5\, \ell_\beta \right)\,,   & \mathcal{O}_{P}^{i  \alpha \beta}&= \frac{\alpha}{4 \pi} (\bar  d_L^i b_R)( \bar \ell_\alpha \gamma_5\,\ell_\beta) \,.
\end{aligned}
\end{align}
Unless specified, the Wilson coefficients $\cC_k^{i\alpha \beta}$ include both the SM and the NP contributions, i.e.\
\begin{align}
& \cC_k^{i\alpha \beta}= \cC_{k, \mathrm{SM}}^{i\alpha \beta} + \cC_{k, \mathrm{NP}}^{i\alpha \beta}\,.
\end{align} 
At tree level, the NP contributions can be matched to the Lagrangian in \eqref{eq:SMEFTLag} via the relations 
\begin{align}
\begin{aligned}
\cC_{9, \rm{NP}}^{i  \alpha \beta} &= -\cC_{10, \rm{NP}}^{i  \alpha \beta} = -   \frac{2 \pi }{\alpha V_{ts}^* V_{tb}  } \, \cC^{ i 3  \beta \alpha}_{LL} \,,\\ 
\cC_{S}^{ i   \alpha \beta} &= - \cC_{P}^{i \alpha \beta} =    \frac{4\pi}{\alpha V_{ts}^* V_{tb}  } \, \cC^{i 3  \beta \alpha }_{LR} \,.
\end{aligned}
\end{align}
Whenever possible, we simplify the notation by omitting the superscript $i$ for $b \to s$ transitions. Also, for lepton-flavor conserving operators we use the superscript $\ell_\alpha$ instead of $\ell_\alpha \ell_\alpha$. 

\subsubsection*{$R_K$ and $R_{K^\ast}$}

The LFU ratios $R_K$ and $R_{K^\ast}$ are defined as
\begin{align}
\label{eq:RKRKs_def} 
R_{K^{(\ast)}}^{\rm [q^2_{\rm{min}},q^2_{\rm{max}}]} &= \frac{ \int_{q^2_{\rm{min}}}^{q^2_{\rm{max}}} d q^2 \frac{d\mathcal{B}}{d q^2}(B \to K^{(*)}\mu^+ \mu^-)}{\int_{q^2_{\rm{min}}}^{q^2_{\rm{max}}} d q^2 \frac{d\mathcal{B}}{d q^2}(B \to K^{(*)}e^+ e^-)}\,.
\end{align}
Their expressions in terms of the Wilson coefficients $\mathcal{C}_{9, \mathrm{NP}}^\mu$ and $\mathcal{C}_{10, \mathrm{NP}}^e$ read
\begin{align}
\begin{aligned}
\label{eq:RKRKs_EFT} 
R_{K}^{\rm [1.1,6]}&=  R_{K, \mathrm{SM}}^{[1.1,6]} \, \frac{1 + 0.24 \, {\rm Re}(\cC_{9, \mathrm{NP}}^\mu) -  0.26\, {\rm Re}(\cC_{10, \mathrm{NP}}^\mu)  + 0.03 \,  \big( |\cC_{9, \mathrm{NP}}^{\mu}|^2 + |\cC_{10, \mathrm{NP}}^{\mu}|^2\big)}{1+ 0.24 \, {\rm Re}(\cC_{9, \mathrm{NP}}^e) - 0.26\, {\rm Re}(\cC_{10, \mathrm{NP}}^e)  + 0.03  \,  \big( |\cC_{9, \mathrm{NP}}^{e}|^2 + |\cC_{10, \mathrm{NP}}^{e}|^2\big)}\,,\\
R_{K^\ast}^{[1.1,6]} & = R_{K^\ast, \mathrm{SM}}^{[1.1,6]} \,   \frac{1 + 0.18 \, {\rm Re}(\cC_{9, \mathrm{NP}}^\mu) -  0.29\,  {\rm Re}(\cC_{10, \mathrm{NP}}^\mu)  + 0.03 \,  \big(|\cC_{9, \mathrm{NP}}^{\mu}|^2 + |\cC_{10, \mathrm{NP}}^{\mu}|^2\big) }{1+ 0.18 \,  {\rm Re}(\cC_{9, \mathrm{NP}}^e) - 0.29\,  {\rm Re}(\cC_{10, \mathrm{NP}}^e)  + 0.03  \,  \big(|\cC_{9, \mathrm{NP}}^{e}|^2 + |\cC_{10, \mathrm{NP}}^{e}|^2\big)}\,, \\
R_{K^\ast}^{[0.045, 1.1]}  & =  R_{K^\ast, \mathrm{SM}}^{[0.045, 1.1]} \,  \frac{1 + 0.040 \,  {\rm Re}(\cC_{9, \mathrm{NP}}^\mu) - 0.080  \,{\rm Re}(\cC_{10, \mathrm{NP}}^\mu)  + 0.010 \,  \big( |\cC_{9, \mathrm{NP}}^{\mu}|^2 + |\cC_{10, \mathrm{NP}}^{\mu}|^2\big) }{1+ 0.037 \,  {\rm Re}(\cC_{9, \mathrm{NP}}^e) - 0.075\,  {\rm Re}(\cC_{10, \mathrm{NP}}^e)  + 0.009  \,  \big( |\cC_{9, \mathrm{NP}}^{e}|^2 + |\cC_{10, \mathrm{NP}}^{e}|^2\big)} \,, 
\end{aligned}
\end{align}
where the superscripts refer to the bins in $q^2\, [\mathrm{GeV}^2]$. As can be seen from these expressions, $R_{K}^{\rm [1.1,6]}$ and $R_{K^\ast}^{[1.1,6]}$ are insensitive to possible lepton-universal NP contributions. A subleading sensitivity to universal corrections is present in the low-$q^2$ bin of $R_{K^\ast}$, due to $m_\ell$-dependent kinematical effects close to threshold (recall that $R_{K^\ast}^{[0.045, 1.1]}\ne 1$ already in the SM). To accurately describe this effect, the expansion coefficients are given to higher accuracy. The latter are estimated taking into account also the non-universal QED corrections 
estimated in~\cite{Bordone:2016gaq}. 

\subsubsection*{Leptonic and semileptonic $B$ decays}

The theoretical expressions for the branching fractions of the relevant leptonic and semileptonic $B$ decays are
\begin{align}
\begin{aligned}
\label{eq:B2ll}
\mathcal{B}(B_i \to \ell^+ \ell^- ) 
&= \mathcal{B}(B_i \to \ell^+ \ell^-)_{\mathrm{SM}}  
\left\{ \abs{ 1 + 
 \frac{\cC_{10,\mathrm{NP}}^{i\ell}}{\cC_{10,\mathrm{SM}}} 
 + \frac{\cC_{P}^{i\ell}}{C_{10,\mathrm{SM}}}\, \frac{m_{B_i}^2}{2 m_\ell (m_b + m_i)} }^2 
\right.\\
& \qquad \qquad \qquad \qquad \qquad \left.+ \left( 1 - \frac{4 m_\ell^2}{m_{B_i}^2}\right) \abs{ 
\frac{\cC_S^{i\ell}}{\cC_{10,\mathrm{SM}}}\, \frac{m_{B_i}^2}{2 m_\ell (m_b + m_i)}}^2 \right\} ,
\end{aligned}
\end{align}
\begin{align}
\label{eq:B2Ktautau}
\mathcal{B}(B \to K \tau^+ \tau^{-} )  
&= 10^{-9}\left( 2.2 \,\abs{\cC_{9}^\tau}^2  + 6.0 \, \abs{\cC_{10}^\tau}^2   + 8.3 \, \abs{\cC_{S}^\tau}^2  + 8.9  \,\abs{\cC_{P}^\tau}^2 \nonumber \right. \\[5pt]
& \left. \quad   + 4.8 \,\mathrm{Re} \left \{ \cC_{S}^\tau \, \cC_{9}^{\tau \ast} \right\} + 5.9 \, \mathrm{Re} \left \{ \cC_{P}^\tau \,  \cC_{10}^{\tau \ast} \right\} \right) \,,\\[10pt]
\label{eq:B2Ktaumu}
\mathcal{B}(B^+ \to K^+ \tau^+ \mu^-)  
& = 10^{-9}\left(  9.6 \,\abs{\cC_{9}^{\tau \mu}}^2  + 10  \,\abs{\cC_{10}^{\tau\mu}}^2   + 13.58 \, \abs{\cC_{S}^{\tau\mu}}^2  + 14.54  \,\abs{\cC_{P}^{\tau \mu}}^2  \right. \nonumber\\[5pt]
& \left. \quad + 12.41\, \mathrm{Re} \left \{ \cC_{S}^{\tau\mu} \,  \cC_{9}^{\tau \mu \ast} \right\} + 15.20 \,\mathrm{Re} \left \{ \cC_{P}^{\tau \mu}\,  \cC_{10}^{\tau  \mu \ast} \right\} \right) ,\\[10pt]
\label{eq:Bs2taumu}
\mathcal{B}(B_s \to \tau^{-} \mu^{+}) & =  \frac{\tau_{B_s} G_F^2 f_{B_s}^2}{16 \pi}\, m_\tau^2\, m_{B_s} \left( 1 - \frac{m_\tau^2}{m_{B_s}^2} \right)^2  \frac{ \alpha^2 }{4 \pi^2}\, \abs{V_{tb} V_{ts}^{\ast}}^2 \\[2pt]
 &\quad \times \left\{ \abs{ \cC_{10}^{\tau \mu} + \frac{m_{B_s}^2}{2 m_\tau (m_b + m_s)}\, \cC_{P}^{\tau \mu}}^2 +  \abs{ \cC_{9}^{\tau \mu} + \frac{m_{B_s}^2}{2 m_\tau (m_b + m_s)}\, \cC_{S}^{\tau \mu}}^2 \right\} . \nonumber
\end{align}
The hadronic coefficients in \eqref{eq:B2Ktautau} and \eqref{eq:B2Ktaumu} have been computed using inputs from  lattice QCD as compiled in \cite{Bouchard:2013eph}. 

\subsection{\texorpdfstring{$\tau$}{tau} decays} 

\subsubsection*{Universality tests in \texorpdfstring{$\tau$}{tau} decays} 

$\tau$ leptonic decay rates provide a powerful test of lepton-flavor universality via the ratios
\begin{align} \begin{aligned} \label{eq:deftauLFUV}
\left(\frac{g_\tau}{g_{\mu (e)}}\right)_\ell &= \left[ \frac{\mathcal{B}(\tau \to e (\mu) \nu \bar \nu)/\mathcal{B}(\tau \to e (\mu) \nu \bar \nu)_{\rm SM}}{\mathcal{B}(\mu \to e \nu \bar \nu)/\mathcal{B}(\mu \to e \nu \bar \nu)_{\rm SM}}  \right]^{\frac12} , \\
\left(\frac{g_\tau}{g_{\mu}}\right)_{\pi} &= \left[\frac{\mathcal{B}(\tau \to \pi \nu)/\mathcal{B}(\tau \to \pi \nu )_{\rm SM}}{\mathcal{B}(\pi \to \mu \bar \nu)/\mathcal{B}(\pi \to \mu \bar \nu)_{\rm SM}} \right]^{\frac12} ,\\
\left(\frac{g_\tau}{g_{\mu}}\right)_{K} &= \left[\frac{\mathcal{B}(\tau \to K \nu)/\mathcal{B}(\tau \to K \nu )_{\rm SM}}{\mathcal{B}(K \to \mu \bar \nu)/\mathcal{B}(K \to \mu \bar \nu)_{\rm SM}} \right]^{\frac12} .
\end{aligned}
\end{align}
In terms of the high-energy Wilson coefficients, they are given by
\begin{align}\label{eq:tauLFUV}
\left(\frac{g_\tau}{g_{\mu}}\right)_{\ell,\pi,K} \approx \left(\frac{g_\tau}{g_{e}}\right)_\ell \approx1-0.079\,C_{LL}^{33\tau\tau} (\Lambda)\,,
\end{align}
where the running is computed assuming $\Lambda = 2 \, \mathrm{TeV}$. 

\subsubsection*{LFV \texorpdfstring{$\tau$}{tau} decays}

In our setup, the radiative decay $ \tau \to \mu \gamma$ is conveniently described via the effective Lagrangian
\begin{align}
\mathcal{L}_{\tau \to \mu \gamma} =  -  \frac{2}{v^2} \left[ \cC_{e \gamma}^{\mu \tau} \cO_{e \gamma}^{\mu \tau}   +  \cC_{LR}^{33 \tau \mu}    \cO_{LR}^{33 \tau \mu} + \mathrm{h.c.}  \right] ,
\label{eq:dipLag}
\end{align}
where $ \cO_{LR}^{33 \tau \mu}$ has been defined in \eqref{eq:opbasis}, and $O_{e \gamma}^{\beta \alpha} = e \,(\bar \ell_L^\beta \sigma^{\mu \nu} e_R^\alpha ) H F_{\mu \nu}$. The running of the Wilson coefficients from $\Lambda=2~{\rm TeV}$ to $m_\tau$ is given by 
\begin{align}\label{eq:taumugRGE}
\begin{aligned}
 \cC^{33\tau\mu}_{LR} (m_\tau)  & \approx  \, \eta_S \,  \cC^{33\tau\mu}_{LR} (\Lambda)  \,, \\
 \cC_{e \gamma}^{\alpha \beta}(m_\tau) &\approx  0.92 \, \cC_{e \gamma}^{\mu \tau} (\Lambda) \,,
\end{aligned}
\end{align}
and the Wilson coefficients at the high scale in the $U_1$ model are defined in~\eqref{eq:4Fmatching} and~\eqref{eq:ta2mugaMatching}, respectively. Starting from \eqref{eq:dipLag}, we find the decay amplitude (with $q=p-p'$)
\begin{align}
i \mathcal{A} (\tau \to \mu \gamma)= - \frac{4e}{v^2}\,  
\bar u(p')\,\sigma^{\alpha\nu} q_\nu\,
\big( \mathcal{A}_R P_R + \mathcal{A}_L P_L \big)\,
u(p)\,\varepsilon_\alpha^\ast(q) \,,
\end{align}
with
\begin{align}
\begin{aligned}
 \mathcal{A}_R & =    \frac{v}{\sqrt{2}}\,\, \cC_{e \gamma}^{\mu \tau}(m_\tau)  +   \frac{ m_b}{16 \pi^2}\,\,   \cC^{33\tau\mu}_{LR} (m_\tau)   \,, \\
 \mathcal{A}_L & =      \frac{v}{\sqrt{2}}\,\,  \cC_{e \gamma}^{\tau \mu \ast}(m_\tau)   \,.
\end{aligned}
\end{align}
Neglecting the muon mass, the branching ratio is given by
\begin{align}\label{eq:tau2mugamma}
\mathcal{B}(\tau \to \mu \gamma) &=   
\frac{8\hspace{0.3mm} G_F^2\hspace{0.3mm} \alpha\hspace{0.3mm} m_\tau^3}{\Gamma_\tau} \left( \abs{\mathcal{A}_R}^2 +  \abs{\mathcal{A}_L}^2  \right) .
\end{align}
For the decay $\tau\to\mu\phi$, we find
\begin{align}
\label{eq:tau2muphi} 
\mathcal{B}( \tau \to \mu \phi)& = \frac{1}{\Gamma_{\tau}} \frac{G_{F}^{2} f_{\phi}^{2}}{16\pi}\,m_{\tau}^{3} \left(1-  \frac{m_\phi^2}{m_\tau^2}\right)^{2} \left(1 + 2 \frac{m_\phi^2}{m_\tau^2}\right)\,
\big|C_{LL}^{22\tau\mu}\big|^{2} \,,
\end{align} 
where $f_\phi\approx225~{\rm MeV}$.

\bibliographystyle{JHEP}

{\footnotesize
\bibliography{paper}

\providecommand{\href}[2]{#2}\begingroup\raggedright\begin{thebibliography}{100}

\bibitem{Aaij:2021vac}
{\bf LHCb} Collaboration, R.~Aaij et~al., {\it {Test of lepton universality in
  beauty-quark decays}},  \href{http://arxiv.org/abs/2103.11769}{{\tt
  arXiv:2103.11769}}.

\bibitem{Aaij:2014ora}
{\bf LHCb} Collaboration, R.~Aaij et~al., {\it {Test of lepton universality
  using $B^{+}\rightarrow K^{+}\ell^{+}\ell^{-}$ decays}},  {\em Phys. Rev.
  Lett.} {\bf 113} (2014) 151601, [\href{http://arxiv.org/abs/1406.6482}{{\tt
  arXiv:1406.6482}}].

\bibitem{Aaij:2017vbb}
{\bf LHCb} Collaboration, R.~Aaij et~al., {\it {Test of lepton universality
  with $B^{0} \rightarrow K^{*0}\ell^{+}\ell^{-}$ decays}},  {\em JHEP} {\bf
  08} (2017) 055, [\href{http://arxiv.org/abs/1705.05802}{{\tt
  arXiv:1705.05802}}].

\bibitem{Aaij:2019wad}
{\bf LHCb} Collaboration, R.~Aaij et~al., {\it {Search for lepton-universality
  violation in $B^+\to K^+\ell^+\ell^-$ decays}},  {\em Phys. Rev. Lett.} {\bf
  122} (2019), no.~19 191801, [\href{http://arxiv.org/abs/1903.09252}{{\tt
  arXiv:1903.09252}}].

\bibitem{Aaij:2013qta}
{\bf LHCb} Collaboration, R.~Aaij et~al., {\it {Measurement of
  Form-Factor-Independent Observables in the Decay $B^{0} \to K^{*0} \mu^+
  \mu^-$}},  {\em Phys. Rev. Lett.} {\bf 111} (2013) 191801,
  [\href{http://arxiv.org/abs/1308.1707}{{\tt arXiv:1308.1707}}].

\bibitem{Aaij:2015oid}
{\bf LHCb} Collaboration, R.~Aaij et~al., {\it {Angular analysis of the $B^{0}
  \to K^{*0} \mu^{+} \mu^{-}$ decay using 3 fb$^{-1}$ of integrated
  luminosity}},  {\em JHEP} {\bf 02} (2016) 104,
  [\href{http://arxiv.org/abs/1512.04442}{{\tt arXiv:1512.04442}}].

\bibitem{CMS:2014xfa}
{\bf CMS, LHCb} Collaboration, V.~Khachatryan et~al., {\it {Observation of the
  rare $B^0_s\to\mu^+\mu^-$ decay from the combined analysis of CMS and LHCb
  data}},  {\em Nature} {\bf 522} (2015) 68--72,
  [\href{http://arxiv.org/abs/1411.4413}{{\tt arXiv:1411.4413}}].

\bibitem{Lees:2012xj}
{\bf BaBar} Collaboration, J.~P. Lees et~al., {\it {Evidence for an excess of
  $\bar{B} \to D^{(*)} \tau^-\bar{\nu}_\tau$ decays}},  {\em Phys. Rev. Lett.}
  {\bf 109} (2012) 101802, [\href{http://arxiv.org/abs/1205.5442}{{\tt
  arXiv:1205.5442}}].

\bibitem{Lees:2013uzd}
{\bf BaBar} Collaboration, J.~P. Lees et~al., {\it {Measurement of an Excess of
  $\bar{B} \to D^{(*)}\tau^- \bar{\nu}_\tau$ Decays and Implications for
  Charged Higgs Bosons}},  {\em Phys. Rev. D} {\bf 88} (2013), no.~7 072012,
  [\href{http://arxiv.org/abs/1303.0571}{{\tt arXiv:1303.0571}}].

\bibitem{Huschle:2015rga}
{\bf Belle} Collaboration, M.~Huschle et~al., {\it {Measurement of the
  branching ratio of $\bar{B} \to D^{(\ast)} \tau^- \bar{\nu}_\tau$ relative to
  $\bar{B} \to D^{(\ast)} \ell^- \bar{\nu}_\ell$ decays with hadronic tagging
  at Belle}},  {\em Phys. Rev. D} {\bf 92} (2015), no.~7 072014,
  [\href{http://arxiv.org/abs/1507.03233}{{\tt arXiv:1507.03233}}].

\bibitem{Aaij:2015yra}
{\bf LHCb} Collaboration, R.~Aaij et~al., {\it {Measurement of the ratio of
  branching fractions $\mathcal{B}(\bar{B}^0 \to
  D^{*+}\tau^{-}\bar{\nu}_{\tau})/\mathcal{B}(\bar{B}^0 \to
  D^{*+}\mu^{-}\bar{\nu}_{\mu})$}},  {\em Phys. Rev. Lett.} {\bf 115} (2015),
  no.~11 111803, [\href{http://arxiv.org/abs/1506.08614}{{\tt
  arXiv:1506.08614}}]. [Erratum: Phys.Rev.Lett. 115, 159901 (2015)].

\bibitem{Hirose:2016wfn}
{\bf Belle} Collaboration, S.~Hirose et~al., {\it {Measurement of the $\tau$
  lepton polarization and $R(D^*)$ in the decay $\bar{B} \to D^* \tau^-
  \bar{\nu}_\tau$}},  {\em Phys. Rev. Lett.} {\bf 118} (2017), no.~21 211801,
  [\href{http://arxiv.org/abs/1612.00529}{{\tt arXiv:1612.00529}}].

\bibitem{Aaij:2017deq}
{\bf LHCb} Collaboration, R.~Aaij et~al., {\it {Test of Lepton Flavor
  Universality by the measurement of the $B^0 \to D^{*-} \tau^+ \nu_{\tau}$
  branching fraction using three-prong $\tau$ decays}},  {\em Phys. Rev. D}
  {\bf 97} (2018), no.~7 072013, [\href{http://arxiv.org/abs/1711.02505}{{\tt
  arXiv:1711.02505}}].

\bibitem{Aaij:2017uff}
{\bf LHCb} Collaboration, R.~Aaij et~al., {\it {Measurement of the ratio of the
  $B^0 \to D^{*-} \tau^+ \nu_{\tau}$ and $B^0 \to D^{*-} \mu^+ \nu_{\mu}$
  branching fractions using three-prong $\tau$-lepton decays}},  {\em Phys.
  Rev. Lett.} {\bf 120} (2018), no.~17 171802,
  [\href{http://arxiv.org/abs/1708.08856}{{\tt arXiv:1708.08856}}].

\bibitem{Hiller:2014yaa}
G.~Hiller and M.~Schmaltz, {\it {$R_K$ and future $b \to s \ell \ell$ physics
  beyond the standard model opportunities}},  {\em Phys. Rev. D} {\bf 90}
  (2014) 054014, [\href{http://arxiv.org/abs/1408.1627}{{\tt
  arXiv:1408.1627}}].

\bibitem{Hiller:2003js}
G.~Hiller and F.~Kruger, {\it {More model-independent analysis of $b \to s$
  processes}},  {\em Phys. Rev. D} {\bf 69} (2004) 074020,
  [\href{http://arxiv.org/abs/hep-ph/0310219}{{\tt hep-ph/0310219}}].

\bibitem{Descotes-Genon:2013wba}
S.~Descotes-Genon, J.~Matias, and J.~Virto, {\it {Understanding the $B\to
  K^*\mu^+\mu^-$ Anomaly}},  {\em Phys. Rev. D} {\bf 88} (2013) 074002,
  [\href{http://arxiv.org/abs/1307.5683}{{\tt arXiv:1307.5683}}].

\bibitem{Altmannshofer:2013foa}
W.~Altmannshofer and D.~M. Straub, {\it {New Physics in $B \to K^*\mu\mu$?}},
  {\em Eur. Phys. J. C} {\bf 73} (2013) 2646,
  [\href{http://arxiv.org/abs/1308.1501}{{\tt arXiv:1308.1501}}].

\bibitem{Hurth:2013ssa}
T.~Hurth and F.~Mahmoudi, {\it {On the LHCb anomaly in B $\to
  K^*\ell^+\ell^-$}},  {\em JHEP} {\bf 04} (2014) 097,
  [\href{http://arxiv.org/abs/1312.5267}{{\tt arXiv:1312.5267}}].

\bibitem{Hurth:2014vma}
T.~Hurth, F.~Mahmoudi, and S.~Neshatpour, {\it {Global fits to $b \to
  s\ell\ell$ data and signs for lepton non-universality}},  {\em JHEP} {\bf 12}
  (2014) 053, [\href{http://arxiv.org/abs/1410.4545}{{\tt arXiv:1410.4545}}].

\bibitem{Altmannshofer:2014rta}
W.~Altmannshofer and D.~M. Straub, {\it {New physics in $b\rightarrow s$
  transitions after LHC run 1}},  {\em Eur. Phys. J. C} {\bf 75} (2015), no.~8
  382, [\href{http://arxiv.org/abs/1411.3161}{{\tt arXiv:1411.3161}}].

\bibitem{Descotes-Genon:2015uva}
S.~Descotes-Genon, L.~Hofer, J.~Matias, and J.~Virto, {\it {Global analysis of
  $b\to s\ell\ell$ anomalies}},  {\em JHEP} {\bf 06} (2016) 092,
  [\href{http://arxiv.org/abs/1510.04239}{{\tt arXiv:1510.04239}}].

\bibitem{Ciuchini:2017mik}
M.~Ciuchini, A.~M. Coutinho, M.~Fedele, E.~Franco, A.~Paul, L.~Silvestrini, and
  M.~Valli, {\it {On Flavourful Easter eggs for New Physics hunger and Lepton
  Flavour Universality violation}},  {\em Eur. Phys. J. C} {\bf 77} (2017),
  no.~10 688, [\href{http://arxiv.org/abs/1704.05447}{{\tt arXiv:1704.05447}}].

\bibitem{Alguero:2019ptt}
M.~Alguer\'o, B.~Capdevila, A.~Crivellin, S.~Descotes-Genon, P.~Masjuan,
  J.~Matias, M.~Novoa~Brunet, and J.~Virto, {\it {Emerging patterns of New
  Physics with and without Lepton Flavour Universal contributions}},  {\em Eur.
  Phys. J. C} {\bf 79} (2019), no.~8 714,
  [\href{http://arxiv.org/abs/1903.09578}{{\tt arXiv:1903.09578}}]. [Addendum:
  Eur.Phys.J.C 80, 511 (2020)].

\bibitem{Ciuchini:2020gvn}
M.~Ciuchini, M.~Fedele, E.~Franco, A.~Paul, L.~Silvestrini, and M.~Valli, {\it
  {Lessons from the $B^{0,+}\to K^{*0,+}\mu^+\mu^-$ angular analyses}},  {\em
  Phys. Rev. D} {\bf 103} (2021), no.~1 015030,
  [\href{http://arxiv.org/abs/2011.01212}{{\tt arXiv:2011.01212}}].

\bibitem{Glashow:2014iga}
S.~L. Glashow, D.~Guadagnoli, and K.~Lane, {\it {Lepton Flavor Violation in $B$
  Decays?}},  {\em Phys. Rev. Lett.} {\bf 114} (2015) 091801,
  [\href{http://arxiv.org/abs/1411.0565}{{\tt arXiv:1411.0565}}].

\bibitem{Bhattacharya:2014wla}
B.~Bhattacharya, A.~Datta, D.~London, and S.~Shivashankara, {\it {Simultaneous
  Explanation of the $R_K$ and $R(D^{(*)})$ Puzzles}},  {\em Phys. Lett. B}
  {\bf 742} (2015) 370--374, [\href{http://arxiv.org/abs/1412.7164}{{\tt
  arXiv:1412.7164}}].

\bibitem{Alonso:2015sja}
R.~Alonso, B.~Grinstein, and J.~Martin~Camalich, {\it {Lepton universality
  violation and lepton flavor conservation in $B$-meson decays}},  {\em JHEP}
  {\bf 10} (2015) 184, [\href{http://arxiv.org/abs/1505.05164}{{\tt
  arXiv:1505.05164}}].

\bibitem{Greljo:2015mma}
A.~Greljo, G.~Isidori, and D.~Marzocca, {\it {On the breaking of Lepton Flavor
  Universality in B decays}},  {\em JHEP} {\bf 07} (2015) 142,
  [\href{http://arxiv.org/abs/1506.01705}{{\tt arXiv:1506.01705}}].

\bibitem{Calibbi:2015kma}
L.~Calibbi, A.~Crivellin, and T.~Ota, {\it {Effective Field Theory Approach to
  $b\to s\ell\ell^{(')}$, $B\to K^{(*)}\nu\overline{\nu}$ and $B\to
  D^{(*)}\tau\nu$ with Third Generation Couplings}},  {\em Phys. Rev. Lett.}
  {\bf 115} (2015) 181801, [\href{http://arxiv.org/abs/1506.02661}{{\tt
  arXiv:1506.02661}}].

\bibitem{Gripaios:2014tna}
B.~Gripaios, M.~Nardecchia, and S.~A. Renner, {\it {Composite leptoquarks and
  anomalies in $B$-meson decays}},  {\em JHEP} {\bf 05} (2015) 006,
  [\href{http://arxiv.org/abs/1412.1791}{{\tt arXiv:1412.1791}}].

\bibitem{Gripaios:2009dq}
B.~Gripaios, {\it {Composite Leptoquarks at the LHC}},  {\em JHEP} {\bf 02}
  (2010) 045, [\href{http://arxiv.org/abs/0910.1789}{{\tt arXiv:0910.1789}}].

\bibitem{Fajfer:2015ycq}
S.~Fajfer and N.~Ko\v{s}nik, {\it {Vector leptoquark resolution of $R_K$ and
  $R_{D^{(*)}}$ puzzles}},  {\em Phys. Lett. B} {\bf 755} (2016) 270--274,
  [\href{http://arxiv.org/abs/1511.06024}{{\tt arXiv:1511.06024}}].

\bibitem{Bauer:2015knc}
M.~Bauer and M.~Neubert, {\it {Minimal Leptoquark Explanation for the
  $R_{D^{(*)}}$, $R_K$, and $(g-2)_\mu$ Anomalies}},  {\em Phys. Rev. Lett.}
  {\bf 116} (2016), no.~14 141802, [\href{http://arxiv.org/abs/1511.01900}{{\tt
  arXiv:1511.01900}}].

\bibitem{Barbieri:2015yvd}
R.~Barbieri, G.~Isidori, A.~Pattori, and F.~Senia, {\it {Anomalies in
  $B$-decays and $U(2)$ flavour symmetry}},  {\em Eur. Phys. J. C} {\bf 76}
  (2016), no.~2 67, [\href{http://arxiv.org/abs/1512.01560}{{\tt
  arXiv:1512.01560}}].

\bibitem{Becirevic:2016yqi}
D.~Be\v{c}irevi\'c, S.~Fajfer, N.~Ko\v{s}nik, and O.~Sumensari, {\it
  {Leptoquark model to explain the $B$-physics anomalies, $R_K$ and $R_D$}},
  {\em Phys. Rev. D} {\bf 94} (2016), no.~11 115021,
  [\href{http://arxiv.org/abs/1608.08501}{{\tt arXiv:1608.08501}}].

\bibitem{Becirevic:2018afm}
D.~Be\v{c}irevi\'c, I.~Dor\v{s}ner, S.~Fajfer, N.~Ko\v{s}nik, D.~A. Faroughy,
  and O.~Sumensari, {\it {Scalar leptoquarks from grand unified theories to
  accommodate the $B$-physics anomalies}},  {\em Phys. Rev. D} {\bf 98} (2018),
  no.~5 055003, [\href{http://arxiv.org/abs/1806.05689}{{\tt
  arXiv:1806.05689}}].

\bibitem{Barbieri:2011ci}
R.~Barbieri, G.~Isidori, J.~Jones-Perez, P.~Lodone, and D.~M. Straub, {\it
  {$U(2)$ and Minimal Flavour Violation in Supersymmetry}},  {\em Eur. Phys. J.
  C} {\bf 71} (2011) 1725, [\href{http://arxiv.org/abs/1105.2296}{{\tt
  arXiv:1105.2296}}].

\bibitem{Feruglio:2016gvd}
F.~Feruglio, P.~Paradisi, and A.~Pattori, {\it {Revisiting Lepton Flavor
  Universality in B Decays}},  {\em Phys. Rev. Lett.} {\bf 118} (2017), no.~1
  011801, [\href{http://arxiv.org/abs/1606.00524}{{\tt arXiv:1606.00524}}].

\bibitem{Feruglio:2017rjo}
F.~Feruglio, P.~Paradisi, and A.~Pattori, {\it {On the Importance of
  Electroweak Corrections for B Anomalies}},  {\em JHEP} {\bf 09} (2017) 061,
  [\href{http://arxiv.org/abs/1705.00929}{{\tt arXiv:1705.00929}}].

\bibitem{Faroughy:2016osc}
D.~A. Faroughy, A.~Greljo, and J.~F. Kamenik, {\it {Confronting lepton flavor
  universality violation in B decays with high-$p_T$ tau lepton searches at
  LHC}},  {\em Phys. Lett. B} {\bf 764} (2017) 126--134,
  [\href{http://arxiv.org/abs/1609.07138}{{\tt arXiv:1609.07138}}].

\bibitem{Greljo:2017vvb}
A.~Greljo and D.~Marzocca, {\it {High-$p_T$ dilepton tails and flavor
  physics}},  {\em Eur. Phys. J. C} {\bf 77} (2017), no.~8 548,
  [\href{http://arxiv.org/abs/1704.09015}{{\tt arXiv:1704.09015}}].

\bibitem{Buttazzo:2017ixm}
D.~Buttazzo, A.~Greljo, G.~Isidori, and D.~Marzocca, {\it {B-physics anomalies:
  a guide to combined explanations}},  {\em JHEP} {\bf 11} (2017) 044,
  [\href{http://arxiv.org/abs/1706.07808}{{\tt arXiv:1706.07808}}].

\bibitem{Pati:1974yy}
J.~C. Pati and A.~Salam, {\it {Lepton Number as the Fourth Color}},  {\em Phys.
  Rev. D} {\bf 10} (1974) 275--289. [Erratum: Phys.Rev.D 11, 703--703 (1975)].

\bibitem{Barbieri:2016las}
R.~Barbieri, C.~W. Murphy, and F.~Senia, {\it {B-decay Anomalies in a Composite
  Leptoquark Model}},  {\em Eur. Phys. J. C} {\bf 77} (2017), no.~1 8,
  [\href{http://arxiv.org/abs/1611.04930}{{\tt arXiv:1611.04930}}].

\bibitem{Barbieri:2017tuq}
R.~Barbieri and A.~Tesi, {\it {$B$-decay anomalies in Pati-Salam SU(4)}},  {\em
  Eur. Phys. J. C} {\bf 78} (2018), no.~3 193,
  [\href{http://arxiv.org/abs/1712.06844}{{\tt arXiv:1712.06844}}].

\bibitem{DiLuzio:2017vat}
L.~Di~Luzio, A.~Greljo, and M.~Nardecchia, {\it {Gauge leptoquark as the origin
  of B-physics anomalies}},  {\em Phys. Rev. D} {\bf 96} (2017), no.~11 115011,
  [\href{http://arxiv.org/abs/1708.08450}{{\tt arXiv:1708.08450}}].

\bibitem{Diaz:2017lit}
B.~Diaz, M.~Schmaltz, and Y.-M. Zhong, {\it {The leptoquark
  Hunter\textquoteright{}s guide: Pair production}},  {\em JHEP} {\bf 10}
  (2017) 097, [\href{http://arxiv.org/abs/1706.05033}{{\tt arXiv:1706.05033}}].

\bibitem{Georgi:2016xhm}
H.~Georgi and Y.~Nakai, {\it {Diphoton resonance from a new strong force}},
  {\em Phys. Rev. D} {\bf 94} (2016), no.~7 075005,
  [\href{http://arxiv.org/abs/1606.05865}{{\tt arXiv:1606.05865}}].

\bibitem{Blanke:2018sro}
M.~Blanke and A.~Crivellin, {\it {$B$ Meson Anomalies in a Pati-Salam Model
  within the Randall-Sundrum Background}},  {\em Phys. Rev. Lett.} {\bf 121}
  (2018), no.~1 011801, [\href{http://arxiv.org/abs/1801.07256}{{\tt
  arXiv:1801.07256}}].

\bibitem{Fornal:2018dqn}
B.~Fornal, S.~A. Gadam, and B.~Grinstein, {\it {Left-Right SU(4) Vector
  Leptoquark Model for Flavor Anomalies}},  {\em Phys. Rev. D} {\bf 99} (2019),
  no.~5 055025, [\href{http://arxiv.org/abs/1812.01603}{{\tt
  arXiv:1812.01603}}].

\bibitem{Bordone:2017bld}
M.~Bordone, C.~Cornella, J.~Fuentes-Martin, and G.~Isidori, {\it {A three-site
  gauge model for flavor hierarchies and flavor anomalies}},  {\em Phys. Lett.
  B} {\bf 779} (2018) 317--323, [\href{http://arxiv.org/abs/1712.01368}{{\tt
  arXiv:1712.01368}}].

\bibitem{Greljo:2018tuh}
A.~Greljo and B.~A. Stefanek, {\it {Third family quark\textendash{}lepton
  unification at the TeV scale}},  {\em Phys. Lett. B} {\bf 782} (2018)
  131--138, [\href{http://arxiv.org/abs/1802.04274}{{\tt arXiv:1802.04274}}].

\bibitem{Fuentes-Martin:2020bnh}
J.~Fuentes-Mart\'\i{}n and P.~Stangl, {\it {Third-family quark-lepton
  unification with a fundamental composite Higgs}},  {\em Phys. Lett. B} {\bf
  811} (2020) 135953, [\href{http://arxiv.org/abs/2004.11376}{{\tt
  arXiv:2004.11376}}].

\bibitem{Fuentes-Martin:2020pww}
J.~Fuentes-Martin, G.~Isidori, J.~Pag\`es, and B.~A. Stefanek, {\it {Flavor
  Non-universal Pati-Salam Unification and Neutrino Masses}},
  \href{http://arxiv.org/abs/2012.10492}{{\tt arXiv:2012.10492}}.

\bibitem{Fuentes-Martin:2019ign}
J.~Fuentes-Mart\'\i{}n, G.~Isidori, M.~K\"onig, and N.~Selimovi\'c, {\it
  {Vector Leptoquarks Beyond Tree Level}},  {\em Phys. Rev. D} {\bf 101}
  (2020), no.~3 035024, [\href{http://arxiv.org/abs/1910.13474}{{\tt
  arXiv:1910.13474}}].

\bibitem{Fuentes-Martin:2020luw}
J.~Fuentes-Mart\'\i{}n, G.~Isidori, M.~K\"onig, and N.~Selimovi\'c, {\it
  {Vector leptoquarks beyond tree level. II. $\mathcal{O}(\alpha_s)$
  corrections and radial modes}},  {\em Phys. Rev. D} {\bf 102} (2020), no.~3
  035021, [\href{http://arxiv.org/abs/2006.16250}{{\tt arXiv:2006.16250}}].

\bibitem{Fuentes-Martin:2020hvc}
J.~Fuentes-Mart\'\i{}n, G.~Isidori, M.~K\"onig, and N.~Selimovi\'c, {\it
  {Vector Leptoquarks Beyond Tree Level III: Vector-like Fermions and
  Flavor-Changing Transitions}},  {\em Phys. Rev. D} {\bf 102} (2020) 115015,
  [\href{http://arxiv.org/abs/2009.11296}{{\tt arXiv:2009.11296}}].

\bibitem{Grzadkowski:2010es}
B.~Grzadkowski, M.~Iskrzynski, M.~Misiak, and J.~Rosiek, {\it {Dimension-Six
  Terms in the Standard Model Lagrangian}},  {\em JHEP} {\bf 10} (2010) 085,
  [\href{http://arxiv.org/abs/1008.4884}{{\tt arXiv:1008.4884}}].

\bibitem{Fuentes-Martin:2019mun}
J.~Fuentes-Mart\'\i{}n, G.~Isidori, J.~Pag\`es, and K.~Yamamoto, {\it {With or
  without U(2)? Probing non-standard flavor and helicity structures in
  semileptonic B decays}},  {\em Phys. Lett. B} {\bf 800} (2020) 135080,
  [\href{http://arxiv.org/abs/1909.02519}{{\tt arXiv:1909.02519}}].

\bibitem{Buchalla:1995vs}
G.~Buchalla, A.~J. Buras, and M.~E. Lautenbacher, {\it {Weak decays beyond
  leading logarithms}},  {\em Rev. Mod. Phys.} {\bf 68} (1996) 1125--1144,
  [\href{http://arxiv.org/abs/hep-ph/9512380}{{\tt hep-ph/9512380}}].

\bibitem{Crivellin:2018yvo}
A.~Crivellin, C.~Greub, D.~M\"uller, and F.~Saturnino, {\it {Importance of Loop
  Effects in Explaining the Accumulated Evidence for New Physics in B Decays
  with a Vector Leptoquark}},  {\em Phys. Rev. Lett.} {\bf 122} (2019), no.~1
  011805, [\href{http://arxiv.org/abs/1807.02068}{{\tt arXiv:1807.02068}}].

\bibitem{Bordone:2018nbg}
M.~Bordone, C.~Cornella, J.~Fuentes-Mart\'\i{}n, and G.~Isidori, {\it
  {Low-energy signatures of the $\mathrm{PS}^3$ model: from $B$-physics
  anomalies to LFV}},  {\em JHEP} {\bf 10} (2018) 148,
  [\href{http://arxiv.org/abs/1805.09328}{{\tt arXiv:1805.09328}}].

\bibitem{Bordone:2016gaq}
M.~Bordone, G.~Isidori, and A.~Pattori, {\it {On the Standard Model predictions
  for $R_K$ and $R_{K^*}$}},  {\em Eur. Phys. J. C} {\bf 76} (2016), no.~8 440,
  [\href{http://arxiv.org/abs/1605.07633}{{\tt arXiv:1605.07633}}].

\bibitem{Aaboud:2018mst}
{\bf ATLAS} Collaboration, M.~Aaboud et~al., {\it {Study of the rare decays of
  $B^0_s$ and $B^0$ mesons into muon pairs using data collected during 2015 and
  2016 with the ATLAS detector}},  {\em JHEP} {\bf 04} (2019) 098,
  [\href{http://arxiv.org/abs/1812.03017}{{\tt arXiv:1812.03017}}].

\bibitem{Sirunyan:2019xdu}
{\bf CMS} Collaboration, A.~M. Sirunyan et~al., {\it {Measurement of properties
  of B$^0_\mathrm{s}\to\mu^+\mu^-$ decays and search for B$^0\to\mu^+\mu^-$
  with the CMS experiment}},  {\em JHEP} {\bf 04} (2020) 188,
  [\href{http://arxiv.org/abs/1910.12127}{{\tt arXiv:1910.12127}}].

\bibitem{LHCbBsmumu}
{\bf LHCb} Collaboration, M.~Santimaria, {\it {Measurement of
  $B^0\to\mu^+\mu^-$ decays with Run 1 + Run 2 data}}, . [LHCb seminar. Slides
  available at this
  \href{https://indico.cern.ch/event/976688/attachments/2213706/3747159/santimaria\_LHC\_seminar\_2021.pdf}{url}].

\bibitem{Beneke:2019slt}
M.~Beneke, C.~Bobeth, and R.~Szafron, {\it {Power-enhanced leading-logarithmic
  QED corrections to $B_q \to \mu^+\mu^-$}},  {\em JHEP} {\bf 10} (2019) 232,
  [\href{http://arxiv.org/abs/1908.07011}{{\tt arXiv:1908.07011}}].

\bibitem{Barlow:2004wg}
R.~Barlow, {\it {Asymmetric statistical errors}},  in {\em {Statistical
  Problems in Particle Physics, Astrophysics and Cosmology}}, 6, 2004.
\newblock \href{http://arxiv.org/abs/physics/0406120}{{\tt physics/0406120}}.

\bibitem{Aaij:2021nyr}
{\bf LHCb} Collaboration, R.~Aaij et~al., {\it {Precise measurement of the
  $f_s/f_d$ ratio of fragmentation fractions and of $B^0_s$ decay branching
  fractions}},  \href{http://arxiv.org/abs/2103.06810}{{\tt arXiv:2103.06810}}.

\bibitem{Isidori:2020acz}
G.~Isidori, S.~Nabeebaccus, and R.~Zwicky, {\it {QED corrections in $
  \overline{B}\to \overline{K}{\mathrm{\ell}}^{+}{\mathrm{\ell}}^{-} $ at the
  double-differential level}},  {\em JHEP} {\bf 12} (2020) 104,
  [\href{http://arxiv.org/abs/2009.00929}{{\tt arXiv:2009.00929}}].

\bibitem{CDF:2012qwd}
{\bf CDF} Collaboration, {\it {Precise Measurements of Exclusive $b \to s
  \mu^+\mu^-$ Decay Amplitudes Using the Full CDF Data Set}},  {\em
  CDF-NOTE-10894} (2012).

\bibitem{Aaij:2014pli}
{\bf LHCb} Collaboration, R.~Aaij et~al., {\it {Differential branching
  fractions and isospin asymmetries of $B \to K^{(*)} \mu^+ \mu^-$ decays}},
  {\em JHEP} {\bf 06} (2014) 133, [\href{http://arxiv.org/abs/1403.8044}{{\tt
  arXiv:1403.8044}}].

\bibitem{Aaij:2016flj}
{\bf LHCb} Collaboration, R.~Aaij et~al., {\it {Measurements of the S-wave
  fraction in $B^{0}\rightarrow K^{+}\pi^{-}\mu^{+}\mu^{-}$ decays and the
  $B^{0}\rightarrow K^{\ast}(892)^{0}\mu^{+}\mu^{-}$ differential branching
  fraction}},  {\em JHEP} {\bf 11} (2016) 047,
  [\href{http://arxiv.org/abs/1606.04731}{{\tt arXiv:1606.04731}}]. [Erratum:
  JHEP 04, 142 (2017)].

\bibitem{Khachatryan:2015isa}
{\bf CMS} Collaboration, V.~Khachatryan et~al., {\it {Angular analysis of the
  decay $B^0 \to K^{*0} \mu^+ \mu^-$ from pp collisions at $\sqrt s = 8$ TeV}},
   {\em Phys. Lett. B} {\bf 753} (2016) 424--448,
  [\href{http://arxiv.org/abs/1507.08126}{{\tt arXiv:1507.08126}}].

\bibitem{CMS:2017ivg}
{\bf CMS} Collaboration, {\it {Measurement of the $P_1$ and $P_5'$ angular
  parameters of the decay $\mathrm{B}^0 \to \mathrm{K}^{*0} \mu^+ \mu^-$ in
  proton-proton collisions at $\sqrt{s}=8~\mathrm{TeV}$}},  {\em
  CMS-PAS-BPH-15-008} (2017).

\bibitem{Aaboud:2018krd}
{\bf ATLAS} Collaboration, M.~Aaboud et~al., {\it {Angular analysis of $B^0_d
  \rightarrow K^{*}\mu^+\mu^-$ decays in $pp$ collisions at $\sqrt{s}= 8$ TeV
  with the ATLAS detector}},  {\em JHEP} {\bf 10} (2018) 047,
  [\href{http://arxiv.org/abs/1805.04000}{{\tt arXiv:1805.04000}}].

\bibitem{Aaij:2020nrf}
{\bf LHCb} Collaboration, R.~Aaij et~al., {\it {Measurement of $CP$-Averaged
  Observables in the $B^{0}\rightarrow K^{*0}\mu^{+}\mu^{-}$ Decay}},  {\em
  Phys. Rev. Lett.} {\bf 125} (2020), no.~1 011802,
  [\href{http://arxiv.org/abs/2003.04831}{{\tt arXiv:2003.04831}}].

\bibitem{Aaij:2020ruw}
{\bf LHCb} Collaboration, R.~Aaij et~al., {\it {Angular analysis of the
  $B^{+}\rightarrow K^{\ast+}\mu^{+}\mu^{-}$ decay}},
  \href{http://arxiv.org/abs/2012.13241}{{\tt arXiv:2012.13241}}.

\bibitem{Aaij:2015esa}
{\bf LHCb} Collaboration, R.~Aaij et~al., {\it {Angular analysis and
  differential branching fraction of the decay $B^0_s\to\phi\mu^+\mu^-$}},
  {\em JHEP} {\bf 09} (2015) 179, [\href{http://arxiv.org/abs/1506.08777}{{\tt
  arXiv:1506.08777}}].

\bibitem{Aaij:2015xza}
{\bf LHCb} Collaboration, R.~Aaij et~al., {\it {Differential branching fraction
  and angular analysis of $\Lambda^{0}_{b} \rightarrow \Lambda \mu^+\mu^-$
  decays}},  {\em JHEP} {\bf 06} (2015) 115,
  [\href{http://arxiv.org/abs/1503.07138}{{\tt arXiv:1503.07138}}]. [Erratum:
  JHEP 09, 145 (2018)].

\bibitem{Aaij:2018gwm}
{\bf LHCb} Collaboration, R.~Aaij et~al., {\it {Angular moments of the decay
  $\Lambda_b^0 \rightarrow \Lambda \mu^{+} \mu^{-}$ at low hadronic recoil}},
  {\em JHEP} {\bf 09} (2018) 146, [\href{http://arxiv.org/abs/1808.00264}{{\tt
  arXiv:1808.00264}}].

\bibitem{Straub:2018kue}
D.~M. Straub, {\it {flavio: a Python package for flavour and precision
  phenomenology in the Standard Model and beyond}},
  \href{http://arxiv.org/abs/1810.08132}{{\tt arXiv:1810.08132}}.

\bibitem{Aebischer:2018iyb}
J.~Aebischer, J.~Kumar, P.~Stangl, and D.~M. Straub, {\it {A Global Likelihood
  for Precision Constraints and Flavour Anomalies}},  {\em Eur. Phys. J. C}
  {\bf 79} (2019), no.~6 509, [\href{http://arxiv.org/abs/1810.07698}{{\tt
  arXiv:1810.07698}}].

\bibitem{Alguero:2018nvb}
M.~Alguer\'o, B.~Capdevila, S.~Descotes-Genon, P.~Masjuan, and J.~Matias, {\it
  {Are we overlooking lepton flavour universal new physics in $b\to
  s\ell\ell$?}},  {\em Phys. Rev. D} {\bf 99} (2019), no.~7 075017,
  [\href{http://arxiv.org/abs/1809.08447}{{\tt arXiv:1809.08447}}].

\bibitem{Bernlochner:2021vlv}
F.~U. Bernlochner, M.~F. Sevilla, D.~J. Robinson, and G.~Wormser, {\it
  {Semitauonic $b$-hadron decays: A lepton flavor universality laboratory}},
  \href{http://arxiv.org/abs/2101.08326}{{\tt arXiv:2101.08326}}.

\bibitem{Amhis:2019ckw}
{\bf HFLAV} Collaboration, Y.~S. Amhis et~al., {\it {Averages of $b$-hadron,
  $c$-hadron, and $\tau$-lepton properties as of 2018}},
  \href{http://arxiv.org/abs/1909.12524}{{\tt arXiv:1909.12524}}.

\bibitem{Zyla:2020zbs}
{\bf Particle Data Group} Collaboration, P.~A. Zyla et~al., {\it {Review of
  Particle Physics}},  {\em PTEP} {\bf 2020} (2020), no.~8 083C01.

\bibitem{Alpigiani:2017lpj}
C.~Alpigiani et~al., {\it {Unitarity Triangle Analysis in the Standard Model
  and Beyond}},  in {\em {5th Large Hadron Collider Physics Conference}}, 10,
  2017.
\newblock \href{http://arxiv.org/abs/1710.09644}{{\tt arXiv:1710.09644}}.

\bibitem{DiLuzio:2018zxy}
L.~Di~Luzio, J.~Fuentes-Martin, A.~Greljo, M.~Nardecchia, and S.~Renner, {\it
  {Maximal Flavour Violation: a Cabibbo mechanism for leptoquarks}},  {\em
  JHEP} {\bf 11} (2018) 081, [\href{http://arxiv.org/abs/1808.00942}{{\tt
  arXiv:1808.00942}}].

\bibitem{Amhis:2016xyh}
{\bf HFLAV} Collaboration, Y.~Amhis et~al., {\it {Averages of $b$-hadron,
  $c$-hadron, and $\tau$-lepton properties as of summer 2016}},  {\em Eur.
  Phys. J. C} {\bf 77} (2017), no.~12 895,
  [\href{http://arxiv.org/abs/1612.07233}{{\tt arXiv:1612.07233}}].

\bibitem{Aaij:2017xqt}
{\bf LHCb} Collaboration, R.~Aaij et~al., {\it {Search for the decays
  $B_s^0\to\tau^+\tau^-$ and $B^0\to\tau^+\tau^-$}},  {\em Phys. Rev. Lett.}
  {\bf 118} (2017), no.~25 251802, [\href{http://arxiv.org/abs/1703.02508}{{\tt
  arXiv:1703.02508}}].

\bibitem{Bobeth:2013uxa}
C.~Bobeth, M.~Gorbahn, T.~Hermann, M.~Misiak, E.~Stamou, and M.~Steinhauser,
  {\it {$B_{s,d} \to l^+ l^-$ in the Standard Model with Reduced Theoretical
  Uncertainty}},  {\em Phys. Rev. Lett.} {\bf 112} (2014) 101801,
  [\href{http://arxiv.org/abs/1311.0903}{{\tt arXiv:1311.0903}}].

\bibitem{TheBaBar:2016xwe}
{\bf BaBar} Collaboration, J.~P. Lees et~al., {\it {Search for
  $B^{+}\rightarrow K^{+} \tau^{+}\tau^{-}$ at the BaBar experiment}},  {\em
  Phys. Rev. Lett.} {\bf 118} (2017), no.~3 031802,
  [\href{http://arxiv.org/abs/1605.09637}{{\tt arXiv:1605.09637}}].

\bibitem{Cornella:2019hct}
C.~Cornella, J.~Fuentes-Martin, and G.~Isidori, {\it {Revisiting the vector
  leptoquark explanation of the B-physics anomalies}},  {\em JHEP} {\bf 07}
  (2019) 168, [\href{http://arxiv.org/abs/1903.11517}{{\tt arXiv:1903.11517}}].

\bibitem{Aaij:2019okb}
{\bf LHCb} Collaboration, R.~Aaij et~al., {\it {Search for the
  lepton-flavour-violating decays $B^{0}_{s}\to\tau^{\pm}\mu^{\mp}$ and
  $B^{0}\to\tau^{\pm}\mu^{\mp}$}},  {\em Phys. Rev. Lett.} {\bf 123} (2019),
  no.~21 211801, [\href{http://arxiv.org/abs/1905.06614}{{\tt
  arXiv:1905.06614}}].

\bibitem{Lees:2012zz}
{\bf BaBar} Collaboration, J.~P. Lees et~al., {\it {A search for the decay
  modes $B^{+-} \to h^{+-} \tau^{+-}l$}},  {\em Phys. Rev. D} {\bf 86} (2012)
  012004, [\href{http://arxiv.org/abs/1204.2852}{{\tt arXiv:1204.2852}}].

\bibitem{Miyazaki:2011xe}
{\bf Belle} Collaboration, Y.~Miyazaki et~al., {\it {Search for
  Lepton-Flavor-Violating tau Decays into a Lepton and a Vector Meson}},  {\em
  Phys. Lett. B} {\bf 699} (2011) 251--257,
  [\href{http://arxiv.org/abs/1101.0755}{{\tt arXiv:1101.0755}}].

\bibitem{Bona:2007vi}
{\bf UTfit} Collaboration, M.~Bona et~al., {\it {Model-independent constraints
  on $\Delta F=2$ operators and the scale of new physics}},  {\em JHEP} {\bf
  03} (2008) 049, [\href{http://arxiv.org/abs/0707.0636}{{\tt
  arXiv:0707.0636}}].

\bibitem{Dmix}
{\bf UTfit} Collaboration, L.~Silvestrini, {\it {Flavour constraints on NP}}, .
  [Talk at La Thuile 2018. Slides available at this
  \href{https://agenda.infn.it/event/14377/contributions/24434/attachments/17481/19830/silvestriniLaThuile.pdf}{url}].

\bibitem{Blumlein:1996qp}
J.~Bl{\"u}mlein, E.~Boos, and A.~Kryukov, {\it {Leptoquark pair production in
  hadronic interactions}},  {\em Z. Phys. C} {\bf 76} (1997) 137--153,
  [\href{http://arxiv.org/abs/hep-ph/9610408}{{\tt hep-ph/9610408}}].

\bibitem{Dorsner:2018ynv}
I.~Dor\v{s}ner and A.~Greljo, {\it {Leptoquark toolbox for precision collider
  studies}},  {\em JHEP} {\bf 05} (2018) 126,
  [\href{http://arxiv.org/abs/1801.07641}{{\tt arXiv:1801.07641}}].

\bibitem{Baker:2019sli}
M.~J. Baker, J.~Fuentes-Mart\'\i{}n, G.~Isidori, and M.~K\"onig, {\it {High-
  $p_T$ signatures in vector\textendash{}leptoquark models}},  {\em Eur. Phys.
  J. C} {\bf 79} (2019), no.~4 334,
  [\href{http://arxiv.org/abs/1901.10480}{{\tt arXiv:1901.10480}}].

\bibitem{Mecaj:2020opd}
B.~Mecaj and M.~Neubert, {\it {Effective Field Theory for Leptoquarks}},
  \href{http://arxiv.org/abs/2012.02186}{{\tt arXiv:2012.02186}}.

\bibitem{Sirunyan:2020zbk}
{\bf CMS} Collaboration, A.~M. Sirunyan et~al., {\it {Search for singly and
  pair-produced leptoquarks coupling to third-generation fermions in
  proton-proton collisions at $\sqrt{s} =$ 13 TeV}},
  \href{http://arxiv.org/abs/2012.04178}{{\tt arXiv:2012.04178}}.

\bibitem{Alves:2002tj}
A.~Alves, O.~Eboli, and T.~Plehn, {\it {Stop lepton associated production at
  hadron colliders}},  {\em Phys. Lett. B} {\bf 558} (2003) 165--172,
  [\href{http://arxiv.org/abs/hep-ph/0211441}{{\tt hep-ph/0211441}}].

\bibitem{Hammett:2015sea}
J.~B. Hammett and D.~A. Ross, {\it {NLO Leptoquark Production and Decay: The
  Narrow-Width Approximation and Beyond}},  {\em JHEP} {\bf 07} (2015) 148,
  [\href{http://arxiv.org/abs/1501.06719}{{\tt arXiv:1501.06719}}].

\bibitem{Mandal:2015vfa}
T.~Mandal, S.~Mitra, and S.~Seth, {\it {Single Productions of Colored Particles
  at the LHC: An Example with Scalar Leptoquarks}},  {\em JHEP} {\bf 07} (2015)
  028, [\href{http://arxiv.org/abs/1503.04689}{{\tt arXiv:1503.04689}}].

\bibitem{Haisch:2020xjd}
U.~Haisch and G.~Polesello, {\it {Resonant third-generation leptoquark
  signatures at the Large Hadron Collider}},
  \href{http://arxiv.org/abs/2012.11474}{{\tt arXiv:2012.11474}}.

\bibitem{Buonocore:2020erb}
L.~Buonocore, U.~Haisch, P.~Nason, F.~Tramontano, and G.~Zanderighi, {\it
  {Lepton-Quark Collisions at the Large Hadron Collider}},  {\em Phys. Rev.
  Lett.} {\bf 125} (2020), no.~23 231804,
  [\href{http://arxiv.org/abs/2005.06475}{{\tt arXiv:2005.06475}}].

\bibitem{Greljo:2020tgv}
A.~Greljo and N.~Selimovic, {\it {Lepton-Quark Fusion at Hadron Colliders,
  precisely}},  \href{http://arxiv.org/abs/2012.02092}{{\tt arXiv:2012.02092}}.

\bibitem{Buonocore:2020nai}
L.~Buonocore, P.~Nason, F.~Tramontano, and G.~Zanderighi, {\it {Leptons in the
  proton}},  {\em JHEP} {\bf 08} (2020), no.~08 019,
  [\href{http://arxiv.org/abs/2005.06477}{{\tt arXiv:2005.06477}}].

\bibitem{Schmaltz:2018nls}
M.~Schmaltz and Y.-M. Zhong, {\it {The leptoquark Hunter\textquoteright{}s
  guide: large coupling}},  {\em JHEP} {\bf 01} (2019) 132,
  [\href{http://arxiv.org/abs/1810.10017}{{\tt arXiv:1810.10017}}].

\bibitem{Angelescu:2021lln}
A.~Angelescu, D.~Be\v{c}irevi\'c, D.~A. Faroughy, F.~Jaffredo, and
  O.~Sumensari, {\it {On the single leptoquark solutions to the $B$-physics
  anomalies}},  \href{http://arxiv.org/abs/2103.12504}{{\tt arXiv:2103.12504}}.

\bibitem{Angelescu:2020uug}
A.~Angelescu, D.~A. Faroughy, and O.~Sumensari, {\it {Lepton Flavor Violation
  and Dilepton Tails at the LHC}},  {\em Eur. Phys. J. C} {\bf 80} (2020),
  no.~7 641, [\href{http://arxiv.org/abs/2002.05684}{{\tt arXiv:2002.05684}}].

\bibitem{Greljo:2018tzh}
A.~Greljo, J.~Martin~Camalich, and J.~D. Ruiz-\'Alvarez, {\it {Mono-$\tau$
  Signatures at the LHC Constrain Explanations of $B$-decay Anomalies}},  {\em
  Phys. Rev. Lett.} {\bf 122} (2019), no.~13 131803,
  [\href{http://arxiv.org/abs/1811.07920}{{\tt arXiv:1811.07920}}].

\bibitem{Aad:2020zxo}
{\bf ATLAS} Collaboration, G.~Aad et~al., {\it {Search for heavy Higgs bosons
  decaying into two tau leptons with the ATLAS detector using $pp$ collisions
  at $\sqrt{s}=13$ TeV}},  {\em Phys. Rev. Lett.} {\bf 125} (2020), no.~5
  051801, [\href{http://arxiv.org/abs/2002.12223}{{\tt arXiv:2002.12223}}].

\bibitem{Afik:2018nlr}
Y.~Afik, J.~Cohen, E.~Gozani, E.~Kajomovitz, and Y.~Rozen, {\it {Establishing a
  Search for $b \rightarrow s \ell^{+} \ell^{-}$ Anomalies at the LHC}},  {\em
  JHEP} {\bf 08} (2018) 056, [\href{http://arxiv.org/abs/1805.11402}{{\tt
  arXiv:1805.11402}}].

\bibitem{Marzocca:2020ueu}
D.~Marzocca, U.~Min, and M.~Son, {\it {Bottom-Flavored Mono-Tau Tails at the
  LHC}},  {\em JHEP} {\bf 12} (2020) 035,
  [\href{http://arxiv.org/abs/2008.07541}{{\tt arXiv:2008.07541}}].

\bibitem{Angelescu:2018tyl}
A.~Angelescu, D.~Be\v{c}irevi\'c, D.~A. Faroughy, and O.~Sumensari, {\it
  {Closing the window on single leptoquark solutions to the $B$-physics
  anomalies}},  {\em JHEP} {\bf 10} (2018) 183,
  [\href{http://arxiv.org/abs/1808.08179}{{\tt arXiv:1808.08179}}].

\bibitem{Kumar:2018kmr}
J.~Kumar, D.~London, and R.~Watanabe, {\it {Combined Explanations of the $b \to
  s \mu^+ \mu^-$ and $b \to c \tau^- {\bar\nu}$ Anomalies: a General Model
  Analysis}},  {\em Phys. Rev. D} {\bf 99} (2019), no.~1 015007,
  [\href{http://arxiv.org/abs/1806.07403}{{\tt arXiv:1806.07403}}].

\bibitem{Bordone:2021olx}
M.~Bordone, C.~Cornella, G.~Isidori, and M.~K\"onig, {\it {The LFU Ratio
  $R_\pi$ in the Standard Model and Beyond}},
  \href{http://arxiv.org/abs/2101.11626}{{\tt arXiv:2101.11626}}.

\bibitem{Kou:2018nap}
{\bf Belle-II} Collaboration, W.~Altmannshofer et~al., {\it {The Belle II
  Physics Book}},  {\em PTEP} {\bf 2019} (2019), no.~12 123C01,
  [\href{http://arxiv.org/abs/1808.10567}{{\tt arXiv:1808.10567}}]. [Erratum:
  PTEP 2020, 029201 (2020)].

\bibitem{Bediaga:2018lhg}
{\bf LHCb} Collaboration, R.~Aaij et~al., {\it {Physics case for an LHCb
  Upgrade II - Opportunities in flavour physics, and beyond, in the HL-LHC
  era}},  \href{http://arxiv.org/abs/1808.08865}{{\tt arXiv:1808.08865}}.

\bibitem{Buchalla:1998ba}
G.~Buchalla and A.~J. Buras, {\it {The rare decays $K\to \pi \nu\bar\nu$, $B
  \to X \nu\bar\nu$ and $B \to l^+ l^-$: An Update}},  {\em Nucl. Phys. B} {\bf
  548} (1999) 309--327, [\href{http://arxiv.org/abs/hep-ph/9901288}{{\tt
  hep-ph/9901288}}].

\bibitem{Celis:2017hod}
A.~Celis, J.~Fuentes-Martin, A.~Vicente, and J.~Virto, {\it {DsixTools: The
  Standard Model Effective Field Theory Toolkit}},  {\em Eur. Phys. J. C} {\bf
  77} (2017), no.~6 405, [\href{http://arxiv.org/abs/1704.04504}{{\tt
  arXiv:1704.04504}}].

\bibitem{BelleIIB2Knn}
{\bf Belle II} Collaboration, F.~Dattola, {\it {Search for $B^+\to
  K^+\nu\bar\nu$ decays with an inclusive tagging method at the Belle II
  experiment}}, . [Talk at Moriond EW 2021. Slides available at this
  \href{http://moriond.in2p3.fr/2021/EW/slides/3\_flavour\_10\_dattola.pdf}{url}].

\bibitem{Grygier:2017tzo}
{\bf Belle} Collaboration, J.~Grygier et~al., {\it {Search for $B\to
  h\nu\bar\nu$ decays with semileptonic tagging at Belle}},  {\em Phys. Rev. D}
  {\bf 96} (2017), no.~9 091101, [\href{http://arxiv.org/abs/1702.03224}{{\tt
  arXiv:1702.03224}}]. [Addendum: Phys.Rev.D 97, 099902 (2018)].

\bibitem{Lutz:2013ftz}
{\bf Belle} Collaboration, O.~Lutz et~al., {\it {Search for $B \to h^{(*)} \nu
  \bar{\nu}$ with the full Belle $\Upsilon(4S)$ data sample}},  {\em Phys. Rev.
  D} {\bf 87} (2013), no.~11 111103,
  [\href{http://arxiv.org/abs/1303.3719}{{\tt arXiv:1303.3719}}].

\bibitem{Lees:2013kla}
{\bf BaBar} Collaboration, J.~P. Lees et~al., {\it {Search for $B \to K^{(*)}
  \nu \overline \nu$ and invisible quarkonium decays}},  {\em Phys. Rev. D}
  {\bf 87} (2013), no.~11 112005, [\href{http://arxiv.org/abs/1303.7465}{{\tt
  arXiv:1303.7465}}].

\bibitem{Sirunyan:2019vgj}
{\bf CMS} Collaboration, A.~M. Sirunyan et~al., {\it {Search for high mass
  dijet resonances with a new background prediction method in proton-proton
  collisions at $\sqrt{s} =$ 13 TeV}},  {\em JHEP} {\bf 05} (2020) 033,
  [\href{http://arxiv.org/abs/1911.03947}{{\tt arXiv:1911.03947}}].

\bibitem{Alwall:2014hca}
J.~Alwall, R.~Frederix, S.~Frixione, V.~Hirschi, F.~Maltoni, O.~Mattelaer,
  H.~S. Shao, T.~Stelzer, P.~Torrielli, and M.~Zaro, {\it {The automated
  computation of tree-level and next-to-leading order differential cross
  sections, and their matching to parton shower simulations}},  {\em JHEP} {\bf
  07} (2014) 079, [\href{http://arxiv.org/abs/1405.0301}{{\tt
  arXiv:1405.0301}}].

\bibitem{Alloul:2013bka}
A.~Alloul, N.~D. Christensen, C.~Degrande, C.~Duhr, and B.~Fuks, {\it
  {FeynRules 2.0 - A complete toolbox for tree-level phenomenology}},  {\em
  Comput. Phys. Commun.} {\bf 185} (2014) 2250--2300,
  [\href{http://arxiv.org/abs/1310.1921}{{\tt arXiv:1310.1921}}].

\bibitem{ATL-PHYS-PUB-2015-039}
{\bf ATLAS} Collaboration, {\it {Commissioning of the ATLAS $b$-tagging
  algorithms using $t\bar{t}$ events in early Run-2 data}},  {\em
  ATL-PHYS-PUB-2015-039} (2015).

\bibitem{Sirunyan:2017ezt}
{\bf CMS} Collaboration, A.~M. Sirunyan et~al., {\it {Identification of
  heavy-flavour jets with the CMS detector in pp collisions at 13 TeV}},  {\em
  JINST} {\bf 13} (2018), no.~05 P05011,
  [\href{http://arxiv.org/abs/1712.07158}{{\tt arXiv:1712.07158}}].

\bibitem{Bols:2020bkb}
E.~Bols, J.~Kieseler, M.~Verzetti, M.~Stoye, and A.~Stakia, {\it {Jet Flavour
  Classification Using DeepJet}},  {\em JINST} {\bf 15} (2020), no.~12 P12012,
  [\href{http://arxiv.org/abs/2008.10519}{{\tt arXiv:2008.10519}}].

\bibitem{Aaboud:2018eqg}
{\bf ATLAS} Collaboration, M.~Aaboud et~al., {\it {Measurements of $t\bar{t}$
  differential cross-sections of highly boosted top quarks decaying to
  all-hadronic final states in $pp$ collisions at $\sqrt{s}=13\,$ TeV using the
  ATLAS detector}},  {\em Phys. Rev. D} {\bf 98} (2018), no.~1 012003,
  [\href{http://arxiv.org/abs/1801.02052}{{\tt arXiv:1801.02052}}].

\bibitem{CMS:2020kdq}
{\bf CMS} Collaboration, {\it {Measurement of differential ${\mathrm
  t}\bar{\mathrm t}$ production cross sections for high-$p_{\text{T}}$ top
  quarks in proton-proton collisions at $\sqrt{s} = 13\,\text{TeV}$}},  {\em
  CMS-PAS-TOP-18-013} (2020).

\bibitem{Aaboud:2019zpc}
{\bf ATLAS} Collaboration, M.~Aaboud et~al., {\it {Search for excited electrons
  singly produced in proton\textendash{}proton collisions at $\sqrt{s}=13$ TeV
  with the ATLAS experiment at the LHC}},  {\em Eur. Phys. J. C} {\bf 79}
  (2019), no.~9 803, [\href{http://arxiv.org/abs/1906.03204}{{\tt
  arXiv:1906.03204}}].

\bibitem{Sirunyan:2020awe}
{\bf CMS} Collaboration, A.~M. Sirunyan et~al., {\it {Search for an excited
  lepton that decays via a contact interaction to a lepton and two jets in
  proton-proton collisions at $\sqrt{s} =$ 13 TeV}},  {\em JHEP} {\bf 05}
  (2020) 052, [\href{http://arxiv.org/abs/2001.04521}{{\tt arXiv:2001.04521}}].

\bibitem{Aaboud:2017dmy}
{\bf ATLAS} Collaboration, M.~Aaboud et~al., {\it {Search for supersymmetry in
  final states with two same-sign or three leptons and jets using 36 fb$^{-1}$
  of $\sqrt{s}=13$ TeV $pp$ collision data with the ATLAS detector}},  {\em
  JHEP} {\bf 09} (2017) 084, [\href{http://arxiv.org/abs/1706.03731}{{\tt
  arXiv:1706.03731}}]. [Erratum: JHEP 08, 121 (2019)].

\bibitem{Aad:2020sgw}
{\bf ATLAS} Collaboration, G.~Aad et~al., {\it {Search for a scalar partner of
  the top quark in the all-hadronic $t{\bar{t}}$ plus missing transverse
  momentum final state at $\sqrt{s}=13$ TeV with the ATLAS detector}},  {\em
  Eur. Phys. J. C} {\bf 80} (2020), no.~8 737,
  [\href{http://arxiv.org/abs/2004.14060}{{\tt arXiv:2004.14060}}].

\bibitem{Aad:2021oos}
{\bf ATLAS} Collaboration, G.~Aad et~al., {\it {Search for bottom-squark pair
  production in $pp$ collision events at $\sqrt{s} = 13$ TeV with hadronically
  decaying $\tau$-leptons, $b$-jets and missing transverse momentum using the
  ATLAS detector}},  \href{http://arxiv.org/abs/2103.08189}{{\tt
  arXiv:2103.08189}}.

\bibitem{Sirunyan:2020tyy}
{\bf CMS} Collaboration, A.~M. Sirunyan et~al., {\it {Search for top squark
  pair production using dilepton final states in ${\text {p}}{\text {p}}$
  collision data collected at $\sqrt{s}=13\,\text {TeV} $}},  {\em Eur. Phys.
  J. C} {\bf 81} (2021), no.~1 3, [\href{http://arxiv.org/abs/2008.05936}{{\tt
  arXiv:2008.05936}}].

\bibitem{Sirunyan:2019kgv}
{\bf CMS} Collaboration, A.~M. Sirunyan et~al., {\it {Search for top squark
  pair production in a final state with two tau leptons in proton-proton
  collisions at $ \sqrt{s} =$ 13 TeV}},  {\em JHEP} {\bf 02} (2020) 015,
  [\href{http://arxiv.org/abs/1910.12932}{{\tt arXiv:1910.12932}}].

\bibitem{DiLuzio:2017chi}
L.~Di~Luzio and M.~Nardecchia, {\it {What is the scale of new physics behind
  the $B$-flavour anomalies?}},  {\em Eur. Phys. J. C} {\bf 77} (2017), no.~8
  536, [\href{http://arxiv.org/abs/1706.01868}{{\tt arXiv:1706.01868}}].

\bibitem{Barbieri:2021wrc}
R.~Barbieri, {\it {A view of flavour physics in 2021}},
  \href{http://arxiv.org/abs/2103.15635}{{\tt arXiv:2103.15635}}.

\bibitem{Geng:2021nhg}
L.-S. Geng, B.~Grinstein, S.~J\"ager, S.-Y. Li, J.~Martin~Camalich, and R.-X.
  Shi, {\it {Implications of new evidence for lepton-universality violation in
  $b\to s\ell^+\ell^-$ decays}},  \href{http://arxiv.org/abs/2103.12738}{{\tt
  arXiv:2103.12738}}.

\bibitem{Altmannshofer:2021qrr}
W.~Altmannshofer and P.~Stangl, {\it {New Physics in Rare B Decays after
  Moriond 2021}},  \href{http://arxiv.org/abs/2103.13370}{{\tt
  arXiv:2103.13370}}.

\bibitem{Carvunis:2021jga}
A.~Carvunis, F.~Dettori, S.~Gangal, D.~Guadagnoli, and C.~Normand, {\it {On the
  effective lifetime of $B_s \to \mu \mu \gamma$}},
  \href{http://arxiv.org/abs/2102.13390}{{\tt arXiv:2102.13390}}.

\bibitem{Hiller:2021pul}
G.~Hiller, D.~Loose, and I.~Ni\v{s}and\v{z}i\'c, {\it {Flavorful leptoquarks at
  the LHC and beyond: Spin 1}},  \href{http://arxiv.org/abs/2103.12724}{{\tt
  arXiv:2103.12724}}.

\bibitem{Chivukula:1987py}
R.~S. Chivukula and H.~Georgi, {\it {Composite Technicolor Standard Model}},
  {\em Phys. Lett. B} {\bf 188} (1987) 99--104.

\bibitem{Barbieri:2012uh}
R.~Barbieri, D.~Buttazzo, F.~Sala, and D.~M. Straub, {\it {Flavour physics from
  an approximate $U(2)^3$ symmetry}},  {\em JHEP} {\bf 07} (2012) 181,
  [\href{http://arxiv.org/abs/1203.4218}{{\tt arXiv:1203.4218}}].

\bibitem{Blankenburg:2012nx}
G.~Blankenburg, G.~Isidori, and J.~Jones-Perez, {\it {Neutrino Masses and LFV
  from Minimal Breaking of $U(3)^5$ and $U(2)^5$ flavor Symmetries}},  {\em
  Eur. Phys. J. C} {\bf 72} (2012) 2126,
  [\href{http://arxiv.org/abs/1204.0688}{{\tt arXiv:1204.0688}}].

\bibitem{DAmbrosio:2002vsn}
G.~D'Ambrosio, G.~F. Giudice, G.~Isidori, and A.~Strumia, {\it {Minimal flavor
  violation: An Effective field theory approach}},  {\em Nucl. Phys. B} {\bf
  645} (2002) 155--187, [\href{http://arxiv.org/abs/hep-ph/0207036}{{\tt
  hep-ph/0207036}}].

\bibitem{Bouchard:2013eph}
{\bf HPQCD} Collaboration, C.~Bouchard, G.~P. Lepage, C.~Monahan, H.~Na, and
  J.~Shigemitsu, {\it {Rare decay $B \to K \ell^+ \ell^-$ form factors from
  lattice QCD}},  {\em Phys. Rev. D} {\bf 88} (2013), no.~5 054509,
  [\href{http://arxiv.org/abs/1306.2384}{{\tt arXiv:1306.2384}}]. [Erratum:
  Phys.Rev.D 88, 079901 (2013)].

\end{thebibliography}\endgroup
}

\end{document}